\documentclass[preprint,3p,times]{elsarticle}

\usepackage{amssymb}
\usepackage{amsthm}
\usepackage{amsmath}
\usepackage{epsfig}
\usepackage{subfigure}
\usepackage[usenames]{color}
\usepackage{algorithm}
\usepackage{algorithmic}
\usepackage{nicefrac}
\usepackage{caption}
\epsfverbosetrue
\graphicspath{{fig/}{../fig/}}

\newcommand{\mb}[1]{\ensuremath{\mathbf{#1}}}

\journal{Annual Review of Heat Transfer}

\begin{document}
 \setlength{\parindent}{0.0ex}
 \setcounter{secnumdepth}{4}
 \setcounter{tocdepth}{4}
\begin{frontmatter}



\title{Thermophysical Phenomena in Metal Additive Manufacturing by Selective Laser Melting: Fundamentals, Modeling, Simulation and Experimentation}


\author[mechanosynth]{Christoph Meier\corref{cor1}}
\ead{chrmeier@mit.edu}
\author[mechanosynth]{Ryan W. Penny}
\author[mechanosynth]{Yu Zou}
\author[mechanosynth]{Jonathan S. Gibbs}
\author[mechanosynth]{A. John Hart\corref{cor1}}
\ead{ajhart@mit.edu}

\address[mechanosynth]{Mechanosynthesis Group, Department of Mechanical Engineering, Massachusetts Institute of Technology, 77 Massachusetts Avenue, Cambridge, 02139, MA, USA}

\cortext[cor1]{Corresponding authors}

\begin{abstract}
Among the many  additive manufacturing (AM) processes for metallic materials, selective laser melting (SLM) is arguably the most versatile in terms of its potential to realize complex geometries along with tailored microstructure.  However, the complexity of the SLM process, and the need for predictive relation of powder and process parameters to the part properties, demands further development of computational and experimental methods. This review addresses the fundamental physical phenomena of SLM, with a special emphasis on the associated thermal behavior. Simulation and experimental methods are discussed according to three primary categories.  First, macroscopic approaches aim to answer questions at the component level and consider for example the determination of residual stresses or dimensional distortion effects prevalent in SLM.  Second, mesoscopic approaches focus on the detection of defects such as excessive surface roughness, residual porosity or inclusions that occur at the mesoscopic length scale of individual powder particles. Third, microscopic approaches investigate the metallurgical microstructure evolution resulting from the high temperature gradients and extreme heating and cooling rates induced by the SLM process. Consideration of physical phenomena on all of these three length scales is mandatory to establish the understanding needed to realize high part quality in many applications, and to fully exploit the potential of SLM and related metal AM processes.


%
%
%
%

\end{abstract}
\end{frontmatter}

%

%
%
\section{Introduction}
\label{sec:intro}
%
%

\begin{figure}[h!!!]
\begin{centering}
\includegraphics[width=1.0\textwidth]{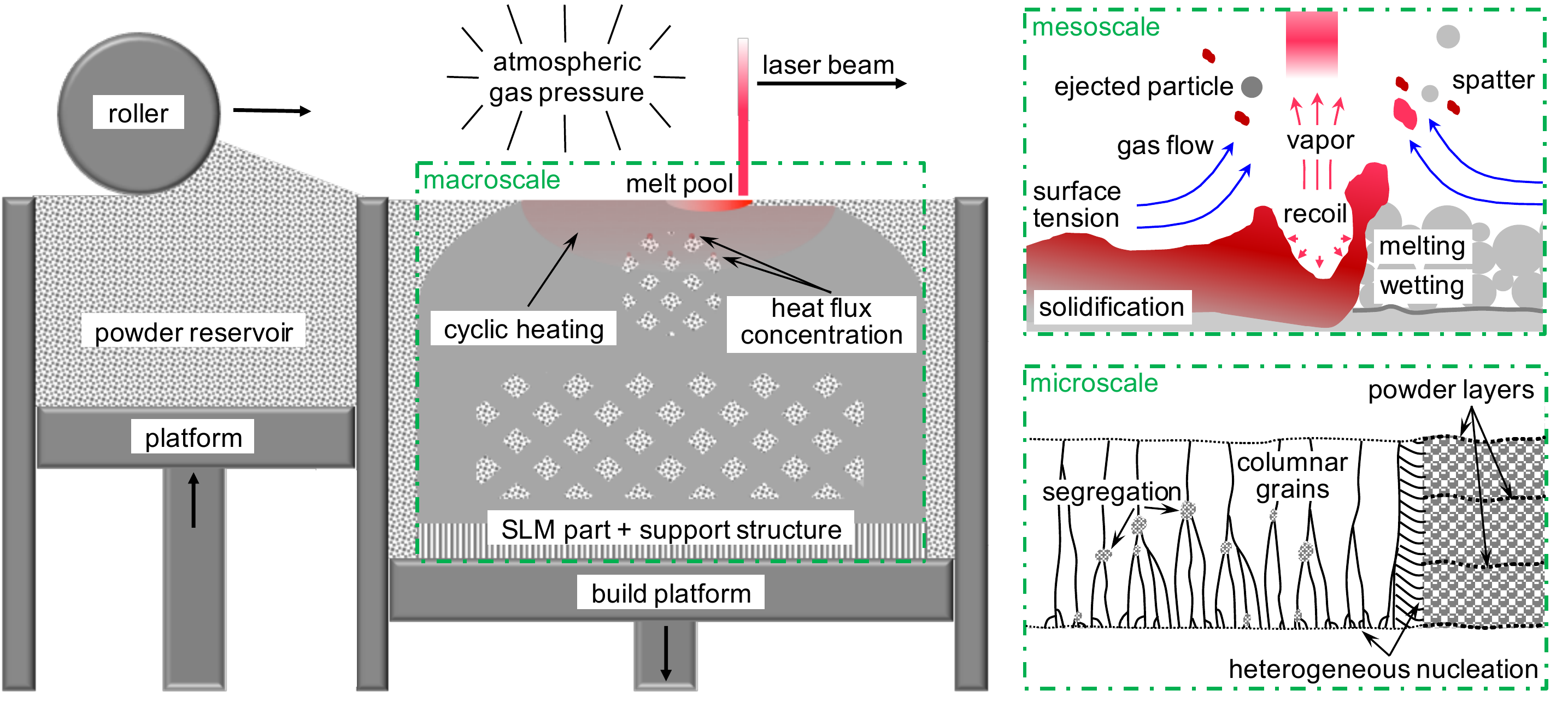}
\caption{Experimental setup of SLM process as well as schematic visualization of macroscale, mesoscale and microscale view.}
\label{fig:Fig_zy1}
\end{centering}
\end{figure}

Additive manufacturing (AM) offers the opportunity to produce parts with high geometric complexity, without the requirement for dedicated tooling. AM processes for polymer parts are quite well established, yet melt-based processes for AM of metals still exhibit severe practical challenges, many of them resulting from the high melting temperatures of metals and their relatively low viscosities~\cite{Lipson2013}. In selective laser melting (SLM) of metals, the arguably most prominent representative of powder bed AM methods, a 3D manufacturing task is digitally segmented into thin 2D layers~\cite{Gibson2010}. A solid part is simply formed by selectively melting pre-defined contours in successive layers of powder using a focused laser beam. After one layer of powder has been scanned, the regions melted by the laser form the cross-section of the final part. Subsequently, the underlying build platform is lowered down and a further layer of powder is deposited by means of a powder coater mechanism. This procedure is successively repeated until the final 3D geometry is completed and the remaining unfused powder is then removed (see Figure~\ref{fig:Fig_zy1}). In this context, the so-called build direction denotes the direction normal to the powder bed. Further, the orientation of the part geometry with respect to this (vertical) build direction has crucial influence on the resulting part properties~\cite{Hanzl2015}.\\

\begin{figure}[h!!]
 \centering
   {
    \includegraphics[width=1\textwidth]{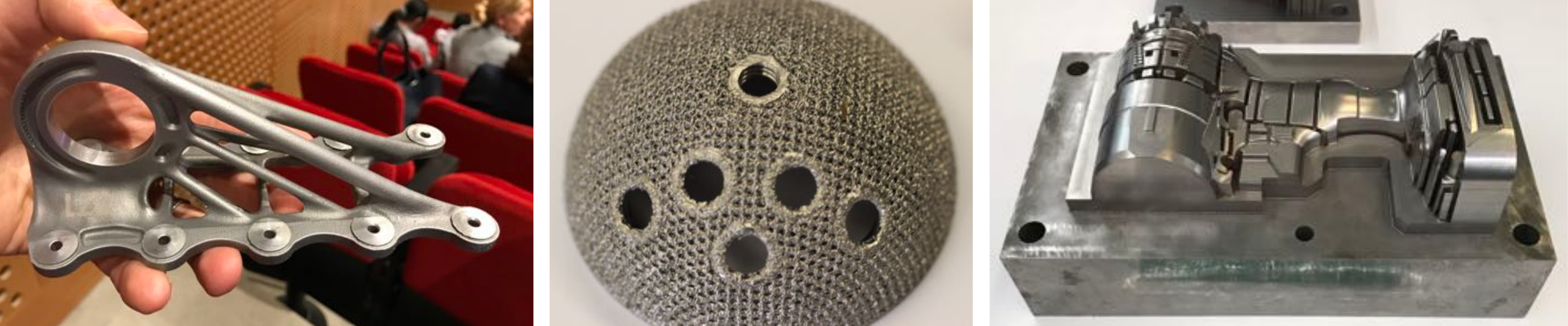}
    \label{fig:lattice}
   }
  \caption{Exemplary metal parts fabricated by SLM/EBM: (a) aircraft bracket (approx. 150 mm width), by Airbus; (b) acetibular cup for human hip implant (approx. 75 mm diameter), by Arcam; (c) injection mold tool with internal cooling passages (approx. 200 mm width), by Sodick. The parts are post-processed, including support removal, machining, and polishing, in addition to the SLM/EBM process as the construction method.}
  \label{fig:SLM_applications}
\end{figure}

SLM of metals offers significant advantages including near-net-shape production without the need for expensive molds or time consuming post-processing, high material
utilization rate and highest production flexibility~\cite{Thijs2010,Stucker2011}. Most essentially, the layer-wise production leads to nearly unlimited freedom of design, which enables the generation of highly complex geometries that cannot be obtained by conventional manufacturing processes. This paradigm shift in mechanical design allows to integrate complex substructures such as lattice-based geometries enabling lightweight yet sufficiently stiff components. Possible fields of application include aerospace or medical engineering and basically all industries requiring highly complex and individualized parts (see~Figure~\ref{fig:SLM_applications}).\\


\begin{figure}[h!!]
 \centering
   {
    \includegraphics[width=0.78\textwidth]{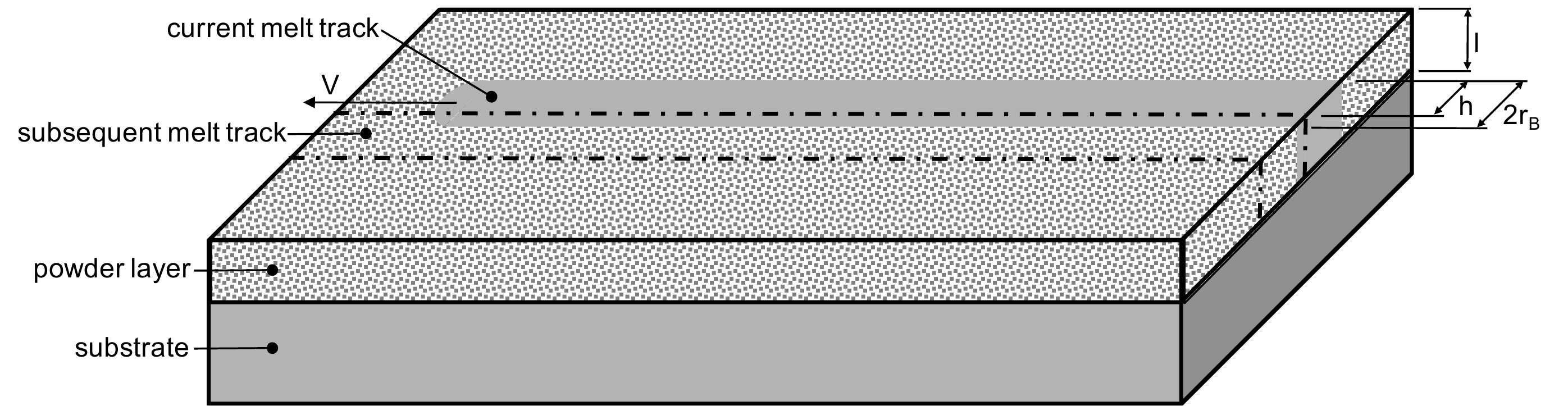}
    \label{fig:lattice}
   }
   {
    \includegraphics[width=0.19\textwidth]{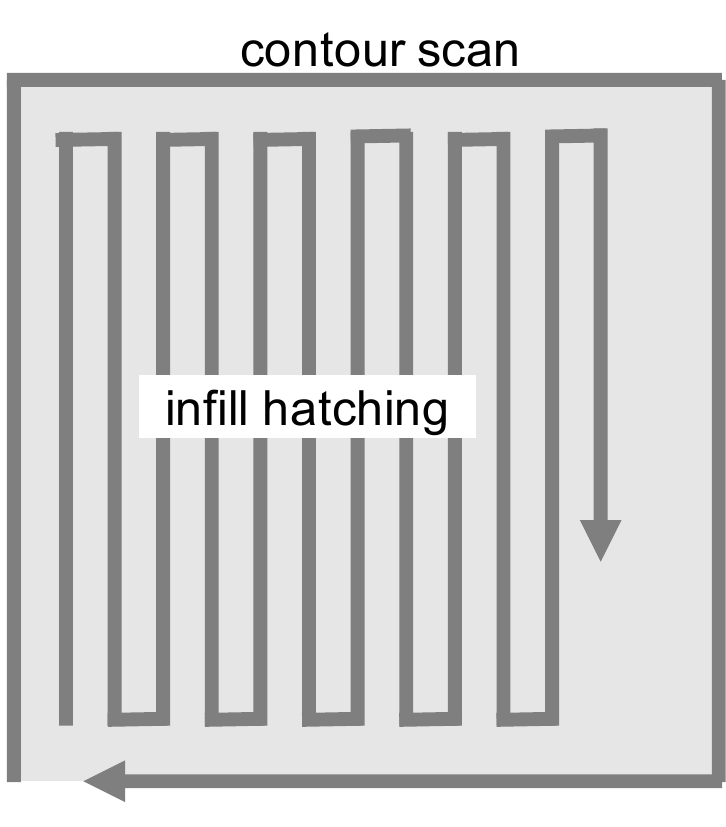}
    \label{fig:hip}
   }
\caption{Typical geometrical dimensions and process parameters (left) as well as typical scan pattern (right) characterizing the SLM process.}
\label{Fig_Thomas_2_2016}
\end{figure}

The overall SLM process is highly complex and governed by a variety of (competing) physical mechanisms. The most important effects and physical phenomena occurring in the powder bed, the melt pool and the solidified phase of typical SLM systems are summarized in the following three subsections, as well as in  Figure~\ref{fig:Fig_zy1}. A typical schematic parametrization of the powder layer and laser beam as well as a typical local scan pattern are shown in  Figure~\ref{Fig_Thomas_2_2016}.

\subsection{Physical phenomena within the powder bed}
\label{sec:intro_powder}

The incident laser beam is a collimated, polarized, monochromatic electro-magnetic wave, with a wavelength in the range of $w \!\approx\! 1 \mu m$. The spatial power density distribution of incident radiation on the powder bed is commonly assumed to follow a Gaussian distribution, with the associated $2 \sigma$ value typically being taken as the laser beam spot size. Typical values for the nominal laser power and the employed laser beam velocities are in the range of $P \!\approx\! 50\!-\!1000W$ and $V \!\approx\! 0.1\!-\!3m\!/\!s$~\cite{Thomas2016}. The effective laser beam absorption within the powder bed is governed by multiple reflections of incident laser rays within the open-pore system of the powder bed, each with partial absorption of the incident radiation. The laser beam can penetrate to considerable depths, which can even reach the range of the powder layer thickness~\cite{Gusarov2009,Gusarov2005}. Thus the net absorptivity of powder beds is considerably higher than the value known for flat surfaces and, moreover, the laser beam energy source must to be thought of a volumetric heat source distributed over the powder bed thickness, as opposed to a surface heat source. The factors influencing overall absorption and local energy distribution are numerous, including the laser beam power, wavelength, polarization, angle of incidence, powder temperature, surface roughness, surface chemistry (e.g. oxidation) and contamination~\cite{Boley2015,Khairallah2014}. Issues of powder bed morphology, determined by particle shape, size distribution and packing density, are also central to radiative transfer. A further important factor is given by the intra- and inter-particle heat transfer within the powder bed. The inter-particle heat transfer is typically governed by the gas in the powder bed pores, with commonly negligible overall conductivity contributions from particle-to-particle contact points as long as loose, i.e. not mechanically compressed, powder layers are considered. Consequently, the thermal conductivity of loose powder is comparable to the conductivity of gas and by orders of magnitude smaller than the conductivity in the solidified phase~\cite{Rombouts2005}. Also when considering the intra-particle heat transfer, it can be observed that the time scales governing this process are typically larger than the time scales governing particle melting. In other words, under typical SLM process conditions, there is not enough time for conductive homogenization of non-uniform energy and temperature distributions across the powder bed but also across individual particles. As consequence, partially molten particles may cause defects such as pores or inclusions~\cite{Boley2015}. A further characteristic of the SLM process is that the powder bed, when considering it as a homogenized continuum, shows \textit{mesoscopic} heterogeneities (in form of individual particles) that are in the same order of magnitude as relevant \textit{macroscopic} process length scales such as powder layer thickness and laser beam spot size. Typically, powder particle sizes in the range of $2R \!\approx\! 10\mu m \!-\! 50\mu m$, layer thicknesses in the range of $l \!\approx\! 20\mu m \!-\! 100\mu m$ and laser beam spot sizes in the range of $2r_B \!\approx\! 20\mu m \!-\! 200\mu m$ are employed. Also, in standard linear scan patterns the distance between two successive laser tracks, denoted as hatch spacing $h$, is an important process parameter. This is typically chosen in the range of $r_B$ such that a sufficient overlap and remelting between two subsequent tracks is guaranteed (Figure~\ref{Fig_Thomas_2_2016}). A good overview of the typically applied range of these process parameters is also given in~\cite{Thomas2016}.\\

The large size of individual powder grains as compared to powder layer thickness and laser beam spot size typically leads to non-uniform energy distributions, across the entire powder bed but also across individual particles, which may have considerable influence on the resulting melting behavior and melt pool hydrodynamics. Furthermore, these comparatively large heterogeneities cause differences in the resulting temperature fields and melt track shapes when considering different samples of stochastically equivalent powder layers (i.e. identical particle shape and size distribution, powder layer density and thickness etc.). Consequently, the variance of process results due to the stochastic nature of the powder layer is considerably greater than for comparable processes such as laser beam welding (LBW, see below). Besides the energy input by the laser beam, also possible energy losses in form of thermal radiation emission, thermal convection or heat conduction from the solidified material to the underlying built platform play an important role in the overall SLM process. For further information on powder bed radiation and heat transfer in the context of SLM, the interested reader is referred to~\cite{Rombouts2005,Gusarov2005,Gusarov2008,Boley2015}.

\subsection{Physical phenomena within the melt pool}
\label{sec:intro_melt}

As soon as the melting temperature is reached at local positions on the powder grain surface, the phase transition from solid to liquid as well as the formation of a melt pool and, ideally, a continuous melt track is induced. Driven by surface tension and capillary forces tending to minimize surface energy, a coalescence of individual melt drops and a reshaping of the resulting melt pool is initiated~\cite{Korner2011,Korner2013}. In addition, the wetting behavior of the low-viscosity melt on the underlying substrate, i.e. the solidified material formed by previous layers, and surrounding powder grains influences the resulting melt pool shape, continuity and adhesion to the previous layer. The wetting behavior crucially depends on the material, temperature, surface roughness, and surface chemistry. Oxidation on the powder grain or substrate surfaces - either due to contaminated primary powder material or due to thermally induced oxidation during the process - is known to considerably decrease the wetting behavior of the melt which might result in instable, balled melt pools and rough surfaces, pores or delamination due to insufficient layer-to-layer adhesion~\cite{Das2003}.\\

The prevalent length and time scales essentially determine which physical effects govern the process and which are negligible. Typically, viscous and gravity forces can be considered as secondary effects while surface tension and capillary forces, wetting behavior but also inertia effects are the primary driving forces that influence the melt pool dynamics and shape as well as the surrounding powder morphology by attracting or rejecting individual grains~\cite{Korner2013}. The heat transfer within the melt pool is governed by convection rather than by heat conduction, with Marangoni convection, i.e. melt flow from hot to cool regions induced by temperature-dependent surface tension, playing a prominent role in this process~\cite{Khairallah2016}. Depending on the amount of absorbed energy density and the surrounding atmospheric pressure, the peak temperature within the melt pool might exceed the boiling temperature and considerable material evaporation may take place~\cite{Gusarov2007}. The evaporation itself as well as the gas flow induced by evaporation may influence the melt pool thermo-hydrodynamics and the overall process as consequence of an evaporative mass loss and additional cooling, of a recoil pressure considerably distorting the melt pool surface and representing a means of transport for potential pollutants, of melt drops spattered out of the pool and even of powder particles ejected away from the direct vicinity of the laser beam~\cite{Khairallah2016}. Driven by the so-called keyholing mechanism, cavitation resulting from evaporated material might considerably contribute to the overall material porosity or even burst the surrounding solidified material due to thermal expansion of trapped gas~\cite{Liu2016}. As soon as the melt pool has solidified, the evolution of possible defects on the mesoscopic scale is virtually established. A discussion of the physical phenomena governing the melt pool thermo-hydrodynamics can e.g. be found in~\cite{Das2003,Kruth2005,Khairallah2016,Korner2013,Lee2015}.

\subsection{Physical phenomena within the solidified phase}
\label{sec:intro_solid}

With the solidification of the melt pool, the development of the metallurgical microstructure, crucially determining the macroscopic properties of the final part, begins. The evolution of the solid-phase microstructure characterized by grain size, grain shape (morphology) and grain orientation (texture) are governed by the prevalent spatial temperature gradients, the cooling rates, as well as the velocity of the solidification front~\cite{Glcksman2010}. In SLM processes, two regimes can be distinguished: The first regime is given by the temperature field in the direct vicinity of the laser beam, in the so-called heat affected zone (HAZ)~\cite{Yadroitsava2015}, which is controlled by highly complex mechanisms such as radiation absorption and heat conduction in the powder bed as well as convective heat transfer within the melt pool with all the individual physical phenomena and process parameters of influence as discussed in the two previous paragraphs. The material in this region is subject to a rapid heating above melting temperature due to the absorption of laser energy by powder grains, a high velocity of the melt pool front induced by the laser beam velocity as well as a rapid solidification of the molten material after the heat source has moved on, which is a direct consequence of the large ratio of solid material to hot molten material. These pronounced non-equilibrium conditions lead to meta-stable microstructures and compositions of the resulting phases, as well as smaller grain sizes - which typically result in higher material strengths - as compared to traditional melt-based manufacturing processes such as casting~\cite{Herzog2016,Hanzl2015}. A second regime of thermal evolution is prevalent in previously deposited material layers located below the current layer and further away from the heat source. Each solidified location experiences repeated heating and cooling cycles with decreasing amplitudes as the laser processes adjacent scan tracks, and as it processes consecutive new layers. The heat transfer in this regime is rather determined by global part properties, e.g. the global laser beam scanning strategy, the build direction, the fixation of the part on the built platform, the temperature of the built platform but also by the part porosity and the metallurgic microstructure distribution itself, which both influence the (effective) thermal conductivity.\\ 

Of course, microstructure also depends on the specific part geometry, e.g. due to heat flux concentration at the transition region from bulk material to slender columns or thin walls (see e.g. Figure~\ref{fig:Fig_zy1}), which are surrounded by the low conductivity unfused powder~\cite{Mercelis2006,Lu2015}. Also the evolution of columnar grain structures oriented in direction of the main temperature gradients, usually in build direction, is typical for SLM processes and often yields a strongly anisotropic macroscopic material behavior with higher material strength in the build direction~\cite{Herzog2016,Gong2015}. With increasing distance from the top powder layer, the maximal temperature values and gradients experienced by a material layer during the repeated thermal cycles decrease and, similar to a heat treatment, these cycles might lead to a coarsening of the microstructure, a reduction of brittle non-equilibrium phases and, consequently, to more ductile material characteristics. As consequence of these effects, the final part will typically exhibit a change of microstructure in built direction: On the one hand, the initial creation of fine grain structures and non-equilibrium phases will be more pronounced in the first material layers deposited in the direct vicinity of the build platform where higher thermal conductivity and faster cooling rates are prevalent. On the other hand, these initially deposited powder layers are exposed to the heat treatment of repeated heating and cooling cycles for longer times, which might lead to longer evolution times for solid phase transformations and grain coarsening.\\

Apart from the microstructure evolution considered so far, the high temperature gradients in the direct vicinity of the melt pool, but also at locations of heat flux concentration, are giving rise to considerable thermal strains induced by a successive thermal expansion and shrinkage of material. These thermal strains result in thermal stresses within the kinematically constrained SLM part. The magnitude of these stresses is essentially determined by the underlying solid microstructure and the resulting macroscopic material behavior. A ductile material can compensate these thermal strain variations by means of local plastic flow. On the contrary, brittle material behavior fostered by small grain sizes, the existence of certain non-equilibrium phases or a local segregation of alloying elements~\cite{Thijs2010} might result in cracks at locations of stress concentration such as residual pores or inclusions. The complexity of this thermo-mechanical coupling is further increased by the fact that the microstructure does not only influence the amount of residual stresses, but that also the prevalence of residual stresses, which are for example known to mechanically stabilize the metastable austenitic phase in steels~\cite{Herzog2016}, will influence the evolution of the microstructure. In order to reduce residual stresses on large surface areas, these surfaces are typically subdivided in smaller islands that are completed successively. Figure~\ref{Fig_Thomas_2_2016} (right) illustrates the processing of a single island consisting of contour scan and infill hatching. During the SLM process, the amplitude of residual stresses might also decrease since stress relaxation is likely to occur during the repeated heating and cooling cycles at lower temperature levels. Furthermore, annealing is typically applied to the final part before removing support structures in order to relief residual stresses. While a neglect of support structures would result in reduced residual stresses, this advantage has to be paid for by dimensional warping as often observed in parts made by SLM. For further information on material aspects and microstructure evolution in the context of SLM, the interested reader is referred to~\cite{King2015, Bourell2016, Hebert2016, Herzog2016, Sames2016}. Exemplary references focusing on the investigation of the residual stresses resulting during the SLM process are~\cite{Hodge2014,Hodge2016,Denlinger2014,Riedlbauer2014,Yadroitsava2015,Lu2015}.\\

All in all, it can be concluded that the entire thermal history between solidification and cooling down to the ambient temperature, governed by many heating and cooling cycles at different temperature levels and time scales, considerably determines the resulting metallurgic microstructure as well as the macroscopically observable material properties such as ductility, micro hardness, yield strength or tensile strength as well as their spatial distribution in a possibly inhomogeneous and anisotropic manner. On the one hand, a restriction of the overall process parameters is required in order to avoid undesirable material characteristics and possible part defects such as excessive residual stresses, dimensional warping, crack propagation or delamination of layers, effects which might destroy the SLM part during the build process or at least reduce the mechanical resilience of the final part considerably. In order to fully exploit the efficiency potential of SLM, this result has to be achieved without the need for time- and cost-intensive post-processing as required by alternative processes such as selective laser sintering (SLS, see below). On the other hand, the flexibility of the SLM process offers the unique opportunity to optimize process parameters in order to manufacture parts with prescribed inhomogeneous and anisotropic microstructures and macroscopic material properties in a controlled manner, contributing to the paradigm shift in design enabled by SLM.

\subsection{Differentiation of related additive manufacturing processes for metals}
\label{sec:intro_relatedprocesses}

References and comparisons to other powder-bed metal AM processes such as electron beam melting (EBM) and selective laser sintering (SLS) will be useful from time to time in the foregoing discussion ~\cite{Kruth2005,Kruth2007,Gong2014,Gong2014a,Herzog2016}. Similar to SLM, the EBM process represents a powder bed-based additive manufacturing process where pre-defined contours are selectively melted in successively deposited powder layers. While SLM applies a laser beam as energy source, EBM is based on an electron beam. EBM is only applicable to electrically conductive materials and has to be performed in a near-vacuum environment in order to avoid an interaction of the electron beam with surrounding gas molecules. On the other hand, SLM is also suitable for dielectric materials and has to be performed in an inert gas atmosphere in order to prevent surface oxidation of (metallic) powder and substrates. Also the way of energy transfer into the powder bed is different: When considering one individual powder grain, the laser beam radiation is (in good approximation) absorbed at the powder grain surface while in EBM, electrons penetrate the powder grain surface, a process in which kinetic energy of the electrons is dissipated leading to melting of the grain. However, the energy of the electron beam is typically completely deposited to the powder grain of first incidence. On the contrary, only a part of the laser beam energy is directly absorbed at the powder grain of first incidence while the remaining part is reflected, leading to considerably higher powder bed penetration depths due to multiple reflections in the open pore system provided by the powder bed. Eventually, the laser beam in SLM is suitable for smaller, more focused spot sizes as well as smaller powder grains sizes, which enables finer geometrical resolutions and an improved surface quality. The higher concentration of incident energy in SLM might yield higher local cooling rates and, thus, smaller grain sizes of the resulting metallurgical microstructure.\\

Similar to SLM, SLS is based on a laser beam energy source. However, in contrast to SLM and EBM, the particles within the powder bed are not fully molten. In this context, solid and liquid state sintering can be distinguished. Solid state sintering is governed by thermally activated diffusion of atoms at temperatures below the melting point leading to slowly growing necks between adjacent powder particles. However, the slow time scales governing this process and the resulting low output rates make it often infeasible from an economic point of view. On the contrary, liquid state sintering aims at a partial melting of powder. The molten liquid will typically spread almost instantaneously between the unmolten particles and acts as binder. In order to achieve defined ratios of molten and unmolten material, either a second material species with lower melt point is employed as binder, bi-modal powder mixtures of one and the same material are used such that the smaller grains melt earlier than the larger ones or the incident energy density has to be adjusted such that only the top surface of powder grains melts while the core remains in solid state. These sintering processes often result in so-called green parts with high porosity, which requires subsequent heat treatment processing.\\

Related processes such as directed energy deposition (DED) and wire-feed laser and electron beam additive manufacturing methods share some similarities with SLM and EBM, yet they generally operate on larger length scales ($mm-cm$ scale molten trajectory). Also the processes of Laser Beam Welding (LBW) and Electron Beam Welding (EBW) will be considered at some points in this work, since some of the underlying physical mechanisms are similar to SLM and EBM. While SLM and EBM aim at additively generating solid material structures out of layers of loose powder, LBW and EBW intend the connection of individual solid parts by partially melting their contact surface.

\subsection{Organization of this article}
\label{sec:intro_summary}

The remainder of this article is structured as follows: In Section~\ref{sec:modeling}, the governing physical mechanisms are further detailed, modeling as well as numerical (and analytical) solution procedures in the context of SLM processes are reviewed and findings derived by these approaches are discussed. Section~\ref{sec:powder_modeling} focuses on the modeling of radiation and heat transfer in powder beds. In Sections~\eqref{sec:macroscopicmodels}-\eqref{sec:microscopicmodels}, a comprehensive overview of existing approaches classified by means of the three main categories of SLM modeling approaches found in the literature, namely macroscopic, mesoscopic and microscopic models, will be given. Section~\ref{sec:experimentalcharacterization} focuses on experimental studies that are especially relevant to understand the thermophysical mechanisms. Following a similar structure, Section~\ref{sec:powder} deals with experimental approaches of powder bed characterization while Sections~\eqref{sec:residualstresses}-\eqref{sec:microstructure} refer to experimental investigations on effects visible on macroscopic, mesoscopic and microscopic level, thus representing the counterpart to the Sections~\eqref{sec:macroscopicmodels}-\eqref{sec:microscopicmodels} considering modeling and simulation. Finally, Section~\ref{sec:practicalimplementationandquestions} provides recommendations concerning practical implementation of the SLM process and gives open questions and potentials for future process improvement.

\section{Modeling and simulation approaches}
\label{sec:modeling}

Approaches to modeling of SLM can be classified in macroscopic, mesoscopic and microscopic models. Macroscopic simulation models typically treat the powder phase as a homogenized continuum resulting in efficient numerical tools capable of simulating the manufacturing of entire parts by SLM. Macroscopic models commonly aim at determining spatial distributions of temperature, residual stresses as well as dimensional warping within SLM parts. Mesoscopic models typically resolve individual powder grains and melt pool thermo-hydrodynamics in order to determine part properties such as adhesion between subsequent layers, surface quality as well as creation mechanisms of defects such as pores and inclusions. Microscopic models consider the evolution of the metallurgical microstructure involving resulting grain sizes, shapes and orientations as well as the creation of thermodynamically stable or unstable phases. The computational effort required for mesoscopic models currently limits the application of these models to single track simulations. However, the insights gained by mesoscopic models might serve as basis to further improve continuum models for powder and melt phase in macroscopic models. Similarly, microscopic models can readily address small areas of the part in the range of the powder layer thickness, and existing approaches are commonly limited to 2D. Nevertheless, they may serve as basis for the development of improved inhomogeneous and anisotropic continuum constitutive models whose quality is essential for the quality of simulation results derived by macroscopic simulation models. Macroscopic as well as mesoscopic models commonly require submodels for the radiation transfer into and the heat transfer within the powder phase. For this purpose, modeling approaches considering powder bed radiation and heat transfer are discussed in Section~\ref{sec:powder_modeling}.\\

A challenge more severe than the actual derivation of physical models, is the application and development of powerful discretization techniques and numerical solution schemes in order to enable a robust and efficient computational solution, two key factors in the simulation-based characterization of SLM processes. While the finite element method (FEM), the finite difference method (FDM) as well as the finite volume method (FVM) represent spatial discretization schemes typically employed in the considered models, temporal discretization is almost exclusively based on explicit or implicit finite difference time integration schemes. For implicit schemes, the fully discretized problem is typically represented by a system of equations that is nonlinear in the unknown discrete primary variables, which requires the application of a nonlinear solver, e.g. a Newton-Raphson scheme. Depending on the characteristics of the nonlinear system of equations to be solved, convergence of the nonlinear solution scheme might not always be guaranteed. On the contrary, explicit schemes allow for a direct extrapolation of the known configuration at time $t_n$ to the unknown configuration at time $t_{n+1}$. The resulting system of equations is linear in the discrete unknowns such that no iterative, nonlinear solution process is required. However, at least in the geometrically linear regime, implicit time integrators can be proven to be unconditionally stable, thus typically allowing for considerably larger time step sizes as compared to explicit schemes in order to preserve system stability, i.e. to keep the total system energy bounded during the simulation. Consequently, implicit schemes are favorable for problems that are dominated by a low frequency response, where the large time step sizes possible for these schemes are sufficient in order to resolve the low-frequent system answer. On the contrary, explicit schemes are rather suited to model high frequency responses and wave-like phenomena such as high velocity impacts. There, small time step sizes are required for both explicit and implicit schemes in order to accurately resolve the high-frequent system dynamics. Since the computational effort per time step is lower for explicit schemes and no nonlinear convergence has to be accounted for, these schemes are preferable in such scenarios. The stability-relevant time step sizes dictated for explicit schemes in the context of mesoscopic SLM models are typically considerably smaller than the time step sizes required to capture the relevant physical phenomena. Consequently, implicit schemes would have the potential to achieve substantial computational savings. However, the complexities arising from the multiple field and domain couplings as well as the geometrical characteristics prevalent in SLM make the implicit treatment of fully resolved mesoscopic models a challenging task.\\

A further important question to be asked from a numerical point of view is the way different physical fields (e.g. thermal and mechanical fields) and domains (e.g. powder, melt and solid phase) are coupled. Here, three concepts of treating phase boundaries shall be distinguished: The first category models a sharp boundary between the phases resulting in a jump of material properties and physical fields under consideration of displacement / velocity continuity conditions and mechanical equilibrium at the interface. Numerical realizations of such models are typically based on explicit interface tracking, e.g. via level set schemes, and discrete solution spaces allowing for discontinuity in the primary variables, e.g., the extended finite element method (XFEM). The second category of schemes still identifies the interface in an explicit manner, e.g., by means of an additional phase variable $\phi$ taking on the value $\phi\!=\!1$ for the one phase and $\phi\!=\!0$ for the other. However, the interface is not represented by a sharp boundary, but rather by a transition range of finite thickness characterized by phase variable values $\phi \in ]0;1[$. An example for such a scheme is the volume of fluid method (VOF). A third category is given by methods that do not introduce additional variables in order to distinguish the phases. These schemes typically treat the phase boundary implicitly by means of specific values of existing primary fields, e.g., by defining the phase boundary between liquid and solid as the isothermal contour $T\!=\!T_m$ representing the melt temperature $T_m$. The rapid change of physical properties at the phase boundary is considered by these schemes in terms of high gradients in the applied material parameters. Thus, again, the interface is extended to a finite thickness. A too small transition region might lead to excessive gradients in the material parameters and, as consequence, to ill-conditioned discrete problems that are challenging to be solved numerically. The SLM modeling approaches reviewed in this work almost exclusively rely on the second and third category of approaches.\\ 

Numerical schemes can also be distinguished concerning the succession of solving different physical fields and domains. Namely, monolithic schemes solve the entire multi-physics problem at once, and partitioned schemes solve the different physical fields/domains subsequently. The partitioned schemes can be further subdivided into so-called iterative or strongly coupled partitioned schemes that iterate between the different fields several times within a time step until achieving the solution of the monolithic problem statement and so-called staggered or weakly coupled partitioned schemes that solve for each field only once per time step without additional iterations, eventually leading to a solution that differs from the monolithic one. Monolithic and strongly coupled partitioned schemes are typically combined with implicit time integrators because these coupling schemes can preserve the desirable stability properties of implicit integrators at large time step sizes. On the other hand, weakly coupled schemes are often combined with explicit time integrators since these schemes also lead to low computational costs per time step. However, for weakly coupled partitioned schemes, even when combined with implicit time integrators on the individual fields, the stability requirement often becomes very restrictive typically dictating very small time step sizes and in some cases even prohibiting stability at all (see e.g.~\cite{Forster2007,Causin2005} for a discussion in the context of fluid structure interaction (FSI) simulations). Since time step size correlates with computational costs, these considerations are of high practical interest. 

\subsection{Modeling of optical and thermal properties of powder phase}
\label{sec:powder_modeling}

The optical and thermal properties of the powder bed, in combination with the laser characteristics, crucially determine the heat distribution in the powder bed and the subsequent melt pool dynamics. This section addresses approaches to modeling the optical and thermal properties of the powder bed as well as the radiation-dominated energy transfer from the laser beam source into the powder bed and the conduction-dominated heat transfer within the powder bed. The considered approaches are typically employed as powder bed submodels within the three main categories of macroscopic, mesoscopic and microscopic models. Often, the laser beam can be described in good approximation by means of a Gaussian power distribution~\cite{Dong2009,Verhaeghe2009,Roberts2009}. In a first step, the question of how the laser energy transfers into the SLM powder bed, a process that can be described via the principles of thermal radiation, geometrical optics and electro-magnetic wave propagation, will be considered. When it is not mechanically compressed after spreading, the powder bed has as high porosity as freely poured powder, which is in the range of $40\% - 60\%$ for typical SLM powders~\cite{Rombouts2005}. As shown in~\cite{Gusarov2009,Gusarov2005}, laser radiation penetrates into powder through pores to a depth of several particle diameters because of multiple reflections~\cite{Wang2002}. This depth is comparable with the powder layer thickness. Thus, laser energy is deposited not on the surface but in the bulk of the powder layer. The incident laser energy consists of portions that are reflected, absorbed and transmitted when impinging upon the surface of powder particles. The emission of radiation from the powder particles themselves can often be neglected. In terms of radiation reflection, it can be distinguished between specular (mirror-like) and diffuse reflection, with the latter exhibiting a reflection intensity that is equally distributed over all possible directions. Concerning radiation transmission, typically the classification of opaque, semi-transparent or transparent particles is made.\\

The multiple reflection/absorption processes of the light in the powder bed is additionally influenced by powder characteristics such as mixture ratio, mean particle size and shape, size distribution, packing density, powder bed depth but also by the laser beam spot size. Also the polarization of the laser beam relative to the powder particle surfaces can lead to non-uniform energy absorption even across individual powder particles. Since the thermal conduction in the powder bed is typically weak due to the prevalent porosity and melting time scales are typically smaller than heat conduction time scales for individual grains, non-uniform energy absorptions across the powder bed but also across individual powder particles will in general have considerable influence on the melt pool shape and the properties of the solidified track. A further factor of influence is given by possible surface oxidation and contamination effects.\\

An extensive review on the topic of radiation transfer within heterogeneous/porous/dispersed media can be found in~\cite{Viskanta2016} as well as~\cite{Baillis2000}. The following section will mainly focus on approaches in the context of SLM. According to~\cite{Baillis2000}, the modeling approaches for radiation transfer in heterogeneous media can be classified as models based on a continuum formulation of the radiation transfer equation (RTE) well-known for homogeneous (participating) optical media, and models based on a discrete formulation of the RTE, which typically leads to ray tracing schemes. Both approaches are based on the simplifying assumptions of geometrical optics which considers light rays that: propagate in rectilinear paths as they travel in a homogeneous medium; bend, and in particular circumstances may split in two at the interface between two dissimilar media; follow curved paths in a medium in which the refractive index changes; and may be absorbed or reflected.

\subsubsection{Continuum model for powder bed radiation transfer}
\label{sec:powder_modeling_continuum}

The general radiation transfer equation (RTE) known for homogeneous media and underlying the homogenized, continuum models for powder bed radiation transfer is (see also~\cite{Baillis2000,Gusarov2005,Chandrasekhar2013}):
\begin{align}
\label{gusarov2005_RTE}
\boldsymbol{\Omega} \bigtriangledown \! I(\mb{x},\boldsymbol{\Omega}) = -(\sigma+\kappa) I(\mb{x},\boldsymbol{\Omega}) + \kappa I_{eb} + \frac{\sigma}{4 \pi} \int \limits_{0}^{4 \pi} I(\mb{x},\boldsymbol{\Omega}^{\prime}) S_c(\boldsymbol{\Omega}^{\prime} \rightarrow \boldsymbol{\Omega}) d \boldsymbol{\Omega}^{\prime}.
\end{align}
The RTE describes the rate of the direction- and position-dependent radiation energy flux density $I(\mb{x},\boldsymbol{\Omega})$ based on an energy balance considering radiation absorption (first term on the right-hand side), radiation emission (second term on the right-hand side) and scattering (first and third term on the right-hand side). In this context, $\boldsymbol{\Omega}$ represents a directional unit vector, $d \Omega$ is an infinitesimal solid-angle increment whose magnitude is per definition identical to an infinitesimal surface element $dA$ on a unit sphere, and the product $\boldsymbol{\Omega} I(\mb{x},\boldsymbol{\Omega}) d \Omega$ represents the vector of energy flux density transferred by photons in direction of the unit vector $\boldsymbol{\Omega}$ within a solid angle increment $d \boldsymbol{\Omega}$ at position $\mb{x}$. The constants $\sigma$ and $\kappa$ are the scattering and absorption coefficients, which are, however, often replaced by the alternative constants extinction coefficient $\beta = \sigma+\kappa$ and albedo (portion of reflected light) $\omega = \sigma/\beta$. The mapping $S_c(\boldsymbol{\Omega}^{\prime} \rightarrow \boldsymbol{\Omega})$ is the (normalized) scattering phase function stating the probability that radiation in $\boldsymbol{\Omega}^{\prime}$-direction is scattered to the $\boldsymbol{\Omega}$-direction and $I_{eb}$ is Planck's blackbody function representing radiation emission at a certain position:
\begin{align}
\label{blackbodyemission}
I_{eb} = k_{SB} (T^4-T_{ref}^4).
\end{align}
Here, $T$ is the powder temperature, $T_{ref}$ is the temperature of the ambient gas atmosphere and $k_{SB}$ is the Stefan-Boltzmann constant. While the radiation transfer equation has originally been derived for homogeneous media, it is well-established to apply~\eqref{gusarov2005_RTE}  also to heterogeneous systems, where the continuous quantities and parameters introduced so far have to be replaced by their \textit{effective} counterparts, determined via spatial averaging over a reference volume that has to be much greater than the length scales of prevalent heterogeneities (e.g., of individual particles in powder beds). When considering radiation heat transfer in powder beds, the extinction coefficient $\beta$ is typically determined by the structure of powder, i.e., by size and shape of particles and by their arrangement, but is independent of optical properties of the material of particles like reflectance. On the contrary, the scattering characteristics such as the albedo $\omega$ and phase function $P(\boldsymbol{\Omega}^{\prime} \rightarrow \boldsymbol{\Omega})$ commonly depend on reflective properties of the material of particles.\\ 

In the following, a coordinate system is chosen such that the positive $z$-axis points into powder layer thickness direction with $z=0$ representing the upper surface of the powder bed and $z=L$ the boundary between powder bed (with thickness $L$) and the underlying solidified phase (substrate). The net radiation heat flux density $\mb{q}_r$, following the typical definition as heat flux per unit area measured in $[W/m^2]$, results from the energy flux density per unit angle increment $I(\mb{x},\boldsymbol{\Omega})$ via integration over all directions of incident radiation. Expressed by means of solid-angle increments $d \Omega$ this is equivalent to an integration over the surface $4\pi$ of a unit sphere. Assuming that the $x-$ and $y-$components of the heat flux density $\mb{q}_r$ resulting from a solution of~\eqref{gusarov2005_RTE} are negligible, also the incident power density per unit volume $u_s$ can be determined based on the following relations:
\begin{align}
\label{gusarov2007_U}
\mb{q}_r = \int \limits_{0}^{4 \pi} I \boldsymbol{\Omega} d \Omega \approx {q}_{rz} \mb{e}_z \quad \quad \rightarrow \quad \quad u_s:= -\frac{\partial q_{rz}}{\partial z}.
\end{align}
Gusarov et al.~\cite{Gusarov2005} considered the general problem of normal incidence of collimated radiation on a thin powder layer consisting of opaque (i.e. optically non-transparent) particles of arbitrary shapes and dimensions placed on a reflecting substrate in order to derive resulting net absorptivities and deposited energy profiles. It is assumed that the particle size is sufficiently small as compared to the laser spot size and powder bed thickness as to allow for the employed homogenization procedure. Further, in SLM the particle size is typically much greater than the wavelength of radiation, thereby justifying the applicability of the geometrical optics theory. The contribution $\kappa I_{eb}$ representing the emission from individual powder grains has been neglected due to the comparatively high energy density of the incident laser radiation. Gusarov et al.~\cite{Gusarov2005} propose a statistical model based on the powder porosity and specific powder surface areas in order to determine the model parameters required in~\eqref{gusarov2005_RTE}. This statistical model accounts for dense powder beds where particles touch each other and radiation transfer cannot be treated as scattering by independent particles, an effect that is referred to as dependent scattering (see e.g. \cite{Drolen1987,Chen1963,Kamiuto1991,Jones1996,Kamiuto1990,Tancrez2004}). Moreover, an analytical solution for the radiation transfer equation has been derived via the so-called two-flux method with the boundary conditions of normally collimated incident flux $\mb{q}_{r0}\!:=\!\mb{q}_{r}(z\!=\!0)$ and specular reflection at $z\!=\!L$. Among others, the additional assumption of the laser spot size being much larger than the absorption depth has been made. Thus, the radiation transfer problem could be considered as 1D-problem in $z-$direction with an incident flux $\mb{q}_{r0}=q_{r0}(x,y)\mb{e}_z $.\\

\begin{figure}[h!!]
 \centering
    \includegraphics[width=1.0\textwidth]{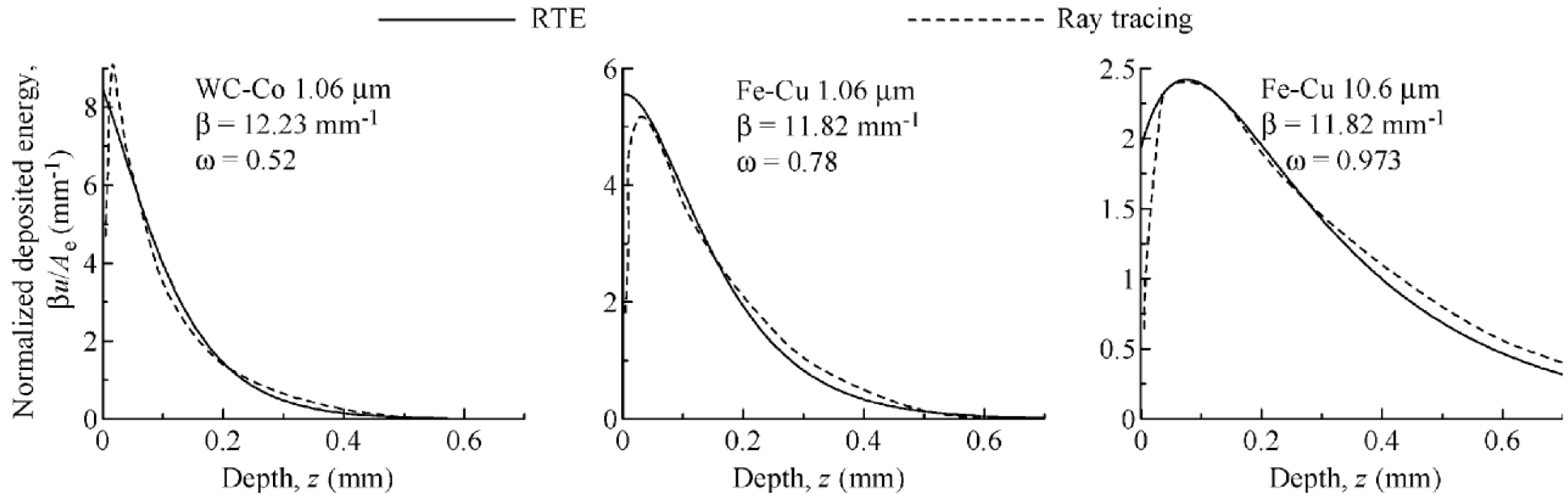}
  \caption{Comparison of deposited energy profiles resulting from the RTE (full lines) and ray tracing simulation (broken lines) for WC-Co or Fe-Cu powders with particles sizes $2R\!=\!50 \mu m$ (Fe and WC), $2R\!=\!30 \mu m$ (Cu) and $2R\!=\!20 \mu m$ (Co) and laser beam wave lengths of $1.06 \mu m$ or $10.6 \mu m$,~\cite{Gusarov2005}.}
  \label{fig:gusarov}
\end{figure}

In Figure~\ref{fig:gusarov}, the resulting energy deposition profiles derived by the RTE solution as proposed in~\cite{Gusarov2005} and by means of ray tracing simulation are compared for three examples considering either WC-Co or Fe-Cu mixed powders with particles sizes $2R\!=\!50 \mu m$ (Fe and WC), $2R\!=\!30 \mu m$ (Cu) and $2R\!=\!20 \mu m$ (Co) as well as laser beam wave lengths of $1.06 \mu m$ or $10.6 \mu m$. Accordingly, these two modeling approaches are in good agreement since a sufficiently deep powder bed and a sufficiently large laser beam size has been chosen. The total/integrated laser energy absorbed in a thin powder layer on a reflective substrate increases with its thickness while the deposited energy density per unit volume decreases. The absorption depths in Figure~\ref{fig:gusarov} seem to be comparatively high as compared to typical values known from SLM powder layers, which might be an indication for rather low packing densities considered in this study. Subsequently, the model of~\cite{Gusarov2005} has e.g. been applied in the independent works~\cite{Verhaeghe2009},~\cite{Khairallah2014} and~\cite{Hodge2014}. In~\cite{Gusarov2008}, the model of~\cite{Gusarov2005} has been further extended and refined. There, it has been shown that the application of the model derived in~\cite{Gusarov2005} is only reasonable if either one of the two phases (powder particles and pores) are opaque, one phase has a much larger volume fraction, or the radiation properties of the two phases are at least similar. On the contrary, an application of this model, which simply averages the radiation intensities across both phases, to problems where the two phases capture comparable volume fractions but show considerably different optical properties at the same time (e.g., mixed powders), seems unjustified.\\ 

For mixed materials, a model based on partial homogenized radiation intensities and denoted as Vector Radiation Transfer Equation (VRTE) has been proposed in~\cite{Gusarov2008}. In this model, the partial values are obtained by averaging over each individual phase. It consists of two transport equations for the partial homogenized radiation intensities. The equations are similar to the conventional RTE, but contain additional terms that take into account the exchange of radiation between the individual phases. The structure of the medium is specified by volume fractions of phases and by specific surfaces of phase boundaries. The optical properties are described by refractive indices and absorption coefficients of phases. It has been verified that the vector RTE model reduces to the conventional RTE model in case one of the two phases is opaque or one phase prevails in volume. Compared to the RTE in~\cite{Gusarov2005}, an analytical solution was no longer achievable for the VRTE~\cite{Gusarov2008}. Instead, a discrete-ordinate method~\cite{Carlson1965} was exploited. 

\subsubsection{Ray tracing model for powder bed radiation transfer}
\label{sec:powder_modeling_raytracing}

As pointed out in Boley et al.~\cite{Boley2015}, the assumption of a RTE continuum model~\cite{Gusarov2005} is questionable for very thin, low-porosity metal powder layers with a layer thickness in the range of a few powder particles and/or a laser spot size comparable to the size of the powder particles. In such scenarios, which are often prevalent in SLM processes, the averaging process underlying the RTE continuum models lead to considerable model errors as compared to the exact solutions, which typically yield an energy absorption that is highly non-uniform with respect to the spatial position of the laser beam. Ray tracing modeling is one possible approach to resolve such heterogeneities. However, it shall be emphasized that depending on the spatial resolution required by the physical question to be answered (e.g. when global residual stress distributions but not the specific locations of individual pores are of interest), also the application of RTE continuum models might be reasonable, which typically require a considerably lower computational effort as compared to ray tracing simulations.\\ 

\begin{figure}[h!!]
 \centering
   \subfigure[Global view on ray tracing scheme.]
   {
    \includegraphics[height=0.3\textwidth]{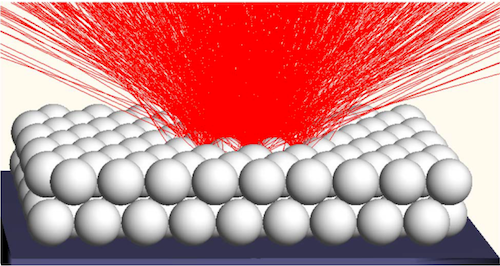}
    \label{fig:boley2015_raytracing_principle}
   }
\hspace{1.0cm}
   \subfigure[Absorptivity across spherical particle,~\cite{Boley2015}.]
   {
    \includegraphics[height=0.3\textwidth]{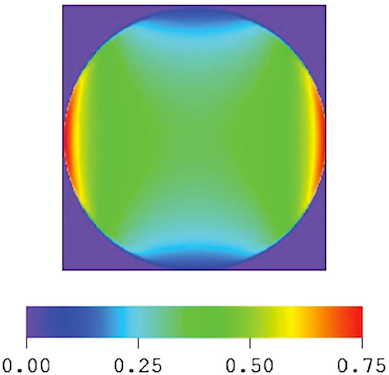}
    \label{fig:boley2015_onesphere}
   }
  \caption{Modeling of laser energy absorption on powder grain scale based on ray tracing approach,~\cite{Boley2015}.}
  \label{fig:raytracing_principle}
\end{figure}

In ray tracing simulations, the total energy emitted by the laser beam in a certain time interval is represented by a discrete  ensemble of rays with defined spatial position, orientation and energy. The position and energy associated with the individual rays is typically chosen such that the overall energy emission but also the spatial energy distribution resulting from the entire ensemble equals the corresponding characteristics of the laser beam (e.g. a Gaussian energy distribution). After defining the individual rays, the path of each ray is traced until striking an obstacle (powder particle). Based on the optical properties of the obstacle surface, part of the ray energy is absorbed whereas the remaining part of the ray energy is represented by a reflected ray with defined energy and orientation, which will further be traced trough the powder bed (Figure~\ref{fig:boley2015_raytracing_principle}). Commonly, several reflections are considered for each ray until the remaining energy drops below some predefined threshold. The individual ray energy contributions absorbed by each particle are accumulated during the simulation. Also the ray-tracing model is based on the principles of geometrical optics, thus, the corresponding requirements mentioned above have to be fulfilled. In particular, the particle radius $R$ has to be considerably larger than the laser wavelength.\\

\begin{figure}[h!!]
 \centering
    \includegraphics[width=0.8\textwidth]{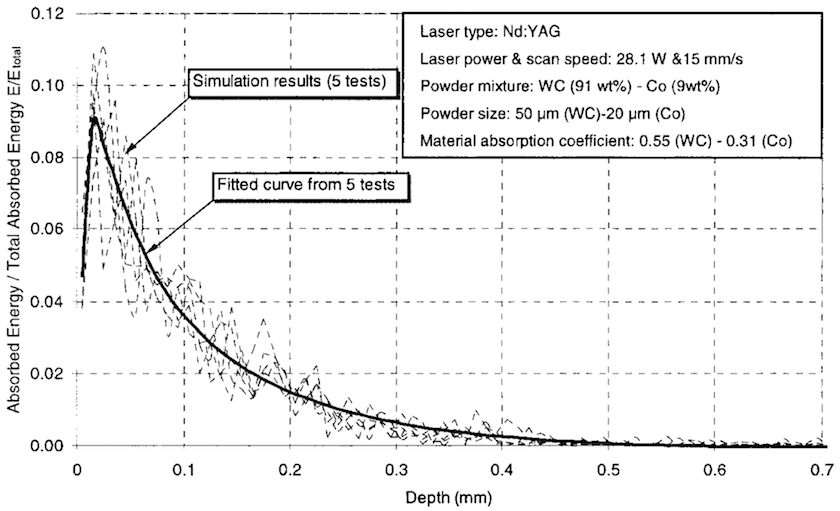}
  \caption{Absorped to incident energy for a mixture of $50 \mu m$ Tungsten Carbide (WC) and $20 \mu m$ Cobalt (Co) particles,~\cite{Wang2002}.}
  \label{fig:wang2002}
\end{figure}

Ray tracing is a well-established approach for studying radiation problems in countless scientific disciplines. Several contributions have applied this approach in order to study the radiation transfer trough porous media and packed beds~\cite{Chan1974,Yang1983,Singh1991,Singh1992,Singh1994,Argento1996}. In one of the first ray-tracing approaches in the context of SLM/SLS, Wang et al.~\cite{Wang2000,Wang2002} studied a powder mixture consisting of two species of spherical particles differing in particle size and material. In~\cite{Wang2002}, the dimensions of the sintering/melting zone are estimated by determining the accumulated energy absorbed by each particle. However, no inter- or intra-particle mechanisms of heat transfer (such as heat conduction or convection) besides radiation has been considered. Based on a powder bed consisting of $50 \mu m$ Tungsten Carbide (WC) particles and $20 \mu m$ Cobalt (Co) particles and typical SLS system parameters, the ray tracing simulations as well as accompanying experiments came to the consistent result that $>96\%$ of the absorbed energy is concentrated within a depth of $400 \mu m$ from the powder bed surface. The resulting spatial distribution of the absorbed energy $E$, normalized by the total incident energy $E_{total}$, is plotted over the powder bed depth coordinate in Figure~\ref{fig:wang2002}. While the highest amount of primary laser beam radiation obviously arrives directly at the top surface of the powder bed, the maximum of the absorbed energy distribution occurs slightly beneath the top surface since at this location also secondary radiation contributions, stemming from multiple reflections in the powder bed, are prevalent.\\

In~\cite{Boley2015}, the model of~\cite{Wang2000,Wang2002} has been further refined by additionally considering the angular- and polarization-dependence of the laser absorption. Thereto, the absorptivity of the laser beam components in perpendicular (index $S$) and parallel (index $P$) polarization states have been distinguished according to the Fresnel formulas (see~\cite{Landau1984}):
\begin{align}
\label{boley2015_alpha}
\alpha_S(\theta)=1-
\left[
\frac{\cos{\theta} - (n^2-\sin^2{\theta})^{1/2}}
{\cos{\theta} + (n^2-\sin^2{\theta})^{1/2}}
\right]^2, \quad \quad
\alpha_P(\theta)=1-
\left[
\frac{n^2\cos{\theta} - (n^2-\sin^2{\theta})^{1/2}}
{n^2\cos{\theta} + (n^2-\sin^2{\theta})^{1/2}}
\right]^2.
\end{align}
This description takes into account the model of laser beam radiation as an electromagnetic wave with speed $c$ and direction of propagation $\mb{k}$. This wave is assumed to be (linearly) polarized with the constant unit vector $\mb{e}$ representing the direction of the electric field normal to $\mb{k}$. In~\eqref{boley2015_alpha}, the complex refraction index $n\!=\!n_c+i n_{\alpha}$, describing the speed of electromagnetic wave propagation (real part $n_c$), absorption of the electromagnetic wave (imaginary part $n_{\alpha}$) and the angle of incidence $\theta$ between $\mb{k}$ and the normal vector $\mb{n}$ onto the powder particle surface have been employed. The perpendicular (index $S$) and parallel (index $P$) components of  the polarized electromagnetic wave are typically determined by means of a projection of the polarization vector $\mb{e}$ into a reference plane spanned by the vectors $\mb{k}$ and $\mb{n}$. In Figure~\ref{fig:boley2015_onesphere}, the resulting spatial absorptivity distribution across one isolated sphere of stainless steel irradiated by a horizontally polarized laser beam is depicted. Correspondingly, the distribution of energy absorption can vary considerably across one individual particle.\\

The high importance and practical relevance of these fluctuations shall be illustrated by the following time scale estimate. The time scales governing the homogenization of energy on a sphere with radius R due to thermal conduction can be estimated by $\tau_c=R^2/a$, where $a$ is the thermal diffusivity of the metal. The time scales governing the melting of a spherical particle can be approximated by  $\tau_m=R H_m/(\alpha_0 I)$, where $H_m$ is the melting enthalpy per volume, $I$ is the laser irradiance, and $\alpha_0$ is the flat-surface absorptivity. For parameter values typical for SLM processes, the melting time is often considerably shorter as compared to the diffusion time. Consequently, non-uniformity of energy absorption results in only partial melting of the particle, which might in turn considerably influence melt pool dynamics or creation mechanisms of pores~\cite{Boley2015}. However, in cases where melting takes place on larger time scales than heat conduction, the initial energy and temperature distributions might homogenize before melting occurs, and, consequently, approaches considering a constant energy distribution across a particle (see e.g.~\cite{Wang2002}) might yield reasonable results. The fact that thermal conductivity of loose powder is typically governed by the gas filling the powder bed pores~\cite{Rombouts2005} yields even larger time scales for the inter-particle heat conduction.\\

\begin{figure}[h!!]
 \centering
   \subfigure[Laser beam path across powders with Gaussian (average size $27 \mu m$, top) and bimodal distribution (particle ratio $1$:$7$,bottom.)]
   {
    \includegraphics[width=0.45\textwidth]{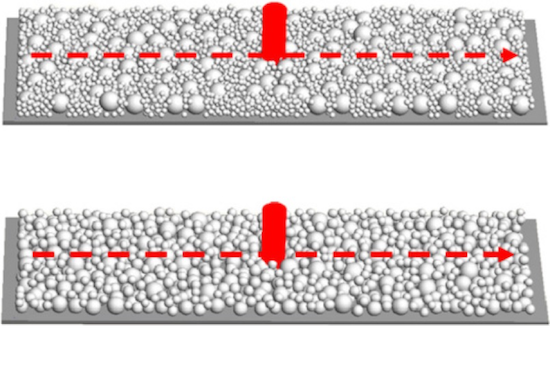}
    \label{fig:boley2015_spatialdistribution2}
   }
   \hspace{0.05\textwidth}
   \subfigure[Variation of absorptivity along laser path for two different laser beam sizes with a radius of $a=8\mu m$ and $a=24\mu m$.]
   {
    \includegraphics[width=0.45\textwidth]{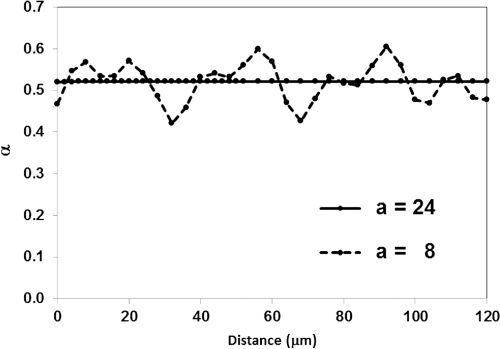}
    \label{fig:boley2015_spatialdistribution1}
   }
  \caption{Dependence of absorptivity on laser beam size and particle size distribution investigated for a stainless steel alloy,~\cite{Boley2015}.}
  \label{fig:absorption_dependence2}
\end{figure}

According to~\cite{Boley2015}, there are two general factors of influence leading to absorption non-uniformity, one related to the non-uniformity of absorption within a single particle as considered above, and the other related to the non-uniformity of the powder bed. In order to investigate this second factor of influence, ray tracing simulations on laser tracks across powder beds (thickness $\approx 50 \mu m$) resting on a solid substrate and consisting of two different powder mixtures, a Gaussian distribution (average size $27 \mu m$, Figure~\ref{fig:boley2015_spatialdistribution2}, top) and a bimodal distribution with maximal packing density (particle ratio $1$:$7$, Figure~\ref{fig:boley2015_spatialdistribution2}, bottom), have been conducted in~\cite{Boley2015}. For both powder mixtures, typical distributions of the total absorbed energy along the scan track length as illustrated in Figure~\ref{fig:boley2015_spatialdistribution1} can be observed. While large laser beam sizes (see curve $a=24\mu m$ in Figure~\ref{fig:boley2015_spatialdistribution1}) lead to a comparatively homogeneous spatial energy absorption, laser beam sizes in the range of the particle diameter (see curve $a=8\mu m$ in Figure~\ref{fig:boley2015_spatialdistribution1}), lead to strongly fluctuating spatial absorption values. Furthermore, the overall absorptivity turned out to be higher for the bimodal distribution, although the degree of fluctuations was comparable for both mixtures. Again, these fluctuations are not considered in homogenized continuum models such as~\cite{Gusarov2005}. Thus, their applicability requires sufficiently small particle sizes as compared to laser spot size and powder layer thickness.\\

\begin{figure}[h!!]
 \centering
   {
    \includegraphics[height=0.4\textwidth]{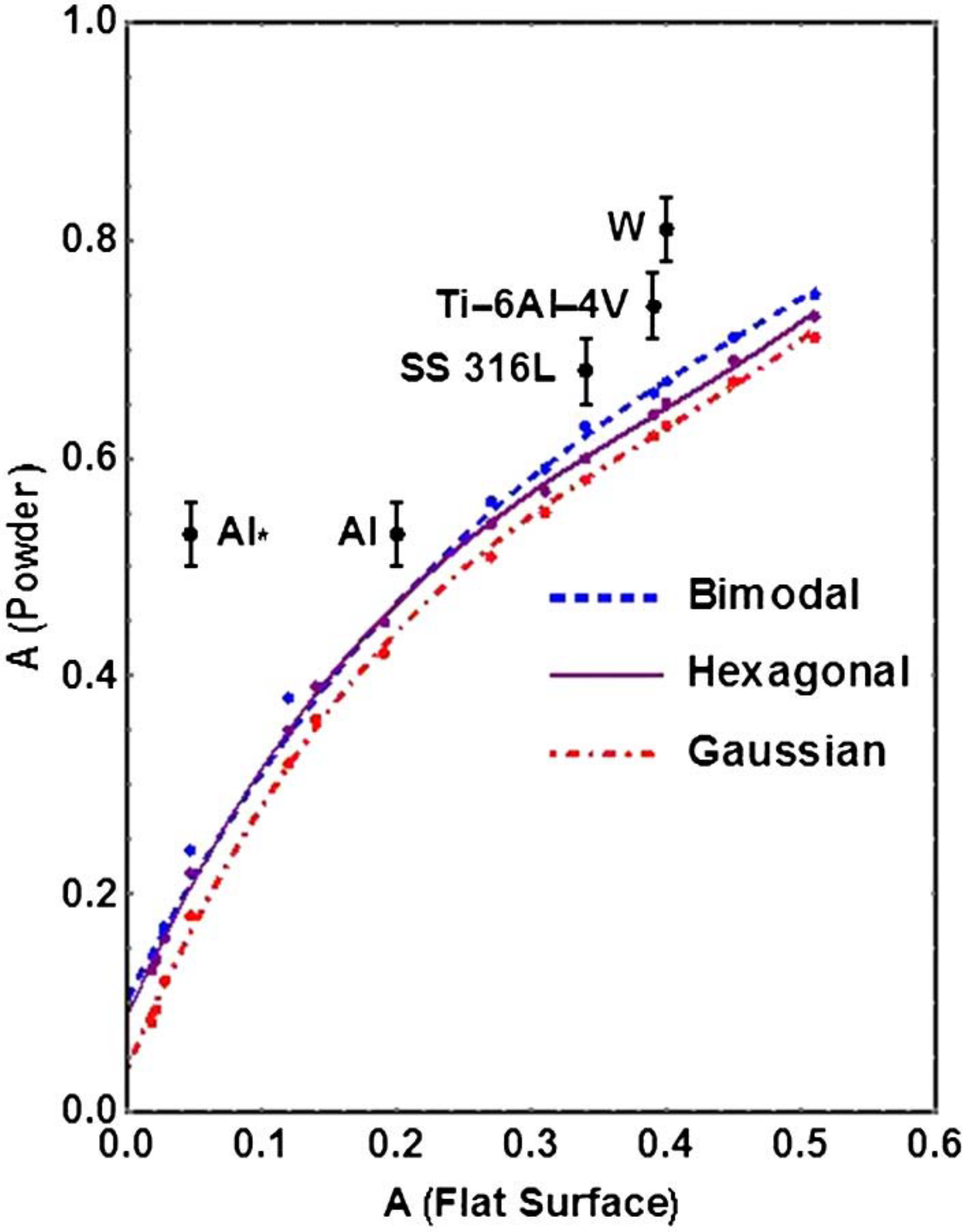}
   }
   \hspace{0.1\textwidth}
   {
    \includegraphics[height=0.4\textwidth]{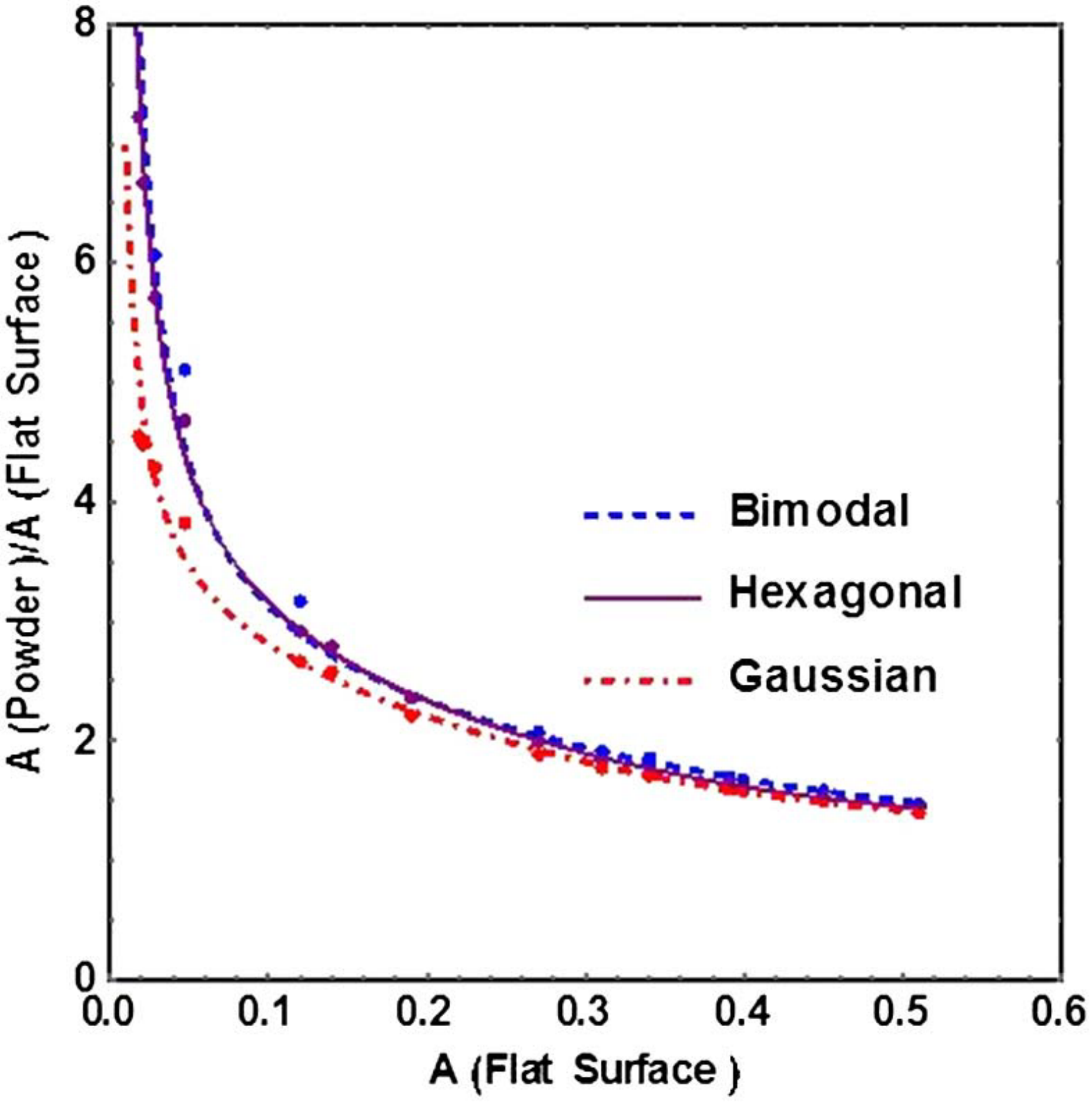}
   }
  \caption{Dependence of powder layer absorptivity on flat surface absorptivity for different materials,~\cite{Boley2016}.}
  \label{fig:absorption_dependence3}
\end{figure}

In~\cite{Boley2016}, the results derived in~\cite{Boley2015} have been extended to a considerable range of different materials and verified experimentally. Concretely, the simulated power bed absorptivity and the flat surface absorptivity of the considered materials have been compared. Plotting the powder bed absorptivity over the flat surface absorptivity has revealed a smooth functional relation between these two quantities as already observed in~\cite{Gusarov2006,Gusarov2010}. This interesting observation allows to extract the assumed powder bed absorptivity for a material with given flat surface absorptivity, a procedure that can be considered to be of highest practical relevance (Figure~\ref{fig:absorption_dependence3}). In~\cite{Zhou2009}, a comparable ray tracing simulation model has been proposed in the context of SLS.

\subsubsection{Continuum models for powder bed heat conduction}
\label{sec:powderheattransfer}

The macroscopic simulation models discussed later in Section~\ref{sec:macroscopicmodels} regard the powder bed as a homogenized continuum. Consequently, these approaches typically rely on a model for an effective, homogenized powder bed conductivity.  On the contrary, mesoscopic simulation models as considered in Section~\ref{sec:mesoscopicmodels}  resolve individual powder grains by spatial discretization. Consequently, each powder particle can be considered as a separate solid body, which simplifies the powder bed heat conduction problem to the two subproblems of intra-solid heat conduction as well as inter-solid heat conduction across the contact interfaces. \\

Study of thermal transport in two phase systems has long been a topic of scientific inquiry, with Maxwell and Rayleigh each publishing models of the problem in 1873 (see~\cite{Maxwell1873}) and 1892 (see~\cite{Rayleigh1892}), respectively. These early contributions have been based on comparatively strong model assumptions. For example, Rayleigh's treatment idealizes the geometry to regularly spaced spherical particles in non-contact within the dispersed phase. An extension of the spherical particles to a larger spectrum of geometric primitives has been proposed in the contribution by Bruggeman~\cite{Bruggeman1935}. In general, the aim of more recent work has been to further relax such assumptions and match the model to expected behavior in limit conditions, often in the context of interpreting experimental results~\cite{Tsotsas1987}. For example, an often cited model has been published by Meredith and Toblas to explain the conductivity of water-propylenecarbonate emulsions. This approach is assumed to yield more realistic predictions than Maxwell's or Bruggeman's models at high volume fractions of the dispersed phase~\cite{Meredith1961}. In the review by Tostsas and Martin~\cite{Tsotsas1987}, models are further divided by how they treat primary parameters of bulk conductivities and porosity, and secondary parameters including radiative transfer, pressure dependence, contact between particles and deformation as well as convective effects.\\

The effective thermal conductivity of loose metallic powders, i.e. of material with \textit{negligible} particle-to-particle contact surfaces, is typically controlled by the gas in the pores~\cite{Rombouts2005,Gusarov2007}. On the contrary, in~\cite{Gusarov2003}, a model has been derived for the heat conduction in power beds in case of small but finite contact areas between the particles. Such a scenario typically arises during the sintering process in SLS, at the beginning of the powder particle melting process in SLM or already in the non-melted solid state of the powder bed in case it is not loose but pre-compressed.\\

Now, the effective thermal conductivity $\lambda_e$ within a powder bed in direction of a given unit vector $\mb{n}_{\lambda}$ can be defined on the basis of the heat flux density $\mb{q}$ and the temperature gradient $\bigtriangledown T$ according to:
\begin{align}
\label{gusarov2003_lambda_e1}
|| \mb{n}_{\lambda} \!\cdot\! \mb{q} || = \lambda_e || \mb{n}_{\lambda} \!\cdot\! \bigtriangledown \! T ||.
\end{align}
For illustration, the effective thermal conductivity shall briefly be presented for two well-known, but conceptionally
different, heat transfer models for heterogeneous / particle-based systems, the Maxwell model mentioned above (index
M) and the Reimann-Weber model (index RW). According to these models, the effective thermal conductivities are:
\begin{align}
\label{gusarov2003_lambda_e2}
\lambda_{e,M}= \lambda_g \left( 1-2\Phi \frac{1-\lambda_s/\lambda_g}{2+\lambda_s/\lambda_g} \right) \Bigg/ \left( 1+\Phi \frac{1-\lambda_s/\lambda_g}{2+\lambda_s/\lambda_g} \right), \quad \quad
\lambda_{e,RW}=\lambda \left(\frac{1}{x}+\frac{1}{\pi} \ln{\frac{2}{x}}\right)^{-1} \, \text{with} \, x=\frac{r_c}{R}.
\end{align}
The well-known Maxwell model describes the effective thermal conductivity of systems consisting of randomly packed spheres with conductivity $\lambda_s$ embedded in a continuous medium with conductivity $\lambda_g$ and is applicable for sufficiently small volume fractions $\Phi $ of spheres. It can easily be verified that this model yields the expected results for limit conditions such as $\lambda_g\!=\!0$, $\Phi\!=\!0$ or $\Phi\!=\!1$. On the contrary, the Reimann-Weber model considers the effective thermal conductivity between two isolated spheres (bulk material conductivity $\lambda$, radius $R$) exhibiting a circular contact area of radius $r_c$. The packing densities prevalent in powder bed AM make the application of the Maxwell model to these processes virtually impossible. For that reason, Gusarov et al.~\cite{Gusarov2003} derived formulations for the thermal conductivity of regular and random packings of spherical particles based on models of the "isolated-spheres" type comparable to the Reimann-Weber model. As such, the effective conductivity of ordered powder beds has been determined by summing up the conductivity contributions of all individual sphere-to-sphere pairs. In the case of random packings, the models of "isolated-spheres" type have been homogenized by spatial integration and the geometrical quantities $r_c$ and $R$ are replaced by effective quantities such as the particle volume fraction $\Phi$, the mean coordination number $n$ (describing the average number of closest neighbors) and the average dimensionless contact size ratio $x$ leading to the following effective thermal conductivity (index G for Gusarov):
\begin{align}
\label{gusarov2003_lambda_e3}
\lambda_{e,G}=\lambda \frac{n\Phi}{\pi}x.
\end{align}
While a strong dependence on the contact size $x$, at least qualitatively, has already been known from experiments, experimental characterization approaches typically consider only the powder density $\Phi$, but not the coordination number $n$. However, the derivations in~\cite{Gusarov2003} recommended for future experiments to treat these two parameters independently since for randomly packed powder beds different coordination numbers, and according to~\eqref{gusarov2003_lambda_e3} considerably differing effective conductivities, are possible for the same powder bed density $\Phi$. Besides modeling approaches for the determination of effective powder bed conductivities, often also experimentally determined effective conductivities are employed in macroscopic SLM models as discussed in Section~\ref{sec:macroscopicmodels}. An extensive comparison of experimental and modeling approaches for effective powder bed conductivities can also be found in Yagi et al.~\cite{Yagi1957}. 

\subsubsection{Conclusion}
\label{sec:powder_modeling_conclusion}

In summary, ray tracing models are based on a comparatively simple theory, yield a high degree of detailedness in the obtained solutions, but also require considerable computational resources to be applied to scenarios of practically relevant size. On the other hand, heat transfer continuum models are based on a more complex theory, are in general not capable of accurately resolving details on the length scales of prevalent heterogeneities (i.e., on the length scale of individual powder particles in the case of SLM), but are considerably less expensive for computation. Some simple cases even allow for the derivation of analytical solutions, for example by Gusarov~\cite{Gusarov2005}. Importantly, ray tracing models require specification of the powder bed structure by particle shape, dimension, and their coordinates, which is a non-trivial problem itself and often requires additional assumptions. In principle, the distinction of these two model classes is comparable to the distinction of atomistic and continuum models in other physical disciplines such as electricity or mechanics. Unfortunately, in the characterization of SLM processes, one is often interested in length scales comparable to the powder layer thickness. Because powder bed heterogeneities in form of individual particles are on the same length scale, the application of continuum models can often only be considered as an estimate of the underlying radiation transfer processes. Moreover, the assumption of a laser beam penetrating into a powder bed by means of multiple reflection actually only applies for sintering or for SLM processes based on point-wise modulated scanning strategies as long as no melt pool is prevalent.  When SLM is performed with linear/continuous scanning, however, the resulting melt pool front typically proceeds the laser beam position and, thus, the laser beam impacts directly on the melt pool and not on an undisturbed powder bed. In order to accurately resolve the impact of the laser beam on the melt pool, whose actual surface contour is typically considerably distorted and part of the solution itself, ray tracing techniques seem to be indispensable. Similar conclusions can be drawn by comparing mesoscopic and homogenized macroscopic models for the powder bed heat conduction. Because the local energy and temperature distribution within the powder bed but also within individual powder grains crucially influence the melting behavior on the mesoscopic scale and the creation of defects such as pores or inclusions, also sufficiently resolved mesoscopic models for the laser beam radiation transfer (e.g. via ray tracing) and the inter- and intra-particle heat conduction (e.g. by resolving these particles in the spatial discretization of the thermal problem) are mandatory if questions on this scale are relevant. However, when applying macroscopic SLM models, i.e. continuum approaches suffering from a powder homogenization error anyways, heat transfer continuum models appear to be a reasonable choice.

\subsection{Macroscopic simulation models}
\label{sec:macroscopicmodels}

Macroscopic simulation models in the context of SLM processes typically treat the powder phase as a homogenized continuum described by means of effective, i.e., spatially averaged, thermal and mechanical properties, without resolving individual powder grains. This homogenization procedure yields efficient numerical tools capable of simulating entire SLM builds of practically relevant size across practically relevant time scales. The thermal problem lies typically in the focus of interest when these models are applied. In some works, an additional solid-mechanics problem statement is considered aiming at the assessment of global residual stress distributions or dimensional warping effects based on the full thermo-mechanical interactions.\\

The thermal problem is commonly given by the following set of balance equations, boundary and initial conditions:
\begin{subequations}
\label{gusarov2007_HCE}
\begin{align}
\frac{\partial (\rho c_p T)}{\partial t} + \bigtriangledown \! \cdot \! (\rho c_p T \mb{v}) =  \bigtriangledown \! \cdot \! (\overbrace{\mb{k} \! \cdot \! \bigtriangledown T}^{=:-\mb{q}_k}) + u_s \quad &\forall (\mb{x},t) \in \Omega \times ]0,t_{end}],\label{gusarov2007_HCE1}\\
T=\bar{T} \quad &\forall (\mb{x},t) \in \Gamma_{T} \times ]0,t_{end}],\label{gusarov2007_HCE2}\\
\mb{q}=\bar{\mb{q}} \quad &\forall (\mb{x},t) \in \Gamma_{\mb{q}} \times ]0,t_{end}],\label{gusarov2007_HCE3}\\
 T = T_m \quad \text{and} \quad
\mb{n}_{sl} \cdot \left(\mb{k}_s \frac{\partial T_s}{\partial \mb{x}} -\mb{k}_l \frac{\partial T_l}{\partial \mb{x}} \right)
 = H_m \rho \, \mb{n}_{sl} \cdot \dot{\mb{x}} \quad & \forall (\mb{x},t) \in \Gamma_m \times ]0,t_{end}],\label{gusarov2007_HCE4}\\
  T=T_0 \quad &\forall (\mb{x},t) \in \Omega \times {0}.\label{gusarov2007_HCE5}
\end{align}
\end{subequations}
Here,~\eqref{gusarov2007_HCE1} represents a variant of the energy equation as typically employed in the macroscopic and mesoscopic SLM models on the problem domain $\Omega$ within the considered time interval $]0,t_{end}]$, where $\rho$ is the density, $c_p$ is the specific heat at constant pressure, $T$ is the temperature and $\mb{v}$ represents the velocity field. In~\eqref{gusarov2007_HCE1}, $\mb{k}$ is the thermal conductivity tensor and $u_s$ represents a heat source term. The two terms on the left-hand side of~\eqref{gusarov2007_HCE1} represent the material time derivative of the thermal energy density constituted of the local and the convective time derivative. It has to be mentioned that the thermodynamically consistent statement of thermo-mechanical problems involving \textit{compressible} materials would require additional thermo-mechanical coupling terms in~\eqref{gusarov2007_HCE1}. However, most of the models discussed in the following typically assume either exact or approximate incompressibility and resign these additional terms. Typically, isotropic conductivity is assumed, which yields $\mb{k}=\lambda \mb{I}$. Often, the source term $u_s$ is modeled based on a solution $I(\mb{x},\boldsymbol{\Omega})$ of the RTE according to~\eqref{gusarov2007_U}, e.g., the one proposed in~\cite{Gusarov2005}. Equation~\eqref{gusarov2007_HCE2} accounts for essential boundary conditions with prescribed temperature $\bar{T}$ on the boundary $\Gamma_{T}$ whereas~\eqref{gusarov2007_HCE3} represents natural boundary conditions with prescribed heat fluxes $\bar{\mb{q}}$ on the boundary $\Gamma_{\mb{q}}$ typically accounting for thermal convection $\bar{\mb{q}}_{c}$ and radiation emission $\bar{\mb{q}}_{e}$ on the powder surface with normal $\mb{n}_p$:
\begin{align}
\label{surfaceheatfluxes}
\bar{\mb{q}}_c = c (T-T_{ref})\mb{n}_p, \quad \quad \quad \bar{\mb{q}}_e = \epsilon k_{SB} (T^4-T_{ref}^4) \mb{n}_p.
\end{align}
Here, $T_{ref}$ denotes the temperature of the ambient gas atmosphere, $c$ is the convection coefficient and $\epsilon$ the emissivity. The Stefan-Neumann equation~\eqref{gusarov2007_HCE4} describes the phase change due to melting at the interface between the powder phase and the melt pool. In this equation, the indices $s$ and $l$ refer to the temperature gradients and thermal conductivities in the solid and liquid phase, respectively. Moreover, $T_m$ is the melting temperature, $H_m$ is the latent heat of melting and $\mb{n}_{sl}$ is the normal vector of the solid-liquid interface defining the melt process as a pure interface phenomenon. Eventually,~\eqref{gusarov2007_HCE5} prescribes the initial temperature $T_0$.\\

If the solid mechanics problem is also solved as well, the following system has to be considered in addition to~\eqref{gusarov2007_HCE}:
\begin{subequations}
\label{solid_balance}
\begin{align}
\rho \dot{\mb{v}} = \bigtriangledown \! \cdot \! \boldsymbol{\sigma} + \rho \mb{b} \quad &\forall (\mb{x},t) \in \Omega \times ]0,t_{end}],\label{solid_balance1}\\
\mb{u}(\mb{x}_u,t)=\bar{\mb{u}} \quad & \forall (\mb{x},t) \in \Gamma_{\mb{u}} \times ]0,t_{end}],\label{solid_balance2}\\
\boldsymbol{\sigma} \cdot \mb{n} \big|_{(\mb{x}_t,t)}=\bar{\mb{t}} \quad & \forall (\mb{x},t) \in \Gamma_{\mb{t}} \times ]0,t_{end}],\label{solid_balance3}\\
 \left(\mb{u}_I^+-\mb{u}_I^-\right)\big|_{\mb{x}_I}=\mb{0} \quad \text{and} \quad
 \left(\boldsymbol{\sigma}^+ -\boldsymbol{\sigma}^- \right)\cdot \mb{n}_I \big|_{(\mb{x}_I,t)}
 = \mb{0} \quad & \forall (\mb{x},t) \in \Gamma_I \times ]0,t_{end}],\label{solid_balance4}\\
 \mb{u}(\mb{x},0)=\mb{u}_0 \quad & \forall (\mb{x},t) \in \Omega \times {0},\label{solid_balance5}\\
  \mb{v}(\mb{x},0)=\mb{v}_0 \quad & \forall (\mb{x},t) \in \Omega \times {0}. \label{solid_balance6}
\end{align}
\end{subequations}
Equation~\eqref{solid_balance1} represents the mechanical equilibrium of linear momentum. In this equation, the Cauchy stress tensor $\boldsymbol{\sigma}(\mb{u},T)=\boldsymbol{\sigma}(\epsilon_e(\mb{u}),\epsilon_p(\mb{u}),\epsilon_T(T))$ is related to the primary displacement field $\mb{u}$ and temperature field $T$ by constitutive parameters with contributions from elastic $\epsilon_e$, plastic $\epsilon_p$ and thermal $\epsilon_T$ strains. The mechanical equilibrium of angular momentum is fulfilled by definition due to the symmetry of the Chauchy stress tensor. The total time derivative $\dot{\mb{v}}$ represents the material acceleration vector and $\mb{b}$ is the vector of volume forces acting on the physical domain $\Omega$. Similar to the thermal problem, essential and natural boundary conditions are given by equations~\eqref{solid_balance2} and~\eqref{solid_balance3} prescribing displacements $\bar{\mb{u}}$ and tractions $\bar{\mb{t}}$. Equation~\eqref{solid_balance4} represents the requirement of displacement continuity and mechanical equilibrium at the relevant interfaces, e.g. the interface powder-melt or the interface melt-solid, characterized by a normal vector field $\mb{n}_I$. The superscripts $(.)^+$ and $(.)^-$ denote quantities on the two different sides of the interface. However, most of the considered macroscopic models do typically not resolve these interfaces in the sense of a sharp 2D interface with discontinuous material parameters, but are rather based on a smooth, homogenized transition between the different phases, with the latter variant being easier to realize numerically. Finally, the initial position and velocity field is given by equations~\eqref{solid_balance5} and~\eqref{solid_balance6}.\\

The spatial discretization of the heat equation~\eqref{gusarov2007_HCE1} and the momentum equation~\eqref{solid_balance1} is typically based on the finite element method (FEM), which requires transfer of these equations into the (equivalent) weak form. For time integration, explicit as well as implicit approaches can be found. The coupling between the thermal and the solid-mechanical problem is often realized in a staggered partitioned manner. In the following, some of the most important representatives of macroscopic models available in the literature will briefly be discussed. As long as macroscopic models are considered, melt pool dynamics are not explicitly resolved and porosity-dependent, effective material properties are employed. The principal strategies of deriving macroscopic models  are comparable for SLS, SLM and EBM processes in many aspects. Thus, scientific contributions related to all of these processes will be considered. \\

In recent years, considerable research effort in the development of macroscopic, mesoscopic and microscopic models has been conducted at the Lawrence Livermore National Laboratory (LLNL), including a macroscopic homogenized continuum model for SLM processes described by Hodge et al. \cite{Hodge2014}. In this model, the laser beam heat source term has been taken from~\cite{Gusarov2005}. Heat losses occurring for example due to thermal emission, evaporation or mass ejection are taken into account by reducing the nominal total laser power $P$ to an effective total laser power, $P_e=2/3P$. In order to describe the phase transition, two spatial phase function fields $\phi_p(\mb{x}) \in [0;1]$ and $\phi_c(\mb{x}) \in [0;1]$ with $\phi_p(\mb{x})  \!+\! \phi_c(\mb{x}) \!=\! 1 \, \forall \, \mb{x}$ are introduced representing the powder and the consolidated phase., i.e.  $\phi_p(\mb{x})\!=\!1, \, \phi_c(\mb{x})\!=\!0$ for pure powder and $\phi_p(\mb{x})\!=\!0, \, \phi_c(\mb{x})\!=\!1$ for pure consolidated material. Importantly, the consolidated phase considered in this model is assumed to have a vanishing porosity and captures both, the melting as well as the solidified phase. In other words, only the phase transition from powder to liquid during melting is explicitly considered by means of a phase function, while the phase boundary between melt pool and solidified material is accounted for by means of high gradients in the material properties at these spatial locations. All of the relevant thermal material parameters (e.g. conductivity, heat capacity etc.) are considered to be a function of both temperature and phase (thus, of porosity). In the spatial discretization process based on the finite element method (FEM) and the subsequent numerical implementation, the phase boundary between powder and melt pool is not resolved in a sharp manner. Instead, there is typically a small band of finite elements where both phases are prevalent, a fact that is reflected by non-vanishing values $0 < \phi_p(\mb{x}),\phi_c(\mb{x})<1$. Correspondingly, the Stefan-Neumann equation is not evaluated in its original form~\eqref{gusarov2007_HCE4} at a 2D interface, but rather in an equivalent form within 3D volume elements.\\

While the phase boundary between powder and melt pool is tracked by means of phase variables, the Stefan-Neumann equation~\eqref{gusarov2007_HCE4} (or a spatially average version of it) is not considered for the phase transition from melting to solid. Instead, a high value of the heat capacity is chosen in the range of the melting temperature in order to account for the released latent heat during the solidification process. Thus, the typical temperature plateau observable in enthalpy-temperature diagrams of elementally pure materials at the melting point is replaced by an interval of very slowly increasing temperature. From a physical point of view, this might rather be comparable to the melting behavior of alloys between liquidus and solidus temperatures $T_l$ and $T_s$. From a numerical point of view, it can be interpreted as a (commonly employed) regularization of the constraint equation $T=T_m$ in~\eqref{gusarov2007_HCE4} of elementally pure materials in order to simplify the numerical solution process.\\

Hodge et al.~\cite{Hodge2014} considered the thermo-mechanical problem by coupling the thermal and mechanical simulations in an iteratively partitioned manner. There, an elasto-plastic material law, additionally accounting for thermal expansion and consolidation shrinkage, has been employed while the boundaries between powder, melt and solidified phase have been considered implicitly via strong gradients of the associated mechanical material parameters with respect to temperature and porosity. This means, the interfaces given in equation~\eqref{solid_balance4} are not explicitly resolved in the mechanical simulation. Furthermore, this elasto-plastic material model has been applied to all three phases, i.e. to the powder phase, to the solidified phase, and also to the liquid phase. Thus, a detailed modeling of the melt pool fluid dynamics, as prevalent in the mesoscopic models discussed in Section~\ref{sec:mesoscopicmodels}, is missing. This statement applies in a similar manner to most of the macroscopic SLM models considered in this section.\\

\begin{figure}[h!!]
 \centering
   \subfigure[\cite{Gusarov2007}: low velocity.]
   {
    \includegraphics[height=0.30\textwidth]{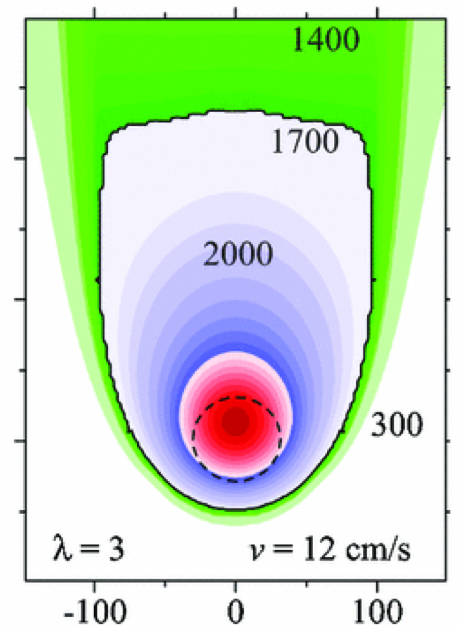}
    \label{fig:hodge2014_meltpool1}
   }
   \subfigure[\cite{Hodge2014}: low velocity.]
   {
    \includegraphics[height=0.30\textwidth]{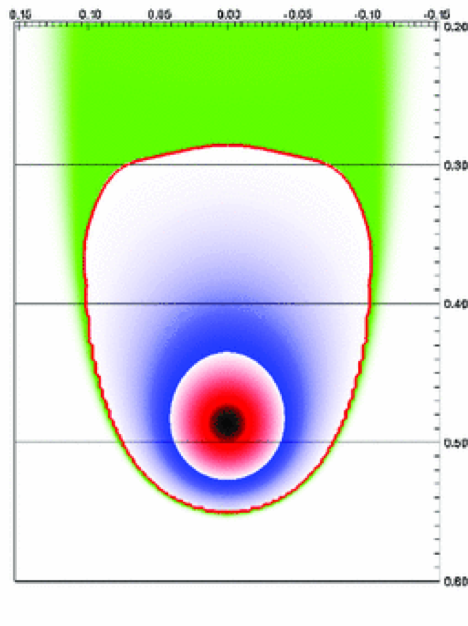}
    \label{fig:hodghe2014_meltpool1}
   }
   \subfigure[\cite{Gusarov2007}: high velocity.]
   {
    \includegraphics[height=0.30\textwidth]{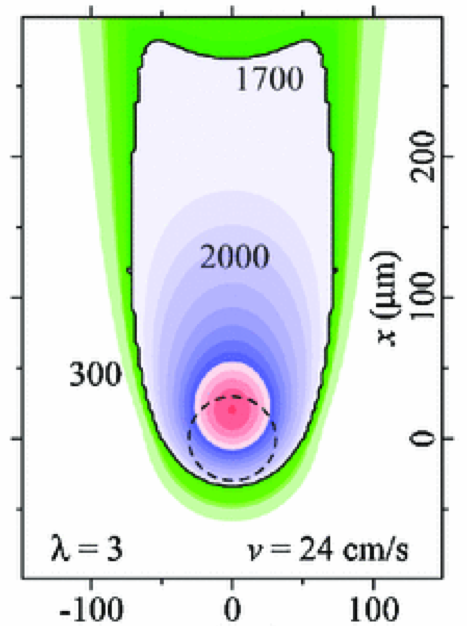}
    \label{fig:hodghe2014_meltpool3}
   }
   \subfigure[\cite{Hodge2014}: high velocity.]
   {
    \includegraphics[height=0.30\textwidth]{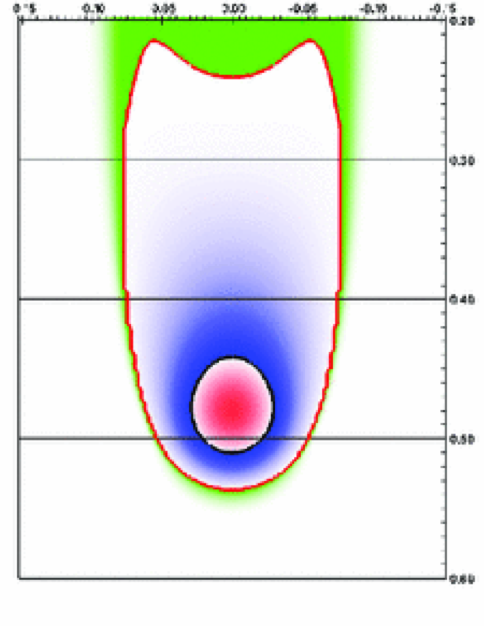}
    \label{fig:hodghe2014_meltpool4}
   }
  \caption{Comparison between melt pool geometries derived via macroscopic SLM models in~\cite{Gusarov2007} and~\cite{Hodge2014} for different scan velocities.}
  \label{fig:hodge2014_meltpool}
\end{figure}

The thermal simulations in~\cite{Hodge2014} confirmed the results of Gusarov et al.~\cite{Gusarov2007} for 316L stainless steel concerning maximal temperature, top surface shape as well as cross-section shape of the melt pool. It was observed that the melt pool shape gets narrower and longer with increasing scan speed and that at some point a concave region at the tail of the melt pool boundary occurs, an effect that can be explained by the higher thermal conductivity of the solidified material behind the melt pool as compared to the powder material to the sides (Figure~\ref{fig:hodge2014_meltpool}). Nevertheless, when considering the similarities between the two models proposed in~\cite{Gusarov2007} and in~\cite{Hodge2014}, it has to be emphasized that both approaches are based on the same radiation transfer model for the incident laser energy, and, that the chosen submodel for the heat source might considerably influence the overall results since the thermal conductivity within the powder phase is strongly limited by the low conductivity of the atmospheric gas filling the powder pores.\\

\begin{figure}[h!!]
 \centering
   \subfigure[Step 1, driving to the left.]
   {
    \includegraphics[height=0.2\textwidth]{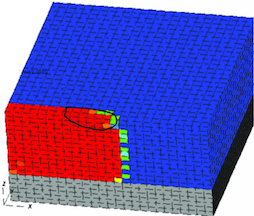}
    \label{fig:hodghe2014_overhang1}
   }
   \subfigure[Step 2, driving to the left.]
   {
    \includegraphics[height=0.2\textwidth]{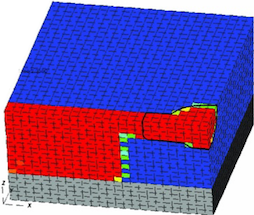}
    \label{fig:hodghe2014_overhang2}
   }
      \subfigure[Step 3, driving to the right.]
   {
    \includegraphics[height=0.2\textwidth]{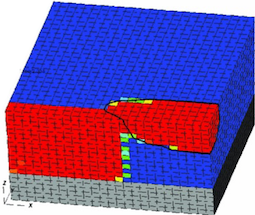}
    \label{fig:hodghe2014_overhang3}
   }
   \subfigure[Step 4, driving to the right.]
   {
    \includegraphics[height=0.2\textwidth]{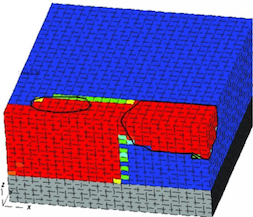}
    \label{fig:hodghe2014_overhang4}
   }
  \caption{Laser beam scanning across an overhang. First, the beam moves to the left (steps 1+2), than back to the right (steps 3+4). Blue color represents loose powder, red color represents consolidated material (melt and solid). The melt pool shape is indicated by black contour lines,~\cite{Hodge2014}.}
  \label{fig:hodge2014_overhang}
\end{figure}

Because the model in~\cite{Hodge2014} explicitly considered the powder-melting phase change via $\phi_p(\mb{x})$ and $\phi_c(\mb{x})$, the inertia of the melting process could be visualized by means of a melt pool boundary lagging behind the isothermal of the actual melting temperature in the range of high scan velocities. Furthermore, \cite{Hodge2014} investigated the melt pool behavior when scanning across overhangs supported by loose (poorly conductive) powder (Figure~\ref{fig:hodge2014_overhang}). Experimental results~\cite{Craeghs2010,Wang2013} could be confirmed that typically larger melt pool sizes, excessive overheating and evaporation occur when applying the same scan parameters in domains of low thermal conductivity such as overhangs or thin-walled features. Commonly, this problem is addressed by adapting the laser parameters and/or increasing the net thermal conductivity in this regions by means of additional support columns.\\

\begin{figure}[h!!!]
\begin{center}
    \subfigure[Experimental measurements.]
    {
    \includegraphics[width=0.31 \textwidth]{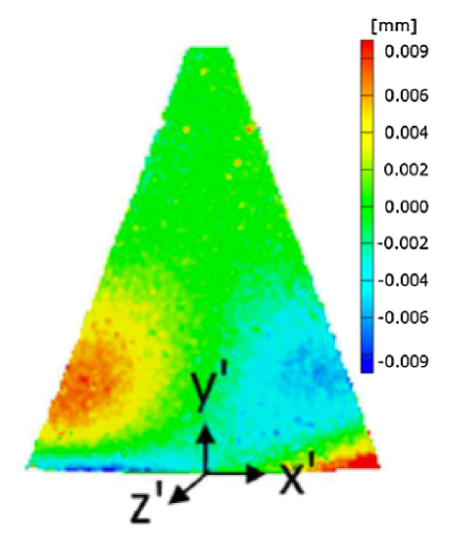}
    \label{fig:hodge2016_1a}
    }
    \subfigure[Simulation: no plastic hardening.]
    {
    \includegraphics[width=0.31 \textwidth]{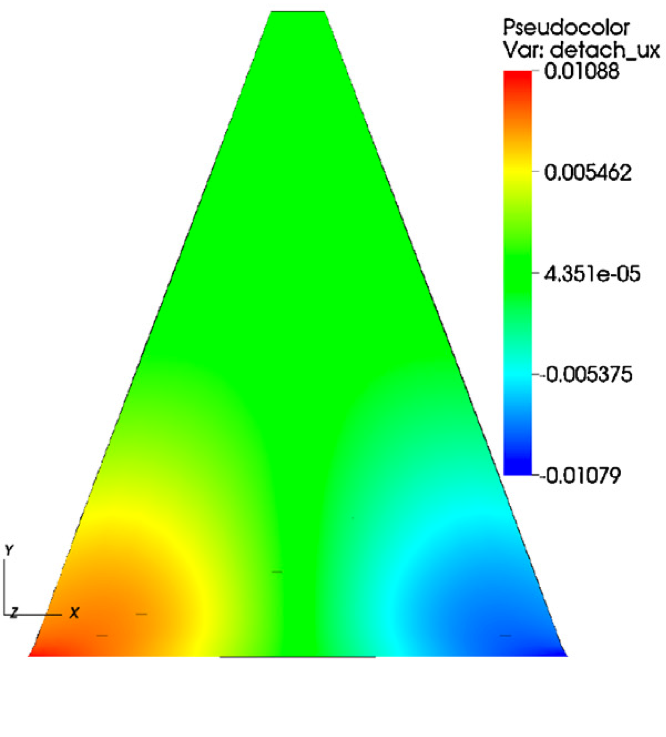}
    \label{fig:hodge2016_1b}
    }
    \subfigure[Simulation: with plastic hardening.]
    {
    \includegraphics[width=0.31 \textwidth]{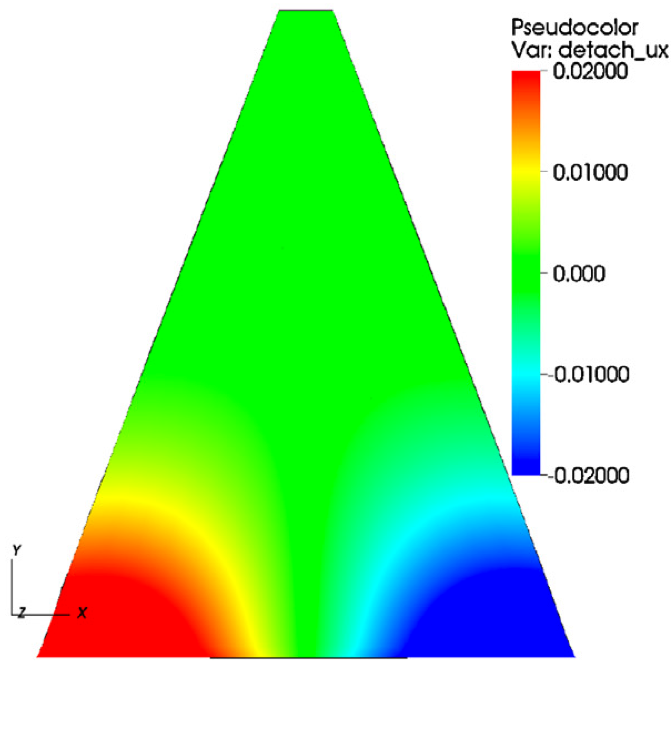}
    \label{fig:hodge2016_1c}
    }
    \caption{Comparison of predicted and actual displacements in $x-$direction arising from dimensional warping of SLM triangular prisms,~\cite{Hodge2016}.}
    \label{Fig_Hodge2016}
\end{center}
\end{figure}

\begin{figure}[h!!!]
\begin{center} 
    \includegraphics[width=0.75 \textwidth]{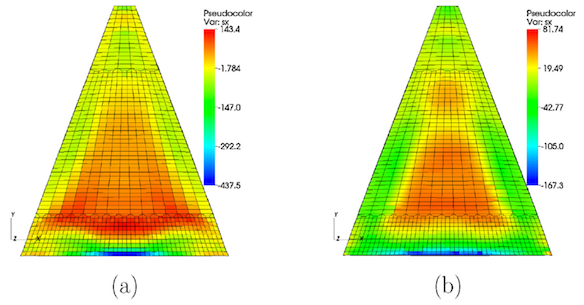}
    \caption{Simulated Cauchy stress in $x-$direction for the cases (a) non-heated and (b) preheated built platform,~\cite{Hodge2016}.}
    \label{fig:hodge2016_2}
\end{center}
\end{figure}

Further validation of the Hodge simulation via experiments is given in the follow-up work~\cite{Hodge2016}, which also investigates the effects of build orientation, infill scan strategy, and preheating on residual stresses and dimensional warping.  Digital Image Correlation (DIC) was used to measure the displacements of 316L triangular prisms fabricated by SLM as test artifacts. Moreover, neutron diffraction experiments were used to measure the stress distribution. Figure~\ref{Fig_Hodge2016} illustrates typical results for a triangular prism printed vertically, with the shortest edge of the triangle forming one edge of the rectangular contact area with the build platform. The prisms are $5mm$ along the bottom edge, $6mm$ tall, and $1mm$ thick, in part to enable comparison to a previous study in~\cite{Wu2014}. In order to measure dimensional warping, the prism is removed from the build platform leading to a considerable distortion of the part geometry due to the previously present residual stresses. In Figure~\ref{fig:hodge2016_1a}, the experimentally measured displacements in $x-$direction, taking on maximal values in the range of $0.01mm$, are illustrated. For comparison, in Figures~\ref{fig:hodge2016_1b} and~\ref{fig:hodge2016_1c} the corresponding simulation results either based on a material law with or without plastic hardening, are illustrated. Besides a qualitative agreement of numerical and experimental results, the relative error of the simulation without plastic hardening lies in the range of $20\%$ in the maximal displacement while the variant with plastic hardening overestimates the maximal displacement approximately by a factor of two. For the displacements in $y-$direction (not illustrated), the behavior turned out to be (almost) the opposite, i.e. rough agreement of the simulation based on a material law with plastic hardening, large deviations for the variant without plastic hardening. This behavior might indicate that the comparatively simple isotropic constitutive laws applied in~\cite{Hodge2016}, and in many other state-of-the-art macroscopic SLM models, are still not accurate enough. Additionally, the simulation results in Figure~\ref{fig:hodge2016_2} show the $x-$component of the Cauchy stresses resulting from the two chosen preheating temperatures of $303K$ and $1000K$. Again, qualitative results can be captured quite well confirming that preheating reduces the peak values of residual stresses.\\ 

A slightly different macroscopic SLM model has been proposed in~\cite{Gusarov2007}, where the radiation transfer model derived in the authors' earlier work~\cite{Gusarov2005} has been supplemented by a thermal simulation framework based on equations~\eqref{gusarov2007_HCE}. In contrast to~\cite{Hodge2014}, no additional phase function field is introduced in order to explicitly solve~\eqref{gusarov2007_HCE4} for the interface between powder and melt pool. Thus, this interface is only given implicitly by the isothermal line $T=T_m$. Because the effective thermal conductivity $\mb{k}$ of loose metallic powders at low temperatures is controlled by gas in the pores~\cite{Rombouts2005}, the effective thermal conductivity of the $10$ - $50 \mu m$ powders considered in~\cite{Gusarov2007} has been chosen to $\mb{k}=k_p\mb{I}$ employing the "gas conductivity" $k_p = 0.3 W/(m K)$ at temperatures below the melting point. Surface contacts between particles are formed when powder melts, so that the thermal conductivity of the powder rapidly approaches the value of the solid material, chosen as $\mb{k}=k_d\mb{I}$ with $k_d= 20 W/(m K)$ for the investigated stainless steel type 316 L. In the model of~\cite{Gusarov2007}, a sharp change from $k_p$ to $k_d$ at the melting point is assumed. Based on the simulated temperature fields and the derived isothermals $T=T_m$, the melt pool geometry has been analyzed for different laser beam scan velocities (see Figure~\ref{fig:hodge2014_meltpool1} and~\ref{fig:hodghe2014_meltpool3} as well as the corresponding discussion in the paragraph above). In~\cite{Gusarov2011}, the pure thermal simulation framework of~\cite{Gusarov2007} has been extended to thermo-mechanics incorporating a simple linear-elastic, plain strain version of~\eqref{solid_balance} with elastic constitutive parameters jumping at the phase boundaries.\\

\begin{figure}[h!!]
 \centering
   \subfigure[Temperature-dependence of specific heat $c_p$.]
   {
    \includegraphics[height=0.3\textwidth]{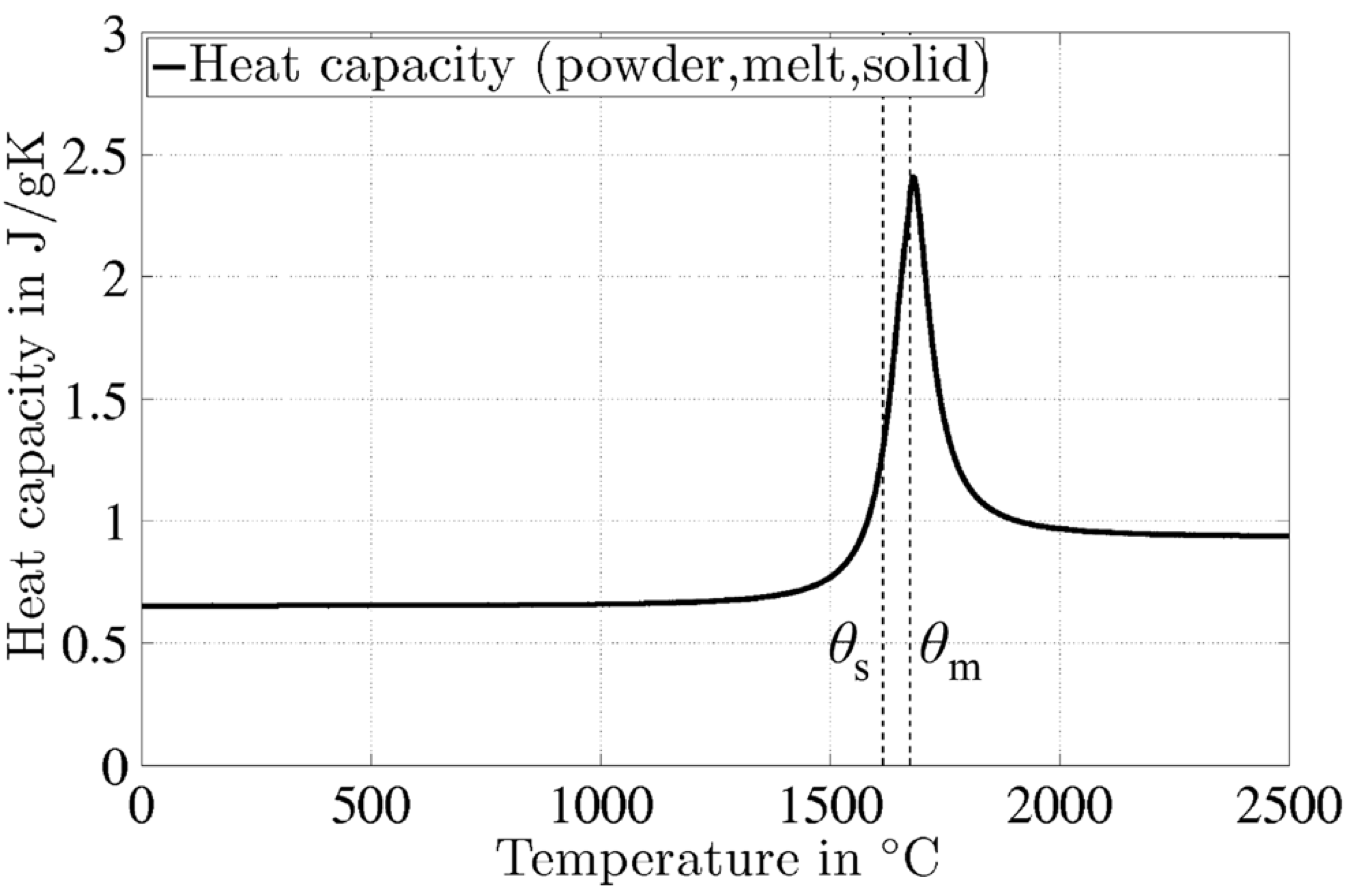}
    \label{fig:riedlbauer2016cp}
   }
   \hspace{0.05\textwidth}
   \subfigure[Temperature-dependence of thermal conductivity $k$.]
   {
    \includegraphics[height=0.3\textwidth]{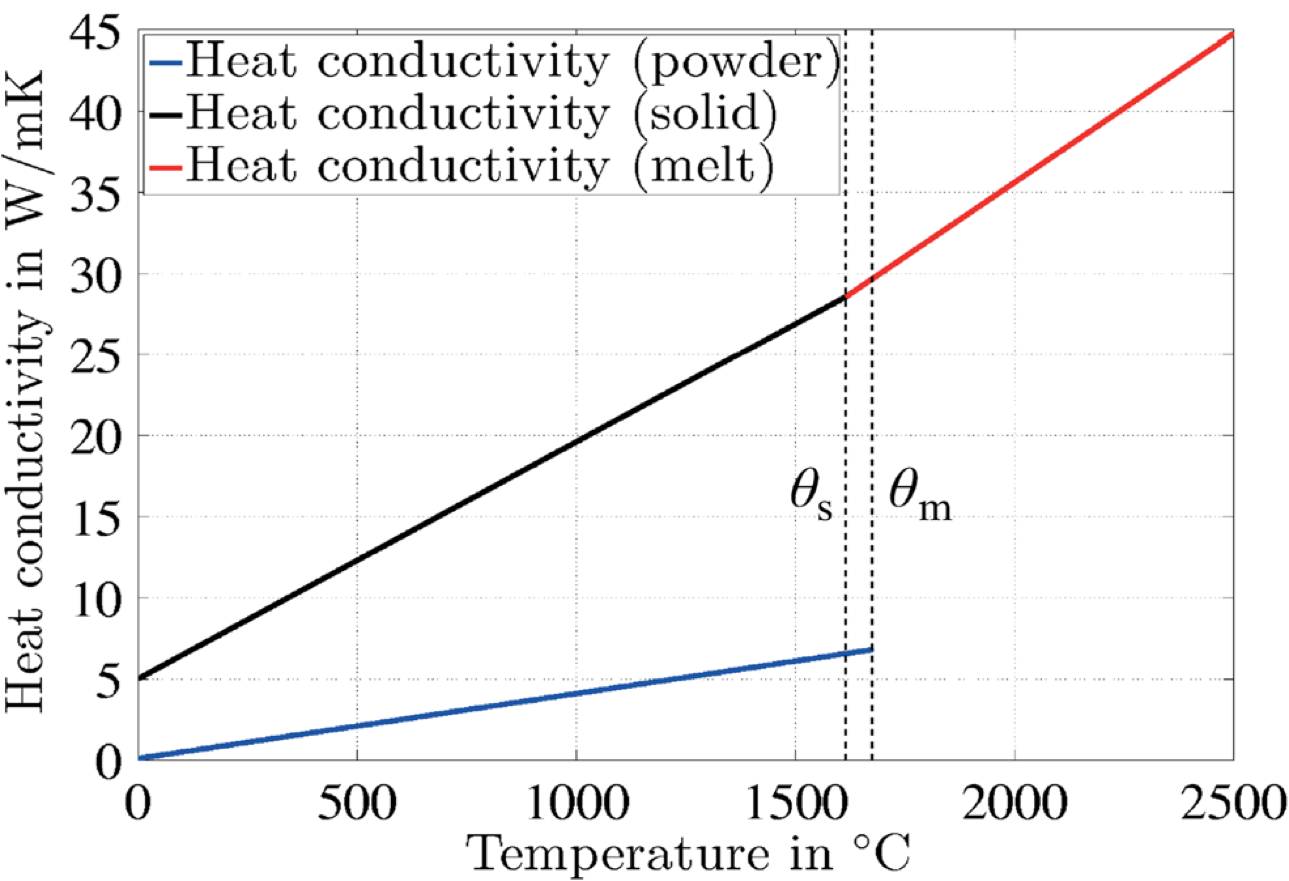}
    \label{fig:riedlbauer2016k}
   }
  \caption{Dependence of thermal material parameters $c_p$ and $k$ of Ti-6Al-4V on current temperature and considered phase,~\cite{Riedlbauer2016}.}
  \label{fig:riedlbauer2016kcp}
\end{figure}

In~\cite{Riedlbauer2016}, the pure thermal problem~\eqref{gusarov2007_HCE} has been solved for the EBM process based on the finite element method for spatial discretization and an implicit Runge-Kutta method for temporal discretization. Additionally, adaptive mesh refinement strategies have been applied in the direct vicinity of the electron beam. In this model, the penetration of the electron beam into the material is described by means of the model~\cite{Klassen2014} and thermal losses due to emission have been considered. The different phases of the investigated Ti-6Al-4V alloy are considered in the thermal model via phase- and temperature-dependence of the thermal material parameters. The heat capacity $c_p$ is prescribed as a nonlinear function of temperature with high gradients around the melting point in order to account for the latent heat of melting (Figure~\ref{fig:riedlbauer2016cp}). The temperature-dependence of the heat conductivity $k$ has been described by a linear law that differs for the three different phases powder, melt and solidified material (Figure~\ref{fig:riedlbauer2016k}). In~\cite{Riedlbauer2014}, the same authors considered the coupled thermo-mechanical problem~\eqref{gusarov2007_HCE} and~\eqref{solid_balance}, wherein the energy equation~\eqref{gusarov2007_HCE1} has been derived in a thermodynamically consistent manner from a free energy density functional leading to the well-known thermo-mechanical coupling term of compressible materials not only in the momentum equation~\eqref{solid_balance} but also in the energy equation~\eqref{gusarov2007_HCE1}. The focus of this work lay in comparing the overall computational efficiency of two different numerical coupling schemes, a monolithic coupling scheme solving the thermal and structural mechanics problem simultaneously as well as a staggered partitioned scheme proposed in~\cite{Armero1992}. The staggered partitioned scheme allowed for lower computational costs per time step due to the smaller system sizes resulting from a separate treatment of the thermal and structural mechanics problem. However, for stability reasons, the partitioned scheme required a considerable smaller time step size resulting in a higher overall computational effort as compared to the monolithic scheme. Consequently, the authors recommend the latter category of coupling schemes for future application to problem classes such as SLM.\\

Much work by the group of Stucker et al.~\cite{Pal2014} has focused on the development of FEM-based simulation tools for SLM processes, and has been commercialized as the software package 3DSIM. In~\cite{Patil2015,Pal2016,Zeng2013}, the thermal problem~\eqref{gusarov2007_HCE} is solved, modeling the laser beam as surface heat flux with Gaussian distribution and accounting for heat losses on the powder bed surface due to thermal convection. In this work, an implicit Crank Nicolson scheme for time integration and a spatial finite element discretization combined with a dynamic adaptive mesh refinement and de-refinement algorithm have been employed. While the physical model is comparatively simple and only treats the pure thermal problem, the dynamic mesh adaption algorithm might allow for considerable efficiency gains as compared to uniform meshing strategies. In~\cite{Zeng2015}, further steps in improving the aforementioned simulation models are conducted by developing homogenization strategies based on representative volume elements (RVE) for specific part geometries with lattice-like support structures. Among others, these tools have been applied for predicting dimensional warping and suggesting strategies of geometrical compensation.\\

For the homogenized, macroscopic continuum models as considered in this section, it can generally be expected that high spatial gradients in the thermo-mechanical solution of SLM problems will predominantly occur in the direct vicinity of the laser beam and at spatial locations where also the individual part geometry is characterized by very fine length scales (e.g. transition from high to very low wall thicknesses etc.). Similarly, high temporal gradients are primarily expected in the direct neighborhood of the laser beam position. Consequently, thermo-mechanical simulation approaches based on time- and space-dependent adaption schemes for the spatial (see e.g.~\cite{Zeng2013} as discussed above) \textit{and} for the temporal discretization are very promising to save computational costs.\\

A further commercial FEM software for SLM simulation is offered by Pan Computing, and was acquired by Autodesk in 2016. For example, the work of~\cite{Denlinger2014} extended the contributions discussed in this section so far by allowing for residual stress relaxation. Specifically, the stress and plastic strain of the underlying thermo-elasto-plastic constitutive model were reset to zero, when the temperature exceeded a prescribed annealing temperature. Based on experimental verifications and thermo-mechanical simulations of Ti-6Al-4V parts, it was claimed that temperature-induced stress-relaxation plays a crucial role in EBM processes and that the error in maximal residual stresses and geometrical distortion could reach a value up to $500\%$ as compared to their model without annealing effects. In~\cite{Denlinger2014a}, the computational efficiency of the model~\cite{Denlinger2014} has been increased by employing an inactive element activation strategy as well as a dynamic mesh coarsening algorithm. In the context of FEM-based SLM process simulations, typically two different strategies of material deposition are distinguished: The notion of quiet elements refers to a scheme were all elements representing the final part are already present in the beginning of the SLM process, whereat elements in power layers that have not been deposited yet are characterized by artificial material properties (e.g. very low thermal conductivity $k$, specific heat $c_p$, Young's modulus $E$ etc.) that are intended to rarely influence the results within the already deposited layers. On the other hand, the notion of \textit{inactive elements} refers  to a scheme that adds new finite elements with every newly deposited layer while elements associated with subsequent layers are not yet present. Of course,  quiet element schemes result in a higher computational effort since a large system of equations (containing all degrees of freedom of the final part) has to be solved in every time step and the resulting system matrices might be ill-conditioned as consequence of the artificial material parameters. On the other hand, inactive element schemes are often not supported by commercial FEM codes.\\

There is an abundant number of further scientific contributions solving the thermal or thermo-mechanical problem prevalent in SLM processes based on macroscopic models and FEM discretization schemes in a manner similar to the examples already given above, with~\cite{Hodge2014} being an exception that explicitly tracks the interface between powder and melting phase.  While the SLM models discussed so far typically assume temperature- and porosity-dependent thermal properties of the powder based on the initial powder bed porosity, in~\cite{Cervera1999} the energy equation~\eqref{gusarov2007_HCE} as well as an evolution equation for the time- and temperature dependent powder bed porosity field during the sintering process based on a model according to~\cite{Scherer1976,Mackenzie1949} have been solved simultaneously. These two equations have been coupled by means of a model for porosity-dependent powder bed conductivity proposed in~\cite{Yagi1957}. In contrast to this SLS model, an explicit solution of the powder sintering state and porosity in order to accurately determine the current powder conductivity is rather untypical for approaches considering SLM. This may be explained by the fact that the time interval separating the states of loose powder and fully molten liquid is rather short and the modeling error by assuming a simpler ad-hoc relation between powder temperature, porosity and conductivity seems not to be essential compared to the basic assumptions underlying typical macroscopic SLM models. In the works of Dai et al.~\cite{Dai2001,Dai2004,Dai2005,Dai2006}, a 3D model for SLS / SLM has been proposed to solve the thermo-mechanical problem~\eqref{gusarov2007_HCE} and~\eqref{solid_balance} defined by a small number of subsequent powder layers with a special emphasis on multi-material parts. The thermal problem accounts for heat losses due to radiation and convection and employs a Gaussian laser energy distribution. The structural mechanics problem is defined via a constitutive law accounting for elastic, plastic and thermal strains. The effect of volume shrinkage during melting of loose powder is considered by means of a rather heuristic model that moves finite element nodes in direction of gravity by a distance defined via the initial powder porosity and mass conservation. In the work~\cite{Matsumoto2002}, a very simple 2D thermo-mechanical model considering one powder layer has been proposed. The mechanical model assumes a linear-elastic state of plane stress and accounts for elastic and thermal strains. The thermal and structural-mechanics problems are solved in a partitioned manner based on the FEM. The contribution~\cite{Childs2005} proposes a finite element model for the thermal problem of SLS / SLM processes. There, two different empiric laws have been proposed for the temperature-dependence of powder porosity during the sintering process. Moreover, a porosity-dependent law for the powder conductivity as observed experimentally in~\cite{Shiomi1999} has been considered. The laser beam has been modeled as volumetric heat source with an intensity decreasing linearly with the penetration depth. In~\cite{Roberts2009}, a computational model for determining the temperature history in a practical SLM process involving the scanning of (up to five) multiple layers has been proposed. The model builds on the previously developed concepts in~\cite{Cervera1999} and~\cite{Kolossov2004} and extends these approaches to the FEM-based simulation of multiple layers. A comparatively rough yet efficient model is proposed in~\cite{Zaeh2010} to estimate residual stresses and dimensional warping effects occurring during the cooling phase of SLM processes. The principal idea is to combine several layers of the SLM process and to apply an equivalent thermal load to this integrated, virtual layer within the heating cycle and to determine residual stresses and dimensional warping effects in a subsequent cooling cycle of the staggered thermo-mechanical scheme. More recent thermal and thermo-mechanical FEM-based SLM models similar to the ones discussed above are e.g. given by~\cite{Shen2012,Shen2012a,Hussein2013,Chen2014,Wits2016}. In~\cite{Shen2012,Shen2012a}, an increased thermal conductivity of the melt pool has been considered to account for the convective melt pool heat transfer, however, without modeling hydrodynamics.\\

In contrast to the finite element models considered so far, in~\cite{Verhaeghe2009} the finite volume method (FVM) has been employed for spatial discretization in combination with an explicit Euler time integration scheme. The underlying model considers the pure thermal problem~\eqref{gusarov2007_HCE}, including the phases of powder, melt, solid \textit{and} evaporated gas, in order to determine melt pool shapes. Furthermore, the radiation transfer model originally proposed by~\cite{Gusarov2005} has been employed. Based on the resulting temperature field, the volume fractions of the phases powder, melt, solid and gas have been determined for each computational cell defined by the FVM. Based on comparisons with experimental data, it was found that the consideration of evaporative heat loss is essential for realistic temperature field results.\\

An interesting approach that crucially differs from the ones discussed so far has been proposed in~\cite{jamshidinia2012,jamshidinia2013}. This model somehow combines the procedures typically applied to macroscopic and mesoscopic SLM models. The powder phase is modeled as homogenized continuum with effective, porosity-dependent properties as typical for macroscopic models, whereas the fluid dynamics within the melt pool are explicitly modeled as it is common for mesoscopic SLM models. The melt pool fluid flow is modeled as a compressible, laminar Newtonian liquid flow governed by the Navier-Stokes equations~\eqref{flow_balance} accounting e.g. for temperature-dependent surface tension, temperature-dependent buoyancy forces as well as frictional dissipation in the mushy zone typical for alloys such as the considered Ti-6Al-4V material at the powder-fluid interface. The thermal problem is described on the basis of~\eqref{gusarov2007_HCE} with a laser beam energy source as proposed in~\cite{Zaeh2010a} as well as heat losses due to thermal radiation. After solving the thermo-hydrodynamics problem based on a commercial, FVM-based computational fluid dynamics (CFD) solver, the resulting temperature fields are used as input data for a subsequent structural mechanics simulation in the sense of a partitioned field coupling. By employing a commercial FEM code, the structural mechanics problem~\eqref{solid_balance} in combination with a thermo-elasto-plastic material law is solved in order to determine residual stress distributions in the solidified phase. All thermal and mechanical properties relevant for the model have been assumed to be temperature- and porosity-dependent. While the employed laser beam model~\cite{Zaeh2010a} is only valid for EBM processes, the remaining model constituents seem to be directly transferable to related processes such as SLM. Although the applied field and domain coupling schemes appear to be comparatively rudimentary, the general idea of employing efficient, homogenized macroscale powder models while simultaneously considering the convection-driven heat transfer in the melt pool on the basis of CFD simulations can be regarded as very appealing. Some basic physical phenomena typically resolved by the mesoscopic models discussed in the next section but not accessible by the models considered above, e.g. increased melt pool widths and decreased melt pool peak temperatures as consequence of an outward-directed Marangoni flow, could already be captured by this model, although at a considerably lower resolution - and numerical cost - as compared to the mesoscopic models. A considerable number of similar approaches that might also be relevant for SLM can e.g. be found in the fields of laser and electron beam welding (see e.g.~\cite{Bachmann2016,Chang2015,Ganser2016,Geiger2009,Ki2002,Ki2002a,Li2004,Rai2008,Semak1999}). Future SLM simulation models could strongly benefit from these works.

\subsection{Mesoscopic simulation models}
\label{sec:mesoscopicmodels}

While macroscopic models do not consider physical phenomena on the length scale of individual powder particles, mesoscopic models explicitly resolve these scales in order to study melt pool dynamics, melt pool heat transport as well as the wetting of melt on substrate and powder particles. Typically, these models aim at the prediction of part properties such as layer-to-layer adhesion, surface quality and defects on the mesoscale (pores, inclusions etc.). In these models, the initial powder grain distribution is either determined on the basis of contact mechanics simulations, e.g. by employing the discrete element method (DEM, see e.g.~\cite{Lee2015}), or by means of more generic packing algorithms such as rain models~\cite{Korner2011,Meakin1987}. Subsequently, thermo-mechanical simulations are performed considering the heat transfer within the powder bed, the melting process as well as the heat transfer and the hydrodynamics in the melt pool. Often, the unmelted powder grains are assumed to be spatially fixed and the powder phase is only solved for the pure thermal problem based on a proper laser beam model (see Section~\ref{sec:powder_modeling_continuum} and~\ref{sec:powder_modeling_raytracing}). The melt pool fluid dynamics are commonly modeled by means of energy equation~\eqref{gusarov2007_HCE} considering the latent heat of melting as well as the following set of balance equations~\eqref{flow_balance}, accounting for conservation of mass and momentum of an incompressible flow:
\begin{subequations}
\label{flow_balance}
\begin{align}
\bigtriangledown \! \cdot \! \mb{v} = 0 \quad &\forall (\mb{x},t) \in \Omega \times ]0,t_{end}],\label{flow_balance1}\\
 \frac{\partial (\rho \mb{v})}{\partial t} + \bigtriangledown \! \cdot \! (\rho\mb{v} \otimes \mb{v})  
 = \bigtriangledown p + \bigtriangledown \! \cdot \! (2 \mu \tilde{\boldsymbol{\epsilon}})  + \rho \mb{b} \quad &\forall (\mb{x},t) \in \Omega \times ]0,t_{end}],\label{flow_balance2}\\
\mb{v}=\bar{\mb{v}} \quad &\forall (\mb{x},t) \in \Gamma_{\mb{v}} \times ]0,t_{end}],\label{flow_balance3}\\
\boldsymbol{\sigma} \!\cdot\! \mb{n} =\bar{\mb{t}} \quad &\forall (\mb{x},t) \in \Gamma_{\mb{t}} \times ]0,t_{end}],\label{flow_balance4}\\
 \left(\mb{u}_I^+ -\mb{u}_I^- \right)=\mb{0} \quad \text{and} \quad
 \left(\boldsymbol{\sigma}^+ -\boldsymbol{\sigma}^- \right) \!\cdot\! \mb{n}_I
 = [\gamma \kappa_I \mb{n}_I + \frac{d \gamma}{d T}(\mb{I} - \mb{n} \otimes \mb{n})\bigtriangledown \! T] \quad &\forall (\mb{x},t) \in \Gamma_I \times ]0,t_{end}],\label{flow_balance5}\\
  \mb{v}=\mb{v}_0 \quad &\forall (\mb{x},t) \in \Omega \times {0}.\label{flow_balance6}
\end{align}
\end{subequations}
In the above system of equations,~\eqref{flow_balance1} is the continuity equation requiring conservation of mass,~\eqref{flow_balance2} is the Navier-Stokes momentum equation,~\eqref{flow_balance3} and~\eqref{flow_balance4} represent essential and natural boundary conditions on the boundaries $\Gamma_{\mb{v}}$ and $\Gamma_{\mb{t}}$, while~\eqref{flow_balance6} prescribes the initial velocity field at $t=0$. The flow is typically assumed to be laminar and incompressible and the fluid to be Newtonian with a stress tensor $\sigma(\mb{v},p)=-p\mb{I} + 2 \mu \tilde{\boldsymbol{\epsilon}}(\mb{v})$. Here, $\tilde{\boldsymbol{\epsilon}}(\mb{v})$ represents the rate-of-deformation tensor, $p$ is the pressure and $\mu$ is the kinematic viscosity. The body force term $\mb{b}$ typically accounts for gravity. This term might contain additional contributions, e.g. related to the phase change from metal powder to liquid  or from temperature-induced buoyancy forces, as discussed below. The equation ~\eqref{flow_balance5} accounts for surface tension effects on the free-surface $\Gamma_{I}$ between melt pool and ambient gas. While the velocity field is assumed to be continuous (left side of~\eqref{flow_balance5}) the surface stress exhibits a jump at the interface (right side of~\eqref{flow_balance5}). The jump in interface-normal direction (first term in square brackets) is already present for the case of spatially constant surface tension $\gamma$ leading to fluid surfaces with curvature $\kappa_I$ and capillary effects already for hydrostatic problems. The jump in the interface-tangential direction (second term in square brackets) only occurs for spatially varying surface tension values, here solely considered due to spatial temperature gradients, and will for Newtonian fluids (with shear stresses being proportional to velocity gradients) always induce flow. The Marangoni effect resulting from such surface tension variations crucially influences the convective heat transfer within the melt pool. All in all, surface tension significantly determines the melt pool shape and the resulting solidified surfaces (see \cite{Khairallah2016,Qiu2015,Lee2015}).\\

For the numerical solution of the mathematical problem defined by~\eqref{gusarov2007_HCE} and~\eqref{flow_balance}, often a staggered solution scheme relying on a sequential solution of the thermal problem in the powder bed and the fluid dynamics problem in the melt pool is applied. In such a partitioned solution procedure, the powder particle zones with $T>T_m$ are typically identified as mass increments per time step that are transferred from the solid to the fluid phase due to melting. Consequently, the contour lines $T=T_m$ define the boundary conditions for the subsequent fluid dynamics solution step. Many of the existing approaches consider a spatial discretization via the finite volume method (FVM) in combination with the volume of fluid (VOF) method in order to track the fluid-gas interface. Temporal discretization of the fluid dynamics problem is predominantly based on explicit time integration schemes. The required resolution of the spatial discretization and the small time step sizes admissible for explicit time stepping schemes lead typically to high computational costs, allowing for an application of these 3D mesoscopic models only to single-/dual-track simulations so far.\\

Arguably one of the best-known mesoscopic models of this kind has been presented by Khairallah et al.~\cite{Khairallah2014}. There, the continuum radiation transfer model~\cite{Gusarov2005} as discussed in Section~\ref{sec:powder_modeling_continuum} has been employed as a laser beam source term, while individual particles are resolved on the mesoscale in order to explicitly consider the conductive heat transfer within the powder bed limited by the atmospheric gas in the pores and the point contacts between the powder grains (see~\cite{Rombouts2005}).  For spatial discretization, the finite element method based on a uniform Cartesian mesh with element size $3 \mu m$ has been applied. The numerical solution process is based on a staggered solution of the thermal and hydrodynamics problem. While the applied explicit time stepping scheme allowed for a robust treatment of this complex simulation problem in case proper stability requirements are fulfilled, the latter restriction limited the achievable time step sizes and consequently the observable spatial and temporal problem dimensions considerably leading to a computational effort in the range of $100.000 \, cpu \, h$ in order to simulate a single laser track with length dimensions in the range of $1mm$ and time spans in the range of several $100 \mu s$~\cite{Khairallah2014}. The model of~\cite{Khairallah2014} includes surface tension effects prevalent in the melt pool flow as well as fluid viscosity and gravity effects, although the latter effect turned out to be negligible given the short time scales and the dominating surface tension effects at the considered length scales. However, the Marangoni effect, i.e., temperature gradient-induced melt flow due to temperature-dependent surface tension characteristics was not considered.\\

\begin{figure}[h!!]
 \centering
   \subfigure[Simulation with surface tension effects.]
   {
    \includegraphics[width=0.48\textwidth]{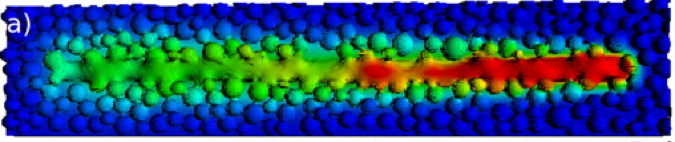}
    \label{fig:khairallah2014_meltpool1a}
   }
   \subfigure[Simulation without surface tension effects.]
   {
    \includegraphics[width=0.48\textwidth]{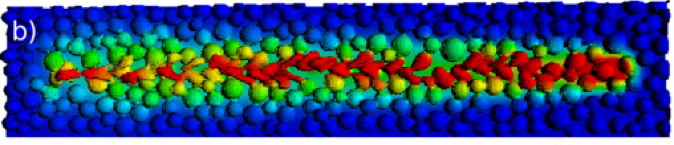}
    \label{fig:khairallah2014_meltpool1b}
   }
   \subfigure[Balling effect attributed to Plateau-Rayleigh instabilities of a liquid cylinder in the range of high scanning velocities.]
   {
    \includegraphics[width=1.0\textwidth]{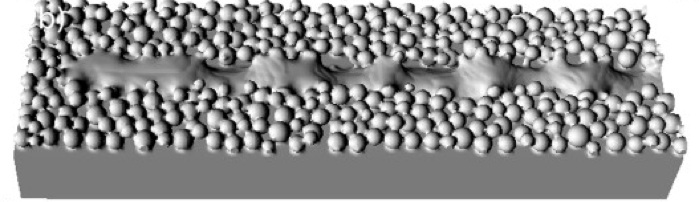}
    \label{fig:khairallah2014_meltpool2}
   } 
   \caption{Single-track mesoscopic simulation of thermo-hydrodynamics prevalent in SLM process of stainless steel 316 powder,~\cite{Khairallah2014}.}
 \label{fig:khairallah2014_meltpool1}
\end{figure}

In order to validate the thermal building block of the simulation framework, effective powder conductivities have been calculated with the developed model and compared to analytical solutions and experimental observations. In agreement to~\cite{Rombouts2005}, the thermal conductivity of stainless steel 316 powder with a typical packing density of $45 \%$ turned out to be only by a factor of three higher than the conductivity of air filling the powder pores. Coupled thermo-hydrodynamics simulations revealed the importance of considering surface tension, which results in a smoother, less granular melt pool geometry and eventually leads to higher effective thermal conductivities within the pool but also from the pool to the underlying substrate. This effect, in turn, led to increased heat transfer, faster cooling rates and consequently smaller melt pool sizes (Figure~\ref{fig:khairallah2014_meltpool1}). The rapid thermal expansion directly below the laser beam turned out to induce high melt flow velocities, especially in backward direction. Also the well-known balling effect~\cite{Levy2003,Kruth2007,Gusarov2007}, attributed to the Plateau-Rayleigh instability of a long cylindrical fluid jet breaking up into individual droplets, has been observed for high scanning velocities (see Figure~\ref{fig:khairallah2014_meltpool2} and~\cite{Gusarov2010} for a theoretical analysis).\\

\begin{figure}[h!!]
 \centering{
 \begin{minipage}{0.75\textwidth}
    \includegraphics[width=1.0\textwidth]{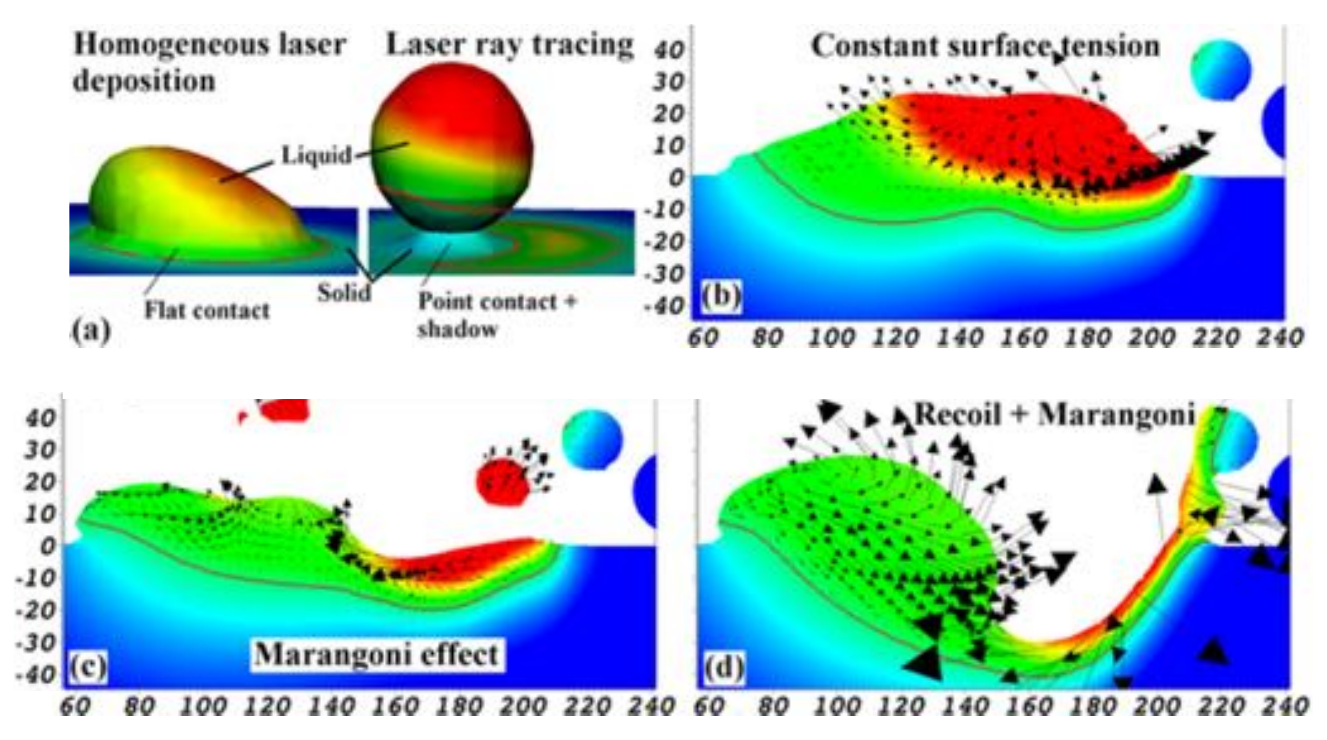}
   \caption{Influence of considering the individual modeling components "laser ray tracing", "constant surface tension", "temperature-dependent surface tension" and "recoil pressure" (from top left to bottom right) on resulting melt pool dynamics. Visualized is the temperature field in the range from $293K$ (blue) to $4000K$ (red) as well as the fluid velocity vector field. The red contour line represents the melt pool front,~\cite{Khairallah2016}.}
 \label{fig:khairallah2016_comparisonmodeleffects}
\end{minipage}}
\end{figure}

\begin{figure}[h!!]
 \centering{
 \begin{minipage}{0.67\textwidth}
    \includegraphics[width=1.0\textwidth]{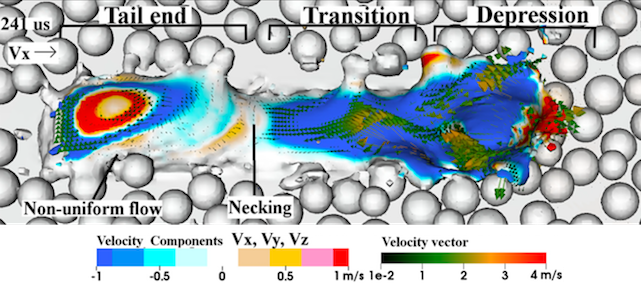}
   \caption{Fluid velocity magnitudes and vectors in three different melt pool regions: depression region, tail end region and transition region. The laser beam has been scanning from the left to the right and is currently located in top of the depression region,~\cite{Khairallah2016}.}
 \label{fig:khairallah2016_laserpoolregions2}
\end{minipage}}
\end{figure}

In~\cite{Khairallah2016}, the model proposed in~\cite{Khairallah2014} has been extended to also (implicitly) account for the effects of recoil pressure in the evaporation zone below the laser beam on the subjacent melt pool flow, Marangoni convection, as well as evaporative and radiative surface cooling (see also~\cite{Khairallah2015} for more details of the model). Moreover, a more detailed resolution of the laser energy absorption has been achieved based on a ray-tracing model as discussed in Section~\ref{sec:powder_modeling_raytracing}. The surface tension has been assumed to decrease linearly with temperature, which accounts for Marangoni effects driving the melt flow from the hot laser spot toward the cold rear. As already observed in~\cite{Gusarov2007}, the surface temperatures below the laser spot can easily reach boiling values. The resulting vapor recoil pressure adds extra forces to the melt pool surface that create a depression below the laser (Figure~\ref{fig:khairallah2016_comparisonmodeleffects}, bottom right). These observations motivated the differentiation of three different melt pool regions, namely the depression region, which is governed by recoil forces, the tail end region, which is governed by surface tension forces as well as the transition region (see Figure~\ref{fig:khairallah2016_laserpoolregions2}).\\

The approach of~\cite{Khairallah2016} does not explicitly resolve vapor flow but employs a traction boundary condition~\eqref{flow_balance4} based on a recoil pressure model proposed in~\cite{Anisimov1995}. Accordingly, the recoil pressure $p(T)$ depends exponentially on temperature:
\begin{align}
\label{khairallah2016_recoilpresssure}
p(T)=0.54 p_{ref} \exp{
\left(
\lambda_{ev}/k_{SB} (1/T-1/T_b)
\right)
},
\end{align}
In this model, $p_{ref}$ is the ambient pressure, $\lambda_{ev}$ is the evaporation energy per particle, $k_B$ is the Stefan-Boltzmann constant, $T$ is the surface temperature of the melting and $T_b$ is the boiling temperature. In order to more accurately model heat losses at the melt pool surface, the evaporation heat flux $\bar{\mb{q}}_v$ (again based on~\cite{Anisimov1995}) as well as the heat flux due to thermal radiation emission (see $\bar{\mb{q}}_e$ in~\eqref{surfaceheatfluxes}) has been considered according to
\begin{align}
\label{khairallah2016_heatlosses}
\bar{\mb{q}}_v=(0.82 A_s p(T)/ \sqrt{2\pi MR_gT}) \mb{n}.
\end{align}
Here, $A_s$ represents a so-called sticking coefficient, $p(T)$ is the recoil pressure according to~\eqref{khairallah2016_recoilpresssure}, $M$ is the molar mass, $R_g$ is the gas constant and $T$ again the surface temperature. However, the mass loss due to evaporation has not been considered in the continuity equation~\eqref{flow_balance1} underlying this model. The effect of considering / neglecting these different model components is illustrated in Figure~\ref{fig:khairallah2016_comparisonmodeleffects}. As compared to a continuum radiation transfer model, the employed ray-tracing model can resolve non-uniformities in the radiation absorption across powder particles leading to more narrow, neck-shaped connections to the underlying substrate and in turn to higher heat accumulations within particles as consequence of the reduced thermal conductivity. The ray-tracing model also captures scenarios such as particles that are only partially melted and might contribute to surface and pore defects more accurately. The principal surface tension effect fostering strongly curved melt pool surfaces with minimized surface energy is observable in Figure~\ref{fig:khairallah2016_comparisonmodeleffects}b), where additional melt flow is generated by temperature-induced buoyancy forces. The consideration of temperature-dependent surface tension allows for Marangoni convection, which induces a melt flow from the hot laser spot toward the cold rear (see Figure~\ref{fig:khairallah2016_comparisonmodeleffects}c)). This effect, in turn, increases the melt pool depth, and contributes to melt flow recirculation and melt spattering when colder liquid metal with lower viscosity is ejected from the pool. The additional consideration of recoil pressure due to evaporation leads to a considerably deeper melt pool, increased convective heat transfer within the melt pool and, in combination with the additional heat sinks due to evaporative and radiative surface cooling, to an increased overall cooling rate. Consequently, this case exhibits the lowest amount of stored heat.\\

In~\cite{Khairallah2016}, it has been stated that the overall melt pool dynamics in the direct vicinity of the laser beam is strongly influenced by the recoil pressure, whose magnitude increases exponentially with temperature, leading to the aforementioned depression directly below the laser. The depth of depression due to the recoil pressure is considered to increase with increasing laser power, an effect that is closely related to the so-called keyhole mechanism observed in laser and electron beam welding (see e.g.~\cite{Rai2008}). On the other hand, the fluid dynamics of the cooler back end of the melt pool are governed by surface tension fostering e.g. balling of the tail during cooling~\cite{Khairallah2014}.\\

Via this analysis,~\cite{Khairallah2016} enabled also theoretical considerations concerning the creation mechanism of so-called denudation zones. These "low powder ditches" alongside the melt pool result from lateral particles that are dragged into the melt pool by surface tension and are responsible for the so-called elevated "edge effect"~\cite{Craeghs2011} typically observed for the first laser track of a powder layer. Accordingly, the effects of denudation are stronger in the first track of a layer since powder particles can be attracted from \textit{two} lateral sides of the melt pool. In~\cite{Matthews2016}, this simulation model has been combined with experimental approaches in order to further investigate the physical phenomena and mechanisms responsible for the occurrence of the denudation zone. Accordingly, the ambient gas flow induced by vapor rising above the melt pool may eject surrounding particles from the powder layer and contribute to denudation.\\

In an alternative model proposed in~\cite{Lee2015}, the discrete element method (DEM) is applied in order to study the spreading (i.e., recoating) of Inconel 718 powder accounting for frictional particle-to-particle, particle-to-wall and particle-to-coater contact interaction as well as gravity as driving forces. Subsequently, the thermo-hydrodynamics problem of melt pool (free surface) flow is solved by means of a commercial computational fluid dynamics (CFD) solver based on the finite difference method (FDM), which resolves the fluid-gas phase boundary between melt pool and surrounding atmosphere on the basis of the volume of fluid (VOF) method. Also in this model, temperature-dependent surface tension effects, but no evaporation-related phenomena, have been considered. The laser beam energy absorption within the powder bed is modeled in a simplified manner based on prescribed heat flux boundary conditions on the top surface of the powder layer following the Gaussian distribution of the laser beam. Similar to~\cite{Khairallah2016}, a depression of the melt pool directly underneath the laser beam could be observed. Since no evaporation-induced recoil pressure has been considered in~\cite{Lee2015}, this observation has essentially been attributed to backward flow of molten metal due to Marangoni convection (comparable to Figure~\ref{fig:khairallah2016_comparisonmodeleffects}c) of~\cite{Khairallah2016}). Also the balling effect has been observed in these numerical simulations and attributed to Plateau-Rayleigh instabilities occurring in dependence of the melt pool dimensions and the powder particle arrangement. Accordingly, a higher packing density was found to reduce the likeliness of the balling effect and to increase the overall surface smoothness while a faster travel speed and lower laser power have been identified as possible sources fostering this effect. On the one hand,~\cite{Lee2015} claimed to obtain these simulation results on the basis of a comparable spatial resolution at considerably lower computational effort as compared to alternative approaches such as~\cite{Khairallah2014,Korner2013}. On the other hand, this reference does not discuss the employed time integrator and the chosen time step size, which have crucial influence on computational performance.\\

In the review article~\cite{Megahed2016}, Megahed et al. discuss several mesoscopic simulation results taken from previous contributions of the authors. The model of Megahed et al. is comparable to the mesoscopic models discussed in this section so far. Among others, it has been shown that lower energy densities yielded smoother melt pool surfaces within the considered range of processing parameters. This effect has been observed in a similar manner in reference~\cite{Khairallah2016}, where it has been argued that the melt pool associated with the high energy density reaches higher peak temperature values and, in turn, the surface tension can act for longer time spans leading to a more pronounced balling effect. On the contrary,~\cite{Lee2015} has reported about an increasing balling effect with decreasing energy density. The authors of~\cite{Lee2015} have argued that the total absorbed energy density decreases when decreasing the laser power at constant velocity, similar, as the total absorbed energy density decreases when increasing the laser velocity at constant laser power, with the latter scenario being known to foster the balling effect and resulting surface roughness. However, the difference between these two cases is that the latter scenario results in a longer, more slender melt pool shape (since less time is available for the melt pool to cool down while the laser beam moves by certain distance along the track), which fosters hydrodynamic instabilities according to the Plateau-Rayleigh theory. All in all, the observations made in~\cite{Lee2015} and the observations made in~\cite{Megahed2016} and~\cite{Khairallah2016} might not necessarily be contradictory. Instead, these observations could be associated with an upper and a lower bound of an energy density interval that allows for stable processing~\cite{Yadroitsev2007,Gusarov2010}, which is consistent with experimental findings. \\

\begin{figure}[h!!]
 \centering
   \includegraphics[width=0.9\textwidth]{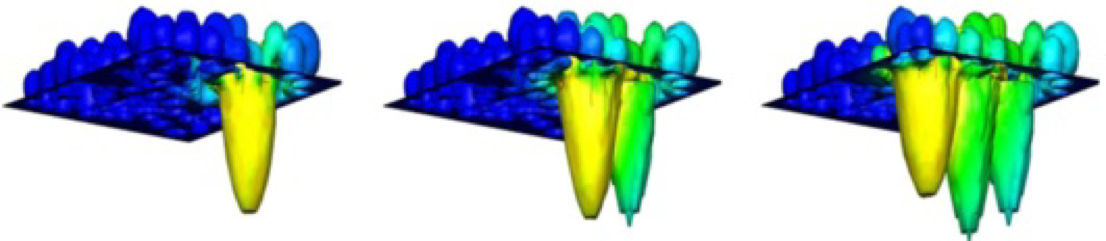}
   \caption{Melt pool shapes for a modulated laser: Temperature at three different points in time ranging from $500K$ (blue) to $2000K$ (yellow),~\cite{Megahed2016}.}
 \label{fig:megahed2016_modulatedlaser}
\end{figure}

In contrast to the continuous laser scan mode as considered so far, some SLM machines use point-wise modulation melting strategies. Megahed et al. have also investigated the melt pool characteristics resulting from such a modulated laser (see Figure~\ref{fig:megahed2016_modulatedlaser} for the resulting melt pool shapes at different points in time). It can be seen that the laser modulation and the corresponding exposure time leads to singular deep melt pools. As the melt pools grow in diameter during the exposure time, they join into one continuous melt track on the upper surface of the powder bed down to a depth as typical for a continuous laser beam. However, as argued in~\cite{Megahed2016}, there are deeper melt pool peaks that might imply additional anchorage for the new layer to the previous ones. In addition to the discussion in~\cite{Megahed2016}, these modulated strategies might also allow for higher energy absorption as the laser beam can penetrate deeply into the pores of the powder layer via multiple reflections instead of directly interacting with the melt pool surface as it is typically the case in continuous scan mode. Potential benefits could arise in terms of lower melt pool surface temperatures and a decreased evaporative mass loss and recoil pressure.\\

In~\cite{Qiu2015}, the melt flow fluid dynamics have been solved by means of the finite volume method (FVM), employing the volume of fluid method (VOF) in order to resolve the interface of the free-surface flow. Besides recoil pressure, temperature-dependent surface tension and temperature-induced buoyancy forces as already considered in~\cite{Khairallah2016},~\cite{Qiu2015} additionally supplemented the Navier-Stokes momentum equations~\eqref{flow_balance2} by drag force contributions due to solid/liquid transition in the mushy zone of the melting Ti-6Al-4V alloy based on Darcy’s term for porous media. In the energy equation~\eqref{gusarov2007_HCE}, the additional heat losses due to evaporation, convection as well as radiation emission have been considered. As laser beam heat source, the model proposed in~\cite{Xu2011} in the context of laser beam welding has been applied. Since the radiation transfer mechanisms prevalent in SLM might strongly depend on the mode of operation, the application of this, but also of other ad-hoc continuum radiation transfer laser models (see Section~\ref{sec:powder_modeling_continuum}), has to be questioned critically. Also in~\cite{Qiu2015}, a strong influence of the recoil pressure on the melt pool dynamic in the direct vicinity of the laser beam as well as thereby induced material spattering could be observed, even though significantly decreased laser track sizes as well as structured instead of random powder packings have been considered in this reference. Similar mesoscopic modeling approaches are for example given in~\cite{Gurtler2013,Yu2016,Yuan2015}.\\

\begin{figure}[h!!]
 \centering
   \includegraphics[width=1.0\textwidth]{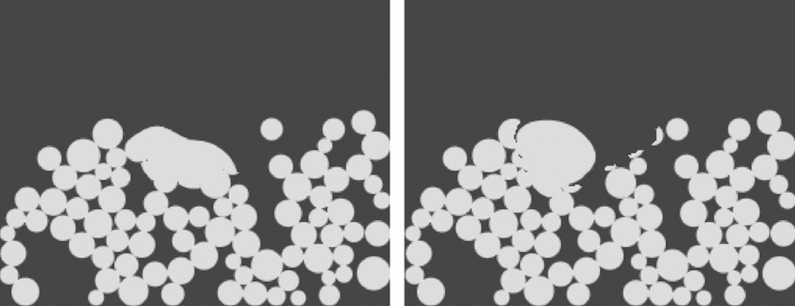}
   \caption{Simulation results based on a 2D Lattice-Boltzmann model with discretization cells of size $5 \mu m$. Illustrated are initial powder grains and melt pool shapes resulting from two different prescribed wetting angles of $\theta_0=10^{\circ}$ (=good wetting, left) and $\theta_0=160^{\circ}$ (worse wetting, left),~\cite{Korner2011}.}
 \label{fig:koerner2011_wettingcomparison}
\end{figure}

While the mesoscopic models considered so far have been discretized by means of either the FEM, FDM or the FVM, an entirely different approach has been proposed in~\cite{Korner2011}. Based on the contributions~\cite{Attar2009} and~\cite{Attar2011}, the Lattice-Boltzmann method in combination with an explicit time integration scheme has been applied in order to discretize a 2D mesoscopic free-surface, incompressible melt flow model incorporating gravity, surface tension and wetting effects as well as the melting and solidification process prevalent in SLM. The initial powder bed generation is based on a random packing algorithm employing a rain model scheme~\cite{Meakin1987}. While the general simulation framework might be applicable to SLM as well as EBM processes, the actual simulations and experiments conducted in this reference focused on the EBM process. Consequently, the energy absorption has been considered to predominantly take place at the powder particle surfaces of first incidence. This fact is reflected by the chosen exponential Lambert-Beer law of light attenuation in dense matter, which is not applicable to the SLM process. The free surface between melt pool and ambient gas is numerically described on the basis of averaged, fluid volume fraction-dependent material properties prevalent in discretization cells at the boundary fluid-gas in a manner similar to the VOM. In~\cite{Korner2011}, temperature-dependent surface tension, and thus Marangoni convection, has been neglected. Figure~\ref{fig:koerner2011_wettingcomparison} shows the 2D melt pool shape resulting from a spatially fixed laser beam position and different wetting angles $\theta_0$, determined by the considered material combination at the triple point solid-fluid-gas, for otherwise identical system parameters.\\

\begin{figure}[t!!!]
 \centering
   \subfigure[Constant laser beam power.]
   {
    \includegraphics[height=0.37\textwidth]{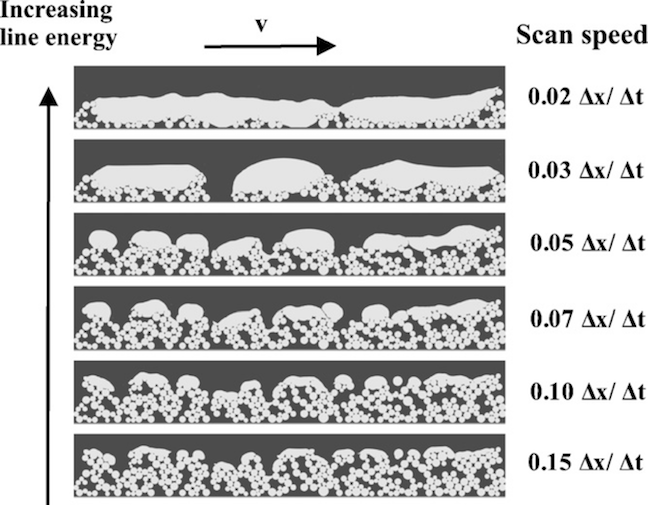}
    \label{fig:koerner2011_constantpowder}
   }
   \hspace{0.05\textwidth}
   \subfigure[Constant energy density.]
   {
    \includegraphics[height=0.37\textwidth]{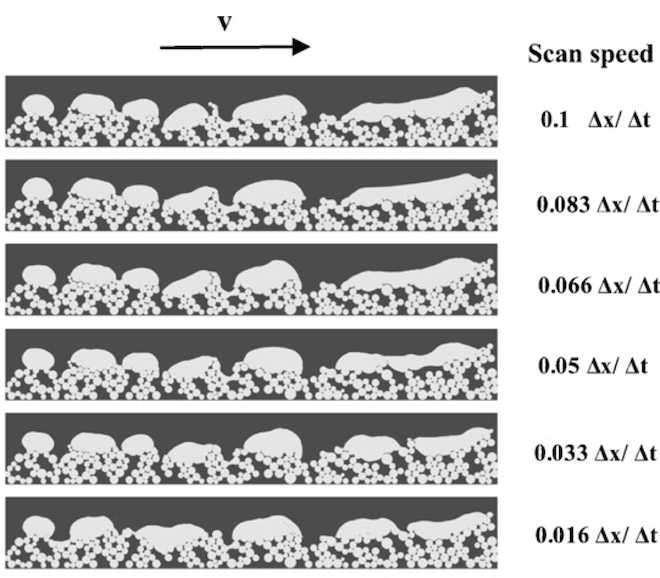}
    \label{fig:koerner2011_constantenergydensity}
   }
   \caption{2D Single-track mesoscopic simulation results: Influence of laser beam velocity on wetting behavior and melt track shape,~\cite{Korner2011}.}
 \label{fig:koerner2011_influencevelocity}
\end{figure}

According to Figure~\ref{fig:koerner2011_wettingcomparison}, the wetting behavior expressed via the wetting angle $\theta_0$ considerably influences the resulting melt pool shape. More detailed investigations revealed, however, that the resulting melt pool shape is considerably influenced by the powder packing density and, at least in the range of lower packing densities, also by the individual stochastic realization of different powder beds with identical packing density. Consequently, the local powder topology as well as the wetting behavior resulting from the prevalent material combination solid-liquid-gas might have considerable influence on balling of the solidifying tail. In addition to these single point simulations, also single-track simulations have been performed in~\cite{Korner2011}. The melt pool shapes resulting from different scan velocities, once at fixed laser beam power and once at fixed spatial energy density are illustrated in Figure~\ref{fig:koerner2011_influencevelocity}. It can be observed that the melt pool stability at constant laser power increases with decreasing laser beam velocity, which, in turn, yields a higher spatial energy density. Small melt pool sizes in the range of the powder grain diameter, which results from low energy input, limits the wetting behavior and fosters round melt pool shapes with minimal surface energy. With increasing size, the relative contribution of surface tension, wetting and gravity effects changes and the melt pool is able to interconnect the powder particles visible in form of a pronounced wetting behavior. On the other hand, comparable melt track topologies are found at constant spatial energy density, advocating this quantity as most relevant factor of influence for the given process conditions. All in all, it has been concluded that the packing density of the powder bed has the most significant effect on the melt pool characteristics.\\

In~\cite{Korner2013}, the model of~\cite{Korner2011} has been applied to study a layer-wise buildup process, again for Ti-Al6-V4 powder material. The model simplifies the actual SLM process to 2D by investigating only the plane spanned by the build and scan direction. Furthermore, dimensionless factors characterizing the different physical phenomena and time scales prevalent in SLM have been analyzed in this reference. Based on a comparison of laser beam interaction time and thermal diffusion time, a recommendation for the optimal laser beam velocity has been made in order to avoid overheating and excessive evaporation on the one hand as well as unnecessarily high energy losses on the other hand. It has been argued that surface tension dominates gravity in terms of driving forces whereas inertia effects dominate viscous damping in terms of resistance forces. Furthermore, the Rayleigh time scale, characterizing the relaxation of an interface perturbation under the action of inertia and surface tension forces, turned out to be more than one order of magnitude smaller than thermal diffusion time scales. That is, surface tension-driven coalescence of powder particles follows the melting process nearly instantaneously.  Further, in~\cite{Bauereiss2014} this model has been employed in order to illustrate the formation of pores. In~\cite{Markl2013,Klassen2014}, the simulation framework proposed in~\cite{Korner2011} has been supplemented by means of alternative models for electron energy dissipation in EBM processes. In reference~\cite{Ammer2014} the method has been extended to 3D problems with a special emphasize on code parallelization strategies to increase computational efficiency.

\subsection{Microscopic simulation models}
\label{sec:microscopicmodels}

\begin{figure}[h!!!]
\begin{centering}
\includegraphics[width=0.55\textwidth]{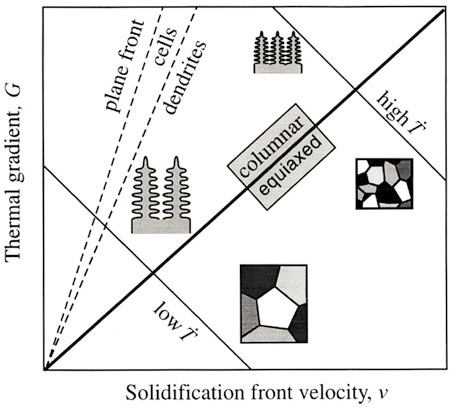}
\caption{Microstructure versus thermal gradients and melt front velocity,~\cite{Glcksman2010}.}
\label{Fig_zy7}
\end{centering}
\end{figure}

The mechanical and physical properties of metals, and therefore of all parts made by SLM, are influenced by their local composition and microstructure characterized by grain size, grain morphology (shape) and grain orientation (texture). In general, the microstructure evolution during solidification processes is governed by the spatial temperature gradients (G), cooling rates $\dot{T}$ and solidification front velocities $v$~\cite{Kurz1986,Glcksman2010}. As visualized in Figure~\ref{Fig_zy7}, a qualitative solidification map can approximately be divided into areas of ‘high’ and ‘low’ cooling rates and into regimes of ‘columnar’ and ‘equiaxed’ structures. Typically, high cooling rates lead to rapid nucleation of grains and non-equilibrium microstructures, characterized by small grain sizes or small dendrite arm spacings. The competition of thermal gradients and solidification front velocities lead to columnar / elongated, dendritic grain morphologies if high thermal gradients are dominating since in this scenario the liquid-solid interface becomes unstable and preferred orientations grow faster than others to finally form dendritic morphologies. On the contrary, equiaxed, rather spherical grain morphologies are expected if high solidification front velocities are dominating as consequence of a more stable liquid-solid interface.\\

In contrast to conventional metallurgy such as casting, SLM is based on a highly localized energy supply yielding extremely high heating rates. As consequence of the localized energy supply, the solidified material of previous melt tracks and layers remains at comparatively low temperatures, which in turn yields extreme cooling rates after the laser has passed by. The magnitude of heating and cooling rates typically lies in the range of $10^3\!-\!10^8 K\!/\!s$. The succession of extreme heating and cooling rates eventually also results in high spatial temperature gradients ($10^3\!-\!10^8 K\!/\!m$) and solidification front velocities ($1\!-\!30 m\!/\!s$) ~\cite{Majumdar2013,Gu2015}. Such a large range of cooling rates and thermal gradients induces a large variety of microstructures of as-deposited materials differing by grain size, morphology, and orientation. The high cooling rates often lead to meta-stable and non-equilibrium phases and comparatively small grain sizes. As a trend, the grain sizes in SLM are typically smaller and the material strength is higher as compared to alternative manufacturing processes such as casting. With the build direction commonly representing the direction of highest temperature gradients, depending on the employed material often elongated, dendritic grain morphologies accompanied by a higher material strength in build direction can be observed. Additionally, solid state transformations and grain growth contribute to crucial microstructure changes during the repeated thermal cycles (while remaining below the melting point)
experienced by previously deposited material after the initial solidification process (see e.g. the transformation from martensite to alpha and beta phases in Ti-6Al-4V as discussed in Section~\ref{sec:microstructure}). \\

The knowledge of the resulting micro structure characteristics might provide important details for the formulation of process-specific continuum constitutive laws, which are crucial for accurate residual stress predictions as intended by macroscopic simulation models. On the other hand, global temperature distributions provided by macroscopic continuum models represent essential input variables for studying the solidification process by means of microstructure models. So-called phase field models based on various spatial and temporal discretization strategies can be considered as one of the most established modeling approaches to study solidification processes~\cite{Wheeler1992,Dorr2010,Krill2002,Chen2002,Steinbach2013,Apel2014,Guo2013,Kobayashi1993,Qin2010,Wang1993} but also diffusionless solid-solid phase transformations occurring for example during the formation of martensitic non-equilibrium phases~\cite{Militzer2011,Arif2014}.\\

Phase field models are based on the definition of a free-energy functional, which typically has the following form:
\begin{align}
\label{phasefield}
\Pi(\phi,c,T,\boldsymbol{\Psi})= \int_{\Omega} \tilde{\Pi} (\phi,c,T,\boldsymbol{\Psi}) d \Omega \quad \text{with} \quad
\tilde{\Pi} (\phi,c,T,\boldsymbol{\Psi})=f(\phi,c,T) + \frac{c_{\phi}^2}{2} || \bigtriangledown \phi ||^2 + \tilde{\Pi}_o(\phi,T,||\bigtriangledown \boldsymbol{\Psi}||).
\end{align}
Here, $\phi$ represents the phase field variable or order parameter taking on the value $\phi\!=\!1$ in the solid phase, $\phi\!=\!0$ in the liquid phase, and a value $\phi\!\in\!]0;1[$ in the finitely thick interface region of smooth phase transition. Furthermore, $c$ represents the (volume / mass) fraction of a certain chemical species prevalent in the problem of interest. For binary systems, the fraction of the second species is given by $1-c$ while for general systems consisting of $n$ species, $n-1$ variables $c_i$ are required. As before, $T$ is the absolute temperature. Also grain orientation is considered in the phase field model~\eqref{phasefield}. In this context, $\boldsymbol{\Psi}$ represents a vector-valued parametrization of grain orientation, e.g. given by rotation vectors, quaternions or Euler parameters. The first term of $\tilde{\Pi} (\phi,c,T,\boldsymbol{\Psi})$ in~\eqref{phasefield} represents the temperature-dependent Gibbs free energy either associated with the liquid or solid phase at a certain location in the liquid or solid domain, or a proper interpolation of these values for locations on the phase boundary region. The second term represents energy contributions stemming from liquid-solid phase boundaries characterized by an existing gradient $\bigtriangledown \phi$ of the phase field. The choice of the parameter $c_{\phi}$ determines the resulting phase boundary thickness and requires a compromise between an accurate resolution of the small boundary layer thickness typically prevalent in physical systems (high value of $c_{\phi}$) and a certain degree of artificially increased thickness for reasons of computational efficiency and robustness (low value of $c_{\phi}$). Eventually, the last term $\tilde{\Pi}_o (\phi,T,||\bigtriangledown \boldsymbol{\Psi}||)$ yields energy contributions due to crystallographic orientation gradients $\bigtriangledown \boldsymbol{\Psi}$. This contribution fosters uniform growth within individual grains and penalizes the misorientations at grain boundaries, making larger grain sizes with reduced overall grain boundary surface favorable in configurations of thermodynamical equilibrium. Variation of the functional~\eqref{phasefield} yields the Euler-Lagrange equations of the variational problem, determining the equilibrium solution $\phi^*,c^*,\boldsymbol{\Psi}^*$ of the variables $\phi,c,\boldsymbol{\Psi}$:
\begin{align}
\label{phasefield2}
\Pi(\phi^*,c^*,T,\boldsymbol{\Psi}^*) = \text{extremum} \rightarrow \frac{\partial \Pi}{\partial \phi}(\phi^*,c^*,T,\boldsymbol{\Psi}^*)=0, \quad 
\frac{\partial \Pi}{\partial c}(\phi^*,c^*,T,\boldsymbol{\Psi}^*)=0, \quad
\frac{\partial \Pi}{\partial \boldsymbol{\Psi}}(\phi^*,c^*,T,\boldsymbol{\Psi}^*)=0.
\end{align}
Starting with a non-equilibrium system, e.g. an undercooled melt, the equilibrium solution defined by the functional~\eqref{phasefield} is found in practical simulations by transforming~\eqref{phasefield2} into a transient problem and searching for the associated steady state solution. Thereto, the so-called time-evolution phase-field approach applies additional rate terms proportional to $\dot{\phi}$, $\dot{c}$ and $\dot{\boldsymbol{\Psi}}$ on the right-hand sides of equations~\eqref{phasefield2}. In SLM, the temperature field is required as input variable for the phase field model and might e.g. be provided by a macroscopic model based on~\eqref{gusarov2007_HCE}.\\

\begin{figure}[h!!]
 \centering
    \includegraphics[width=1.0\textwidth]{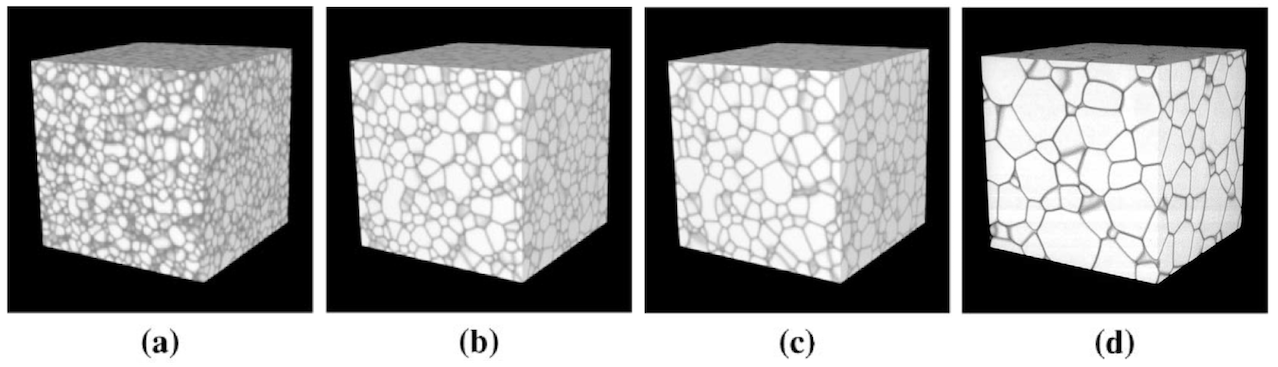}
   \caption{3D phase field simulation of solidification and grain growth process assuming isotropic boundary energy and mobility,~\cite{Krill2002}.}
 \label{fig:krill2002}
\end{figure}

The spatial discretization resolution required for a sufficiently accurate phase boundary representation typically results in a considerable numerical effort. For that reason, many of the available phase field models are applied to 2D scenarios. For illustration of the general capability of phase field models, two existing 3D approaches - even though not applied in the context of SLM - shall briefly be presented. One has been proposed in~\cite{Krill2002}, based on a finite difference discretization scheme in space and in time. In this work, only one chemical species (variable $c$ not required) and only a discrete set of possible grain orientations is considered. Consequently, the variable $\boldsymbol{\Psi}$ representing a continuous spectrum of possible orientations is replaced by additional phase field variables $\phi_i$, one for each possible orientation. Figure~\ref{fig:krill2002} shows different time steps of a solidification problem allowing for $20$ discrete grain orientations. Considering the employed $180x180x180$ grid and $8000$ time steps, the computational effort seems to be large. As visualized in Figure~\ref{fig:krill2002}, the initial number of $6000$ grains reduces to approximately $200$ at the end of the simulation, yielding a decreased overall grain boundary energy in thermodynamical equilibrium.\\

\begin{figure}[h!!]
 \centering
    \includegraphics[width=0.9\textwidth]{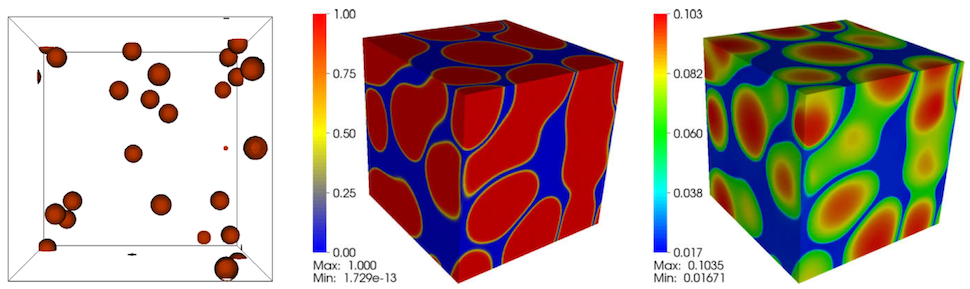}
   \caption{3D phase field simulation considering the grain growth process within the $\epsilon$-$\delta$ solid phase region of a binary alloy phase diagram: Initial distribution of $\delta$-phase nuclei (left), final distribution of $\epsilon$- and $\delta$-phase (middle) as well as concentration of the second species $B$ (right),~\cite{Dorr2010}.}
 \label{fig:dorr2010}
\end{figure}

A further 3D PFM based on a spatial FVM discretization and an implicit backward difference time stepping scheme is given in~\cite{Dorr2010}. There, two different chemical species (reflected by one concentration variable $c$) as well as a continuous grain orientation field $\boldsymbol{\Psi}$ as given in~\eqref{phasefield} has been employed. Figure~\ref{fig:dorr2010} shows the results of a simulation considering grain growth within a two-phase region (solid phases $\epsilon$ and $\delta$) of a binary alloy (chemical species $A$ and $B$) phase diagram under non-equilibrium conditions. Starting with randomly distributed $\delta$-phase nuclei (see Figure~\ref{fig:dorr2010}, left), this phase begins growing within the $\epsilon$-phase matrix (see Figure~\ref{fig:dorr2010}, middle, for the phase distribution at the end of the simulation). Due to the lower diffusion constant of the $\delta$-phase and the high cooling rates chosen in this example, the so-called coring effect, i.e. a thermodynamical non-equilibrium state characterized by an inhomogeneous concentration distribution in the $\delta$-phase (see Figure~\ref{fig:dorr2010}, right), has been observed. The representability of such non-equilibrium configurations is very relevant for the modeling of SLM, given the high cooling rates present. The numerical simulation was based on a $256x256x256$ finite volume grid and $1400$ time steps in order to simulate $2s$ of physical time, leading to $14h$ computation time on $64$ processors.\\

\begin{figure}[h!!]
 \centering
    \includegraphics[width=1.0\textwidth]{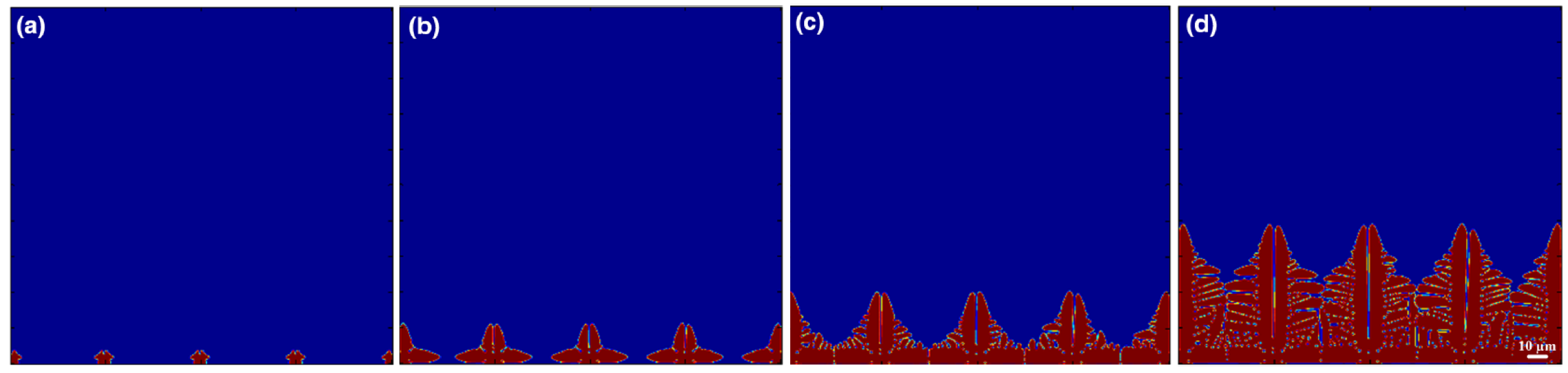}
   \caption{2D phase field simulation of $\beta$-structure growth during EBM of Ti-6Al-4V with nucleation on substrate at different time steps,~\cite{Gong2015}.}
 \label{fig:gong2015_1}
\end{figure}

Only a few approaches to microstructure modeling can be found for SLM/EBM, all of them restricted to 2D. In Gong et al.~\cite{Gong2015}, the solidification and growth of primary $\beta$-phase grains during the EBM processing of Ti-6Al-4V has been studied. Thereto, a phase field model similar to~\eqref{phasefield} has been employed with $c$ representing the concentration of Ti and $1-c$ the concentration of the solute, treated as the combination of Al and V. Thus, the ternary alloy Ti-6Al-4V has been simplified as a binary system, and the influence of grain orientation by means of a parameter $\boldsymbol{\Psi}$ has not been considered. The required temperature field has been provided via a finite element solution of the energy equation~\eqref{gusarov2007_HCE}. Figure~\ref{fig:gong2015_1} shows the simulated growth of columnar structures at different time steps during the solidification process. Initially, a pre-defined number of nuclei is placed on the substrate below the molten powder layer. The pre-defined number of nuclei has been determined via a regression equation accounting for an increasing nucleation density with increasing undercooling. Initial dendrite growth occurred primarily in scan direction (horizontal) and build direction (vertical) until neighboring dendrites started to contact each other. Subsequently, vertical growth was observed, resulting in a grain orientation parallel to the build direction as typically observed in experiments.\\

\begin{figure}[h!!]
 \centering
    \includegraphics[width=0.85\textwidth]{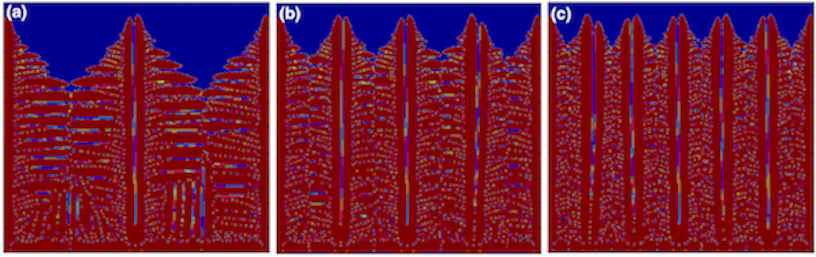}
   \caption{2D phase field modeling of dendrite growth for different scan velocities: (a) low, (b) medium, (c) high,~\cite{Gong2015}.}
 \label{fig:gong2015_2}
\end{figure}

In Figure~\ref{fig:gong2015_2}, the final configurations resulting from different scan speeds are depicted. As expected and in agreement with experiments considered in~\cite{Gong2015}, the higher cooling rates induced by higher scan velocities result in a higher nucleation density and, eventually, in a finer grain structure, which is still oriented in the build direction.\\

A second category of approaches often applied to study solidification processes is given by cellular automaton (CA) schemes. Like PFMs, CA schemes require the solution of the thermal field as input. However, the CA schemes do not rely on an additional phase field variable $\phi$ that explicitly describes the location of the individual phases and phase boundaries. Instead, the evolution of the grain growth is typically constructed in a rather geometrical manner by tracking the solid-liquid interface based on so-called cell capturing schemes under consideration of a given initial nucleus orientation and the current interface geometry. A comprehensive overview of methodologies for the modeling of microstructure evolution, among others considering PFM and CA schemes, is also given by~\cite{Boettinger2000}.\\

\begin{figure}[b!!]
 \centering
    \includegraphics[width=1.0\textwidth]{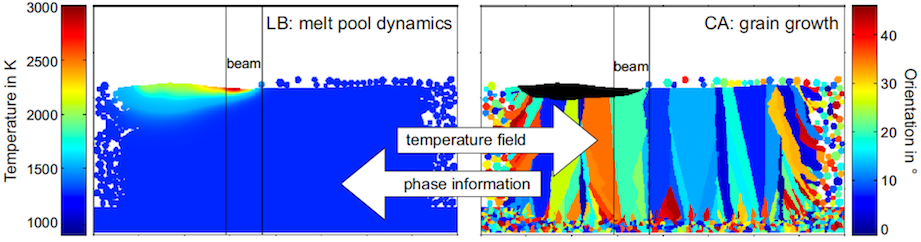}
   \caption{Exchange of temperature field and phase information between mesoscopic and microscopic simulation model,~\cite{Rai2016}.}
 \label{fig:mark2016micro}
\end{figure}

In~\cite{Rai2016}, the columnar grain growth during solidification of Inconel 718 alloy is studied based on a 2D realization of the CA approach. As visualized in Figure~\ref{fig:mark2016micro}, in every time step, the temperature field as well as the phase distribution, are exchanged in a weakly coupled manner. The required temperature field is derived by means of a 2D mesoscopic Lattice Boltzmann simulation model~\cite{Korner2011} already discussed in Section~\ref{sec:mesoscopicmodels}. This study predicts columnar grain growth in the build direction. In contrast to~\cite{Gong2015}, different grain orientations are distinguished by this model and the effect of stray grain formation due to partially molten powder particles on the left and right boundary of the computational domain could be taken into account thanks to the powder bed resolution on the mesoscopic scale. Also an epitaxial grain growth in build direction, i.e. grain growth on a newly deposited layer continuing the crystallographic structures of grains prevalent in the previous layer~\cite{Herzog2016}, is predicted, leading to grain sizes that considerably exceed the powder layer height. The effect of grain coarsening in the solid phase, i.e. competition between solid grains of different orientation leading to a growth of certain grains, has not been addressed by this CA scheme.\\

Also in the recent contribution~\cite{Zhang2013}, first steps are made towards a 2D simulation framework predicting the resulting microstructure evolution during solidification of 316L in SLM processing. The thermal field has been simulated based on a FEM discretization of~\eqref{gusarov2007_HCE} while the microstructure model relies on a CA approach.\\

\begin{figure}[h!!!]
\begin{centering}
\includegraphics[width=0.75\textwidth]{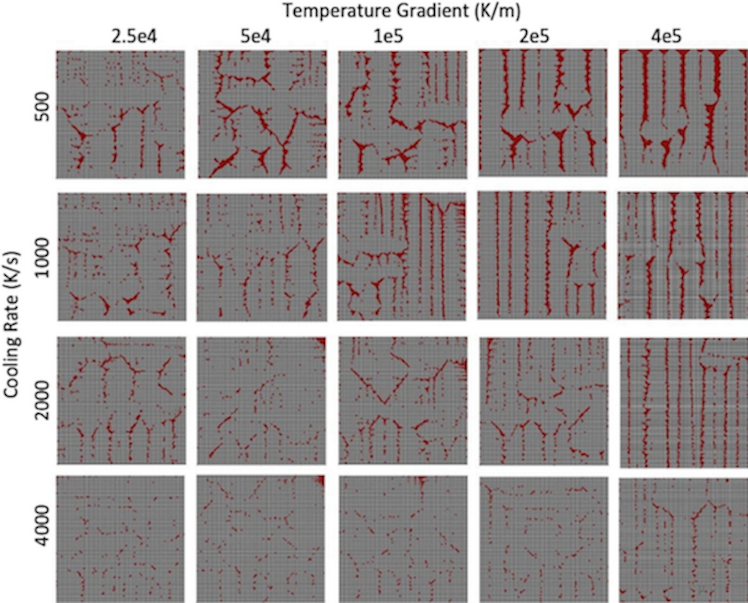}
\caption{Formation of Laves phase particles under different solidification conditions,~\cite{Nie2014}.}
\label{Fig_zy13}
\end{centering}
\end{figure}

A problem during the SLM processing of Inconel 718 is the segregation of Nb and the formation of so-called Laves phase grains in the interdendritic region, which may considerably degrade mechanical properties. Nie et al.~\cite{Nie2014} employed a numerical scheme based on a stochastic approach for studying the nucleation and growth of dendrites, the segregation of niobium (Nb) and the formation of Laves phase particles during the solidification. Additionally, a FEM-based discretization of the heat conduction equation has been employed for solving the thermal problem, which is required as input for the stochastic microstrucutre evolution model. Based on numerical simulations, Nie et al. demonstrated that low cooling rates as well as a high ratio \textit{G/v} of temperature gradients (G) to solidification front velocities (v) yield microstructures tending to large columnar dendrite arm spacing and, consequently, continuously distributed coarse Laves phase particles. Thus, it is suggested to increase cooling rates and to decrease temperature gradients in order to obtain small equiaxed dendrite arm spacings and discrete Laves phase particles (Figure~\ref{Fig_zy13}).

\subsection{Summary of existing macro-, meso- and microscale models and potential for future developments}
\label{sec:modelsummary}

\begin{figure}[h!!]
 \centering
    \includegraphics[width=1.0\textwidth]{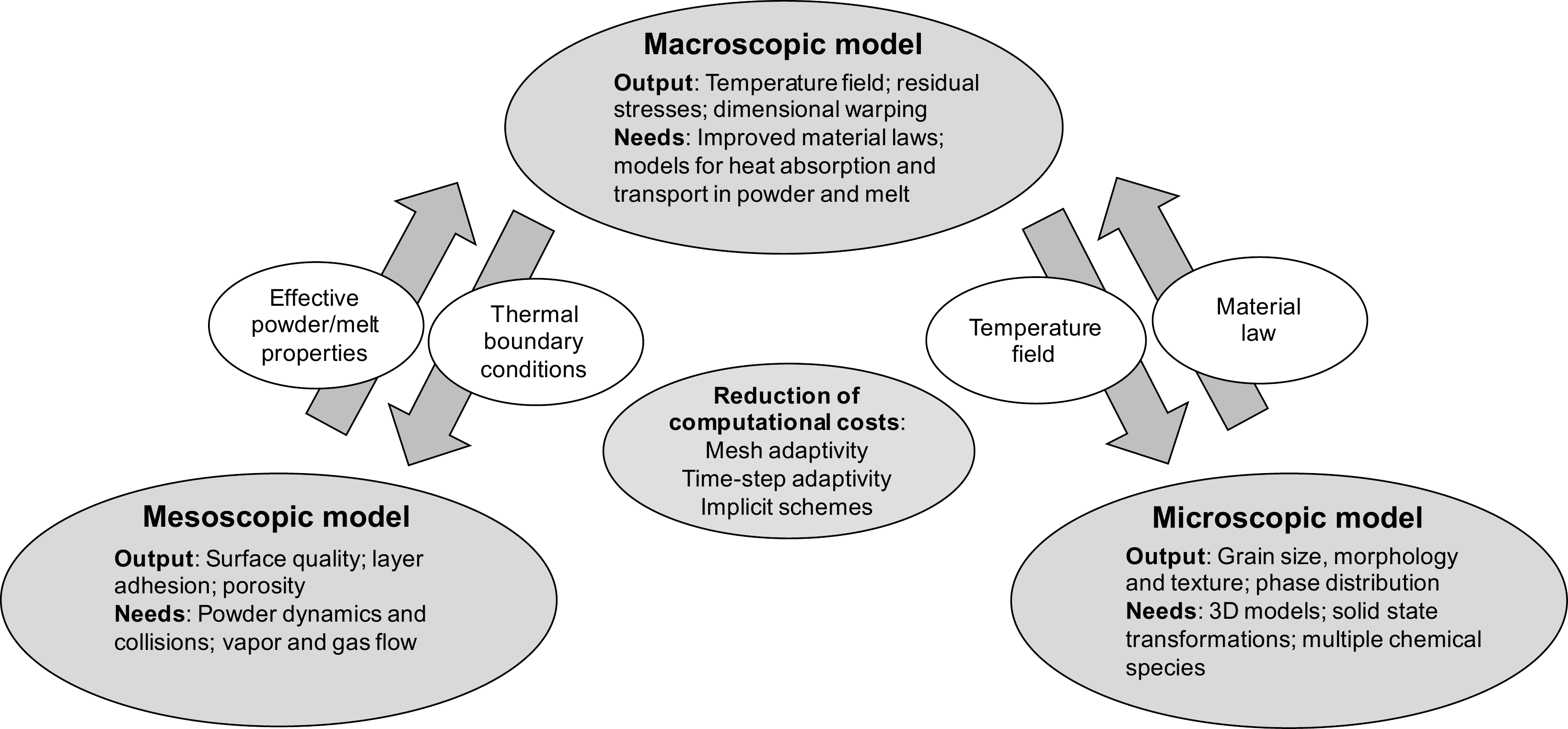}
   \caption{Potential for future improvements and coupling of submodels on macro-, meso- and microscale.}
 \label{fig:sketch5}
\end{figure}

Computational efficiency represents one of the key requirements for SLM simulation tools to be capable of capturing practically relevant time and length scales. Potential improvements of existing simulation frameworks comprise temporally and spatially adaptive time and space discretization schemes, the use of implicit time integrators, efficient code design suitable for high-performance parallel computing as well as the development of efficient and consistent strategies for material deposition and dynamic adaption of the computational domain. Besides computational performance, existing simulation tools require further improvements in terms of model and discretization accuracy. In the following, such improvements but also a potential coupling of the three model classes will be discussed (Figure~\ref{fig:sketch5}).\\

\textbf{Macroscopic models:} Macroscopic SLM models typically consider entire parts in order to predict global temperature fields, residual stresses or dimensional warping. The melt pool shape and the resulting temperature distribution crucially depend on an accurate model of laser radiation transfer into and heat conduction transfer within the powder phase. In this context, macroscopic models considering the powder phase in a spatially homogenized sense could be improved by transferring information from mesoscopic models. Also model information related to the melt pool hydrodynamics, which determine the convection-dominated heat transfer within the melting phase, could either be gained from mesoscopic models or explicitly considered in macroscopic models following existing approaches in the fields of laser and electron beam welding. These measures might considerably improve the accuracy of the predicted temperature field in the direct vicinity of the melt pool. While macroscopic models tend to strongly abstract the complex physics prevalent in this regime, they are expected to provide good predictions of the temperature field in regions further away from the melt pool, where heat flow is predominantly governed by the global part geometry, thermal boundary conditions as well as solid material characteristics. However, one of the ultimate goals of macroscopic models is the prediction of residual stress distributions. While temperature-induced thermal strains can be identified as the kinematic origin of these residual stresses, the actual magnitude of these stresses as well as the magnitude of maximally admissible stresses (given e.g. by the yield strength) is essentially determined by the solid material properties, and, thus, by the underlying metallurgic microstructure, which is, however, part of the unknown solution variables of the SLM process. On the contrary, even if derived on the basis of exact temperature solutions, the significance of residual stress predictions is strongly limited as long as comparatively simple material laws are employed. Supplying macroscopic constitutive models with further details concerning material inhomogeneity, anisotropy and temperature-history-dependence, e.g. by means of microscale SLM models, could drastically improve the quality of predictive macroscopic SLM models.\\

\textbf{Mesoscopic models:} Mesoscopic SLM models resolve length scales of individual powder grains and below in order to accurately describe radiation and heat transfer within the powder bed as well as heat and mass transfer within the melt pool governed by surface tension, wetting effects and evaporation-induced recoil pressure. The ultimate goal pursued by these highly resolved models is to gain an essential understanding of underlying physical phenomena responsible for melt track stability, resulting adhesion between subsequent layers, surface quality as well as creation mechanisms of pores and other types of volume and surface defects. This high spatial resolution naturally goes along with a considerable computational effort. So far, the computational challenge of solving the resulting numerical problems involving complex field and domain couplings has been addressed by explicit time integration limiting the currently observable length scales to single tracks of a few $mm$ length and the observable time scales to several $100 \mu s$ despite massive usage of high performance computing resources. Concerning model accuracy, recent experimental investigations suggest that the evaporation-induced gas flow above the melt pool might considerably affect melt pool and powder bed dynamics~\cite{Matthews2016}. While some first contributions have already considered evaporation in an implicit manner via recoil pressure and heat transfer models, an explicit modeling of the ambient gas flow, at least in the direct vicinity of the melt pool, could allow for considerable insights into the governing physical mechanisms and suggest strategies of influencing these (typically undesirable) gas dynamics. Also effects such as powder particle ejection or denudation are not sufficiently understood and require further investigation. While existing mesoscopic models typically consider spatially fixed powder particles, models that account for powder particle dynamics and collision could allow to study and understand these effects in detail. Eventually, mesoscopic models could benefit from improved thermal and mechanical boundary conditions at the interface between the representative mesoscopic volume and the global SLM part. In this context, macroscopic SLM models might provide useful data.\\

\textbf{Microscopic models:} Microscopic SLM models investigate the metallurgical microstructure evolution resulting from the high temperature gradients and extreme heating and cooling rates during the SLM process and aim at the prediction of resulting grain sizes, morphologies and textures as well as phase distributions. While mesoscopic models intend rather universal statements concerning optimal adjustment of parameters such as laser beam velocity and power, powder layer thickness and packing density as well as grain shape and size distribution for a given powder material, global temperature distributions, residual stress fields and microstructure evolutions as derived by macroscopic and microscopic SLM models strongly depend on the specific part geometry. Consequently, efficient numerical tools are required in order to enable the simulation of real-size parts in acceptable simulation time and a possible exchange of information between the macroscopic and microscopic scale. Macroscopic models typically provide the temperature field solution as input for microscopic SLM models. Compared to the variety of existing macroscopic and mesoscopic SLM models, microscopic modeling approaches that take into account the specific thermal conditions prevalent in SLM processes can be regarded as being less elaborated. The few existing approaches rely on a simplified 2D representation of the problem although many of the physical mechanisms governing metallurgical microstructure evolution are three-dimensional in nature. Current approaches do not consider more than two individual chemical species. However, an explicit modeling of all relevant alloying elements might be desirable in order to describe experimentally observed effects such as the segregation of individual alloying components. Also mechanisms of solid state phase transformation and grain growth, which have been observed in experiments and are likely to considerably change the material properties during repeated thermal cycles after the initial solidification process, might be desirable supplementations for future models.

\section{Experimental studies of thermophysical mechanisms in SLM}
\label{sec:experimentalcharacterization}

In the following, the modeling approaches discussed in Section~\ref{sec:modeling} shall be supplemented by representative methods of experimental characterization. The next Section~\ref{sec:powder} focuses on powder bed radiation and heat transfer as discussed from a rather theoretical point of view in Section~\ref{sec:powder_modeling}. In the subsequent Sections~\ref{sec:residualstresses},~\ref{sec:meltpool} and~\ref{sec:microstructure}, the experimental characterization of melt track stability, surface quality and defects, of residual stresses and dimensional warping effects as well as of metallurgical microstructures and grain morphology will be considered. These sections represent the counterpart to the Sections~\ref{sec:macroscopicmodels},~\ref{sec:mesoscopicmodels} and~\ref{sec:microscopicmodels}, where exactly these aspects, prevalent at different length scales of the SLM-manufactured part, have been described by means of macroscopic, mesoscopic and microscopic models.

\subsection{Optical and thermal characterization of powders}
\label{sec:powder}

\begin{figure}[t!!!]
\begin{centering}
    {
    \includegraphics[height=0.45\textwidth]{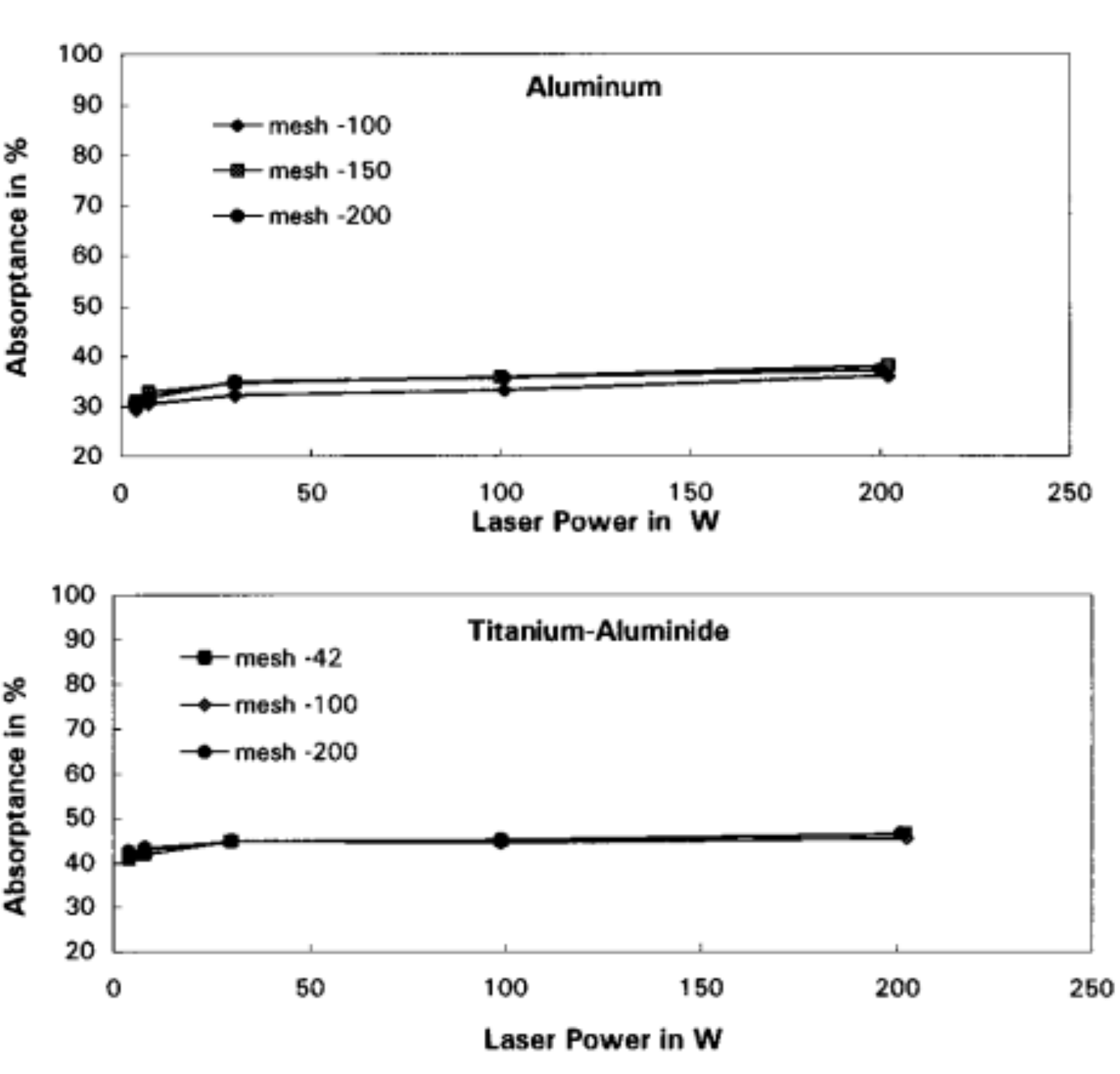}
    }
    {
    \includegraphics[height=0.46\textwidth]{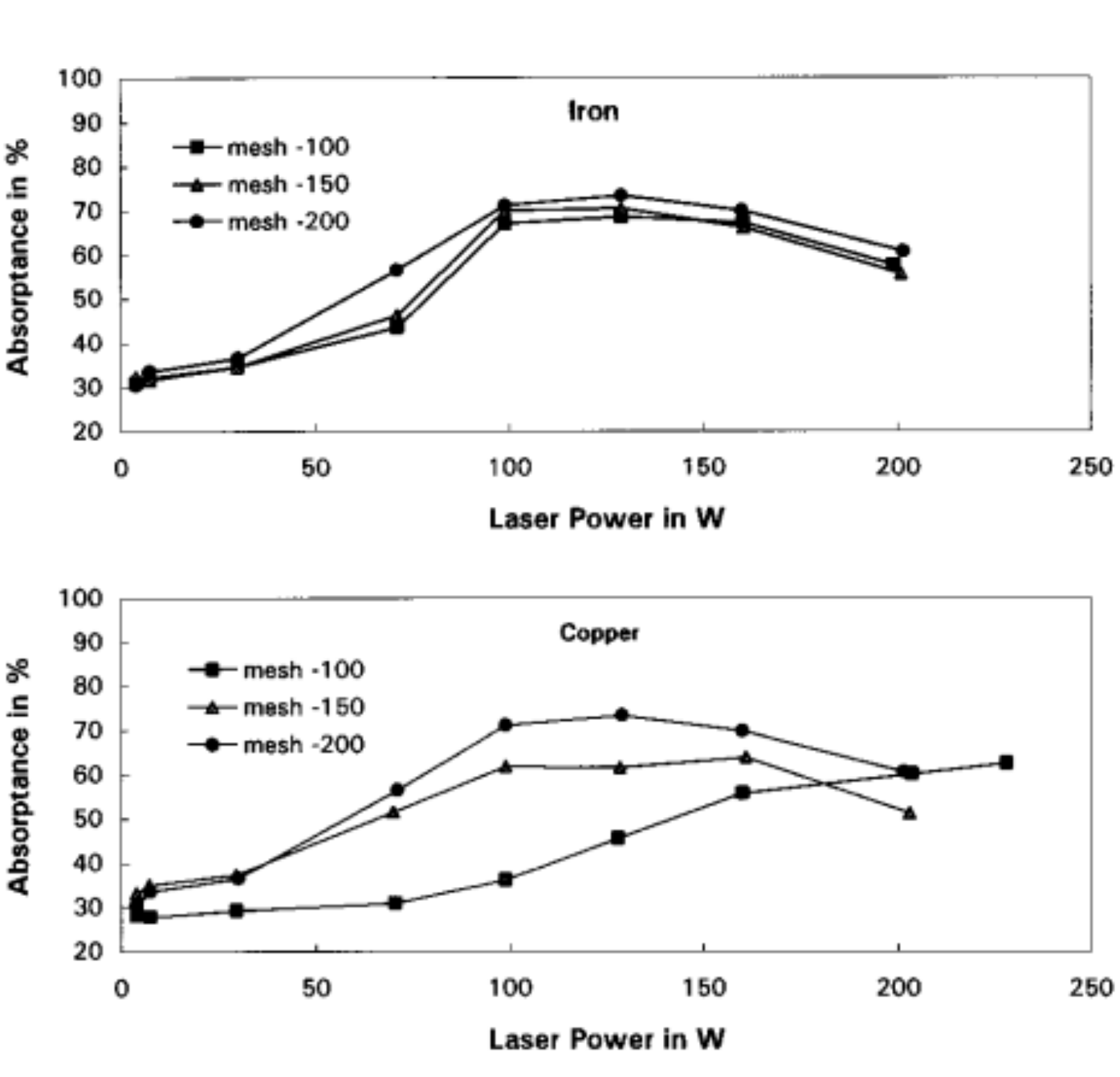}
    }
\caption{Effects of powder grain size, material as well as laser power on relative powder layer absorption ($w\!=\!10.6 \mu m$),~\cite{Haag1996}.}
\label{Fig_Haag1996}
\end{centering}
\end{figure}

Initial research on powder-laser interaction began not with AM in mind, but rather with surface modification techniques such as cladding and hard-facing processes~\cite{Haag1996}. Haag et al.~\cite{Haag1996} describe the archetypal experiment to measure absorption coefficients of metal powder, wherein a laser of known power is used to irradiate a powder bed, with the absorption coefficient calculated from the time-derivative of the powder bed temperature. In this study, a $CO_2$ laser ($w\!=\!10.6 \mu m$) was used to heat aluminum, iron, titanium aluminide, and copper powders.  Selected particle sizes range from $42 \mu m$ to $200 \mu m$, validated by SEM. The key results are summarized in Figure~\ref{Fig_Haag1996}: In the range of low laser power, material-dependent absorption coefficients as well as the consideration of different powder sizes cause only slight variations in the overall powder bed absorption leading to values of $30\%-45\%$ absorption as compared to the incident energy radiation. The increase in absorption of iron and copper at higher laser powers, and correspondingly high temperatures, is attributed to surface oxidation of the material. A practically identical experiment was performed by Tolochko~\cite{Tolochko2000} to catalog nominal absorption coefficients for materials relevant to SLS and SLM processing at wavelengths of $1.06 \mu m$ and $10.6 \mu m$.  Results indicate that the wavelength $1.06 \mu m$ represents a superior choice for processing metals, with absorption coefficients generally lying between $0.6$ and $0.7$, as opposed to values around $0.25$ to $0.45$ for the longer wavelength.  Conversely, absorption of $10.6 \mu m$ radiation is greater for the cataloged polymer materials, and ceramic materials are shown to vary widely.\\


\begin{figure}[h!!!]
\begin{centering}
    \includegraphics[width=0.8\textwidth]{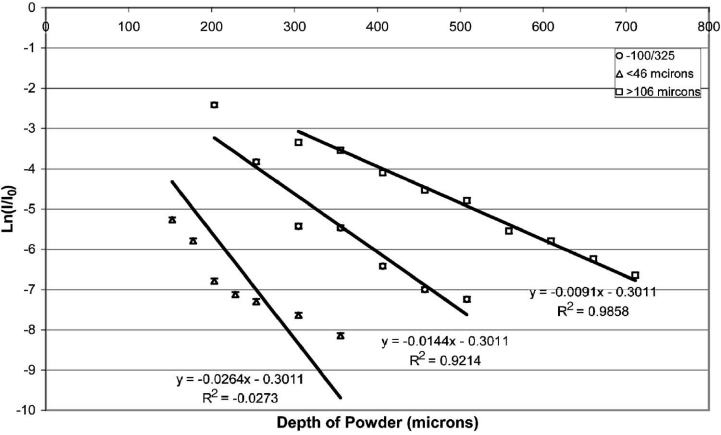}
\caption{Optical depth versus powder layer thickness for iron powder of different grain size  ($w\!=\!1.064 \mu m$),~\cite{McVey2007}.}
\label{Fig_McVey2007}
\end{centering}
\end{figure}

McVey presented an approach to measure the optical power delivered, reflected, and transmitted through a layer of powder using integrating spheres above and below the powder bed~\cite{McVey2007}. In Figure~\ref{Fig_McVey2007}, the effective optical depth (the logarithm of the ratio of measured / non-absorbed to incident light), is plotted versus the powder bed thickness for an applied laser beam of wavelength $w\!=\!1.064 \mu m$.  Accordingly, the absorption of $1.064 \mu m$ light has a strong dependence on particle size, with finer particles yielding stronger absorption.\\

\begin{figure}[htb]
\begin{centering}
\includegraphics[scale=.3]{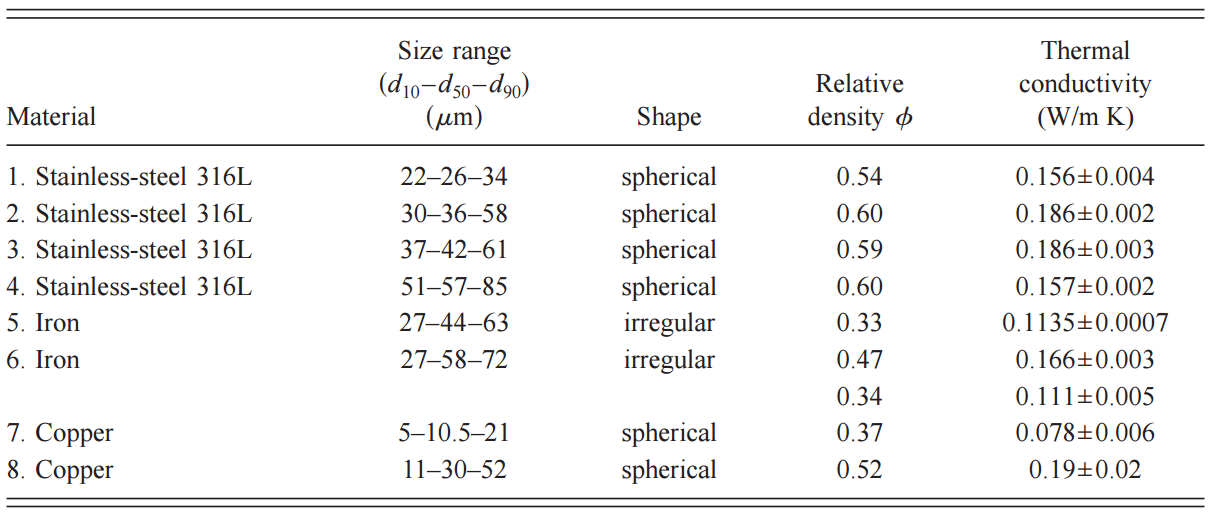}
\caption{Thermal conductivities of metallic powder beds for different materials and grain sizes distributions,~\cite{Rombouts2005}.}
\label{Fig_Rombouts_2005}
\end{centering}
\end{figure}

More recently, Rombouts et al.~\cite{Rombouts2005} have studied the thermal conductivity of powdered 316 stainless steel, iron, and copper. Their apparatus relies on a modulated laser beam to heat an optically opaque sample container, which subsequently conducts heat into a powder bed within the container. Under these circumstances, heat flows through the powder bed to pyro-electric detectors, which are read out using lock-in amplifiers indexed to the laser's modulation. Experimental results are shown in Figure~\ref{Fig_Rombouts_2005}. Accordingly, the powder conductivities are two to three orders of magnitude lower than in the fused material or build platform with bulk conductivities of $15,\; 74$ and $395\; W/m\;K$ for stainless steel, iron, and copper, respectively. It is argued that the thermal transport in powder beds of such fine particles occurs primarily through the gaseous phase (Knudsen diffusion) while direct particle-to-particle conduction is the governing effect for larger particle sizes (millimeter scale).

\subsection{Measurement of residual stresses and dimensional warping}
\label{sec:residualstresses}
    
As already stated earlier, basically, two thermal regimes responsible for the creation of residual stresses can be distinguished in SLM. First, stresses might be induced in the solid substrate just underneath the melt pool due to high thermal gradients and cooling rates in this region. These residual stresses are influenced by the powder structure and melt pool thermo-hydrodynamics and might result in a delamination of the current material layer. Secondly, residual stresses might also result as consequence of the repeated thermal cycles prevalent in material layers in larger distance from the top layer. The main factors of influence in this region are given by the specific geometry of the part (e.g. slender columns, thin walls and overhanging structures) as well as thermal boundary conditions e.g. in form of fixations on the build platform. Even for primitive geometries such as cubes this second category of thermal stresses might occur due to the temperature gradient in build direction, which typically results in compressive stresses in lower layers induced when the layers of higher temperatures above cool down. In all cases, these thermal stresses need to be reduced, or better avoided, in order to fully exploit the mechanical strength of a part when exposed to external loads, in order to avoid dimensional warping after removing the parts from the build platform and eventually in order to avoid crack propagation typically initiated at locations with residual stress peaks. On the one hand, residual stresses can be relieved by annealing after the SLM process (before build plate removal). On the other hand, such a heat treatment can compromise dimensional accuracy and impose grain growth demanding a tradeoff between these diverse requirements.\\

\begin{figure}[h!!!]
\begin{centering}
\includegraphics[width=0.9\textwidth]{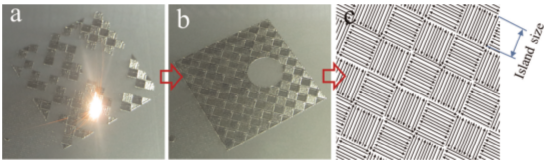}
\caption{Determination of optimal island sizes for an island-based SLM scanning strategy,~\cite{Lu2015}.}
\label{Fig_Lu2015}
\end{centering}
\end{figure}

Lu et al.~\cite{Lu2015} studied the effect of laser scan pattern on the properties of Inconel 718. The laser beam scanning strategy is depicted in Figure~\ref{Fig_Lu2015}, where part infill is fused by sequentially scanning square regions known as islands. Specifically, dimensions of $2 mm \;x\;2 mm$, $3 mm\; x \;3 mm$, $5mm \;x \;5mm$, and $7mm\; x \;7mm$ have been studied, using a $180W$ laser, $600mm\!/\!s$ scan speed, $30 \mu m$ layer thickness, and a $150 \mu m$ spot size. The study begins with a discussion of part density, in which the $5mm$ and $7mm$ specimens are shown to have $99.1\%$ of their theoretical density.  Both the $2mm$ and $3mm$ specimens featured lower densities at $98.67\%$ and $98.86\%$, respectively (see also the next section for a discussion of mesoscale defects such as pores). These findings are congruent with surface micrographs, which show increased porosity within the smaller island sizes, and even cracking within the $2mm$ specimen. Stress-strain data is also included, which shows comparatively little change in offset yield strength and ultimate tensile strength, yet elongation at break increases from $16.85\%$ to $25.25\%$ with increasing island size.  Again, this is in agreement with the surface micrographs, as the increased density of defects serve as sites to propagate cracks and initiate fracture. Finally, micro-hardness and residual stresses were evaluated. Counterintuitively, the $2mm$ specimen featured the lowest residual stress. This observation is believed to result from stress relaxation due to cracking. As expected, the $3mm$ specimens showed greater residual stresses than the $5mm$ and $7mm$ specimens. The greater thermal gradients of the smaller $3mm$ specimen, arising from a greater degree of heat remaining from the previous pass, can explain the greater residual stresses as compared to the $5\;mm$ specimens.\\

In Hodge et al.~\cite{Hodge2016}, a combined simulation and experimental approach has been applied in order to access residual stresses as well as the amount of dimensional warping observed in AM components, and to find how those stresses are dependent upon the build orientation and infill scan strategy. As described earlier, Digital Image Correlation (DIC) has been used to measure stress-induced dimensional changes in triangular test components. This experimental technique is based upon time-correlating successive images to extract strain measurements as an object undergoes a physical change, such as in temperature or external loading~\cite{Khoo2016}. Hodge uses this technique to monitor the surface of triangular test components as they are removed from the build platform. This process changes the residual stress distribution within the component, thereby inducing a corresponding change in strain. These measurements and previously discussed simulation results are augmented using neutron diffraction experiments to directly measure the residual stress distribution in the test pieces. Like DIC, this experimental method is well established in solid mechanics for mapping stress distributions in crystal-like materials~\cite{Withers2007}. The full strain tensor may be determined by approaching the same volume from a plurality of directions. Of course, this measurement represents the average strain tensor over the volume probed, commonly between $0.2$ and $1000mm^3$. Hodge et al. mainly employed the described experimental techniques for verification of their simulation model (see Section~\ref{sec:macroscopicmodels} for a discussion of the results).\\

Mercelis and Kruth~\cite{Mercelis2006} evaluated residual stresses as dependent on scan strategy, build height, and degree of base plate pre-heating. They relied on sectioning their parts followed by X-Ray diffraction studies to determine residual stresses. X-Ray diffraction determines residual stresses on identical principles to neutron diffraction. However, the comparatively low penetration of the X-Ray technique makes it only useful as a surface technique, hence the need for sectioning. The authors clearly showed that residual stresses increase with component height. As long as the parts stayed connected to the baseplate, very high stress levels, typically in the range of the material’s yield strength, could be observed. Also the exposure strategy that is being used to fuse the powder layers has a large influence on the residual stress levels being developed. It was found that stresses are larger perpendicular to the scan direction than in scan direction. A scanning strategy subdividing the surface in small islands resulted in a lower maximum stress value. Preheating the baseplate to $200^\circ C$ was shown to reduce induced residual stresses by nearly 10\%. Finally, the authors showed that removing the part from the baseplate relaxes approximately 50\% of the residual stresses in the part, which negatively impacts the dimensional accuracy of the component once removed.\\

Havermann et al.~\cite{Havermann2015} presented a unique method in which fiber optics were embedded within SLM parts. Thereto, the authors fabricated rectangular test pieces. At the location where a strain measurement was desired, a fiber optic strain sensor (Bragg grating) was placed into the part, thereby enabling the strains induced by residual stresses in the component to also deform the optical fiber. This changes the optical transmission of the embedded Bragg grating, enabling the quantification of strain. The employed small fiber diameter enabled measurements within the first several layers of SLM parts, where interaction with the baseplate is central to the development of stresses. The authors observed increasing compressive residual stresses as the first 6 layers above the fiber were fused. After building 14 layers above the fiber and allowing the component to cool, residual stresses in the range of $70 MPa$ were measured.\\

\begin{figure}[h!!!]
\centerline{
\begin{minipage}{0.8\textwidth}
\includegraphics[width=1.0\textwidth]{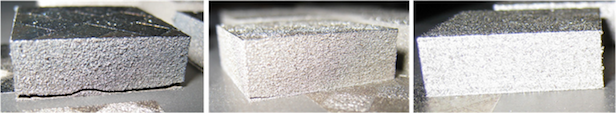}
\caption{Blocks of M2 high speed steel (HSS), manufactured by SLM at build platform temperatures of $90^{\circ} C$ (left), $150^{\circ} C$ (middle), and $200^{\circ} C$ (right), showing the critical influence of preheating temperature on cracking,~\cite{Kempen2014}.}
\label{Fig_zy17}
\end{minipage}}
\end{figure}

Kempen et al.~\cite{Kempen2014} studied the effect of build platform preheating temperature in order to lower thermal gradients and to finally yield crack-free parts with high density. Figure~\ref{Fig_zy17} shows three samples, which were built using three different preheating temperatures ($90^{\circ} C$ (left), $150^{\circ} C$ (middle), and $200^{\circ} C$ (right)). It can be seen that the higher preheating temperature clearly results in lower residual stresses and a lower degree of crack formation. In practical applications, too high preheating temperatures and too long preheating times might however lead to an undesirable sintering of loose powder, which complicates its removal from the final part.

\subsection{Characterization of melt track quality and defect detection}
\label{sec:meltpool}

\begin{figure}[h!!!]
\begin{centering}
\includegraphics[width=1.0\textwidth]{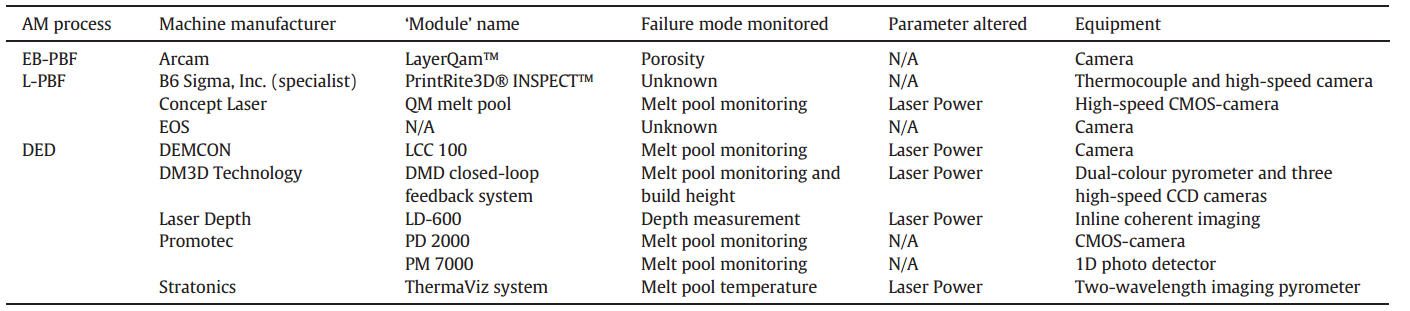}
\caption{Summary of different commercially available monitoring systems for SLM, EBM and DED processing,~\cite{Everton2016}.}
\label{Fig_Everton2016}
\end{centering}
\end{figure}

In this section, some exemplary experimental approaches for characterizing melt track stability and quality as well as possible defects will be discussed. In recent years, much effort has been invested in monitoring SLM and EBM processes for process and quality control~\cite{Everton2016}. Defect detection centers on identifying pores, balling, unfused powder, and cracking as a result of sub-optimal process parameters. In parallel, high speed monitoring has been researched with the aim of real time process control. The results of these efforts have manifested in several commercially available systems for monitoring metal AM processes as shown in Figure~\ref{Fig_Everton2016}. In the following, we will discuss several of the key works behind these products, and refer to the review by Everton et al.~\cite{Everton2016} for complete treatment.\\

Craeghs et al.~\cite{Craeghs2011} studied melt pool heat transfer using a radiometer and high speed camera, arranged coaxially with the laser. In this work, they describe the so-called edge effect that occurs when a track is scanned without an adjacent contour as already discussed in Section~\ref{sec:mesoscopicmodels}. Heat from scan segments adjacent to neighboring contours are shown to conduct both into the underlying layers and into the previously solidified, adjacent track in the same layer. However, heat from the initial contour does not have this second path, meaning the bulk of the heat flux must be directed into the underlying layer. As a result, the melt pool is larger, reaches higher peak temperatures, stays hot longer, and incorporates more material than when subsequent contours are melted. Overhanging features were also studied, where it was found that scan parameters suitable for standard processing conditions delivered too much power to the initial layers of overhanging features fabricated above areas of unfused powder. The extra heat enhances Rayleigh instabilities, leading to balling, poor surface finish, and undesirable mechanical properties. Finally, acute corners were shown to present many of the same challenges with temperature control. Specifically, the case of a scan pattern incorporating a 180 degree U-turn was investigated. Again, due to the reduced area for heat to flow out of the melt pool and the partly scan path re-heating, this extra heat results in blob-like formations at the apex of the U-turn. This same experimental hardware was also used to investigate the influence of support strategies on process temperatures for fabrication of overhanging features~\cite{Krauss2012}. Here, the authors printed T-shaped structures, in which the overhung regions were supported with uniformly spaced and sized column like additions. The spatial density of the columns was shown to greatly influence the maximum temperatures achieved in the overhanging layer, and therefore porosity and surface finish, as the support structure serves as a heat sink.\\ 

Krauss et al.~\cite{Krauss2014} performed long-wave infrared (LWIR) imaging to determine component porosity in Inconel-718 specimens. The study begins by deriving an approximation for the thermal diffusivity $a$, which is given by
\begin{equation}
  a \approx \dfrac{1}{t}\left(\dfrac{1}{T(t)-T_{ref}}\right)^2\left(\dfrac{AP}{Vhc_p\rho\sqrt{\pi}}\right)^2.
  \label{Eqn_ApproxDiffusivity}
\end{equation}
In this expression, $P$ and $V$ represent laser power and scan velocity, $h$ is the hatch spacing and $A$ represents the efficiency of energy absorption in the powder layer (assumed $60\%$). Moreover, $c_p$ and $\rho$ are the specific heat capacity and density of the material, as usual. Finally, $T_{ref}$ is the ambient temperature of the build environment, and $T(t)$ is the radiometrically determined temperature of the pixel as a function of time $t$. The authors assign an imperfection level based upon the observed porosity and degree of binding error. The imperfection level $IP$ was determined to be inversely proportional to the diffusivity $a$ of cubic specimens at a layer height of $4mm$ with a correlation coefficient of $0.7$. Based on a proper calibration of the $IP$-$a$ relation and~\eqref{Eqn_ApproxDiffusivity}, the proposed procedure enables the in-situ determination of part imperfection based on measured temperature profiles.  The method is shown to be sensitive to both delamination and porosity arising from sub-optimal process parameters, with sensitivity demonstrated over a range from nearly fully dense material to a void fraction of 40\%.  Ultimate sensitivity of this method is not calculated. However, the authors note that the spatial and temporal resolution of imaging sensors represents the greatest challenge.  In the follow-up study~\cite{Krauss2015}, more detailed information on laser power and scan rate is provided, and the resulting effects on thermal diffusivity, maximal temperature, and observation of melt sputter have been presented. \\

\begin{figure}[h!!!]
\begin{centering}
\includegraphics[width=1.0 \textwidth]{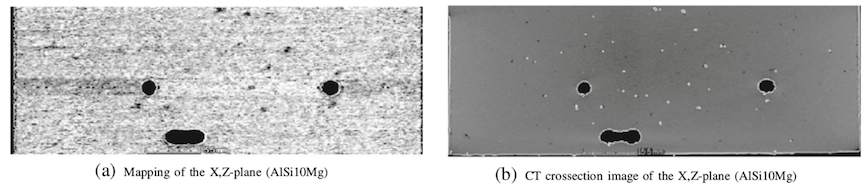}
\caption{Detection of porosity from melt pool size and temperature: Comparison of a) predicted and b) actual defect locations (black areas),~\cite{Clijsters2014}.}
\label{Fig_Clijsters2014}
\end{centering}
\end{figure}

Simultaneously, Clijsters et al.~\cite{Clijsters2014} have studied in-situ monitoring for quality control and are working towards feedback for process control. Using an InGaAs NIR camera and a pair of photo diodes with differing band pass filters, the authors extract signals representing melt pool radiance (intensity of emitted light), area, length, and width. They report success in detection of overheating of the melt pool, specifically at the extremes of fill scan vectors, along with increased porosity in these regions. Using data-fusion techniques, the authors are able to predict the locations of porosity within AM parts, as verified by CT cross sectional images (see Figure~\ref{Fig_Clijsters2014} for illustration).\\

Experimentally, process maps can be constituted by gathering data on part density and other properties as related to the scan rate, hatch spacing, laser power, etc. While ultimate control of SLM will arguably require determination of local scan parameters according to the local part geometry and thermal boundary conditions, the present approach is to determine a generalized "process window" which can be applied to approximately achieve full density. The literature in this field is abundant, considering a wide variety of materials and processes. Thomas et al.~\cite{Thomas2016}, propose a methodology of normalizing process parameters for SLM and EBM. The methodology builds on the previous work~\cite{Ion1992}, where a dimensionless laser beam power $P^{*}$ and velocity $V^{*}$ are defined:
\begin{subequations}
\begin{equation}
  P^{*} := \frac{AP}{r_{B}\lambda(T_{m}-T_{0})},
  \label{Eqn_Thomas_1}
\end{equation}
\begin{equation}
  V^{*} := \frac{Vr_{B}}{a}.
  \label{Eqn_Thomas_2}
\end{equation}
\end{subequations} 
Originally, these dimensionless quantities have been identified from analytical solutions derived for the temperature profiles resulting from a moving heat source on a semi-infinite body, a model that has been applied in order to study heat transport in laser beam welding. In~\eqref{Eqn_Thomas_1}, the effective laser power, defined as laser power $P$ times absorption coefficient $A$, is related to the temperature difference between initial powder bed temperature $T_0$ and melting temperature $T_{m}$, the thermal conductivity $\lambda$ and the laser beam radius $r_{B}$. Equation~\eqref{Eqn_Thomas_2} relates the velocity $V$ of the laser beam heat source to the thermal diffusivity $a$. According to~\cite{Ion1992}, the choice of the parameters $P^{*}$ and $V^{*}$ controls the peak temperature and heating rate. This representation matches Figure~\ref{Fig_Thomas_2_2016}.\\

\begin{figure}[h!!!]
\centerline{
\begin{minipage}{0.8\textwidth}
\includegraphics[width=1.0\textwidth]{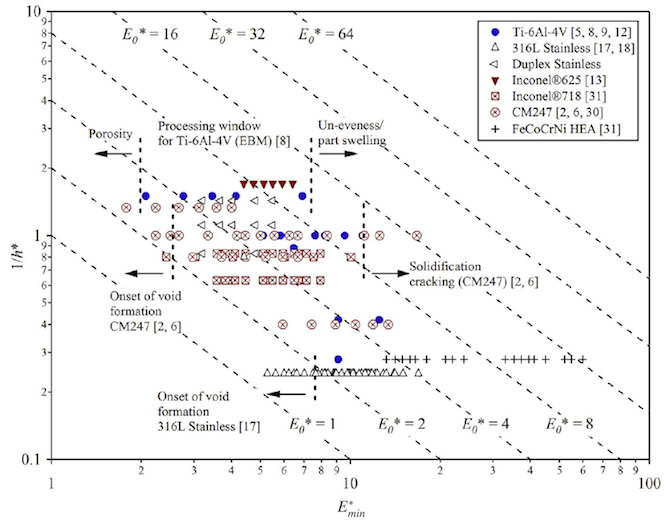}
\caption{Normalized processing diagram showing different SLM/EBM realizations from the literature. The inverse dimensionless hatch spacing $h^{*}$ is plotted over the dimensionless energy ratio $E^{*} _{min}$,~\cite{Thomas2016}.}
\label{Fig_Thomas_1_2016}
\end{minipage}}
\end{figure}

Thomas et al. supplemented the parameters $P^{*}$ and $V^{*}$ proposed in~\cite{Ion1992} by the following dimensionless quantities:
\begin{subequations}
\begin{equation}
  l^{*} := \frac{2l}{r_{B}}
  \label{Eqn_Thomas_3}
\end{equation}
\begin{equation}
  h^{*} := \frac{h}{r_{B}}
  \label{Eqn_Thomas_4}
\end{equation}
\begin{equation}
  E^{*} := \frac{A P}{2 V l r_{B}}\cdot \frac{1}{0.67 \rho c_{p}(T_{m}-T_{0})}
  \label{Eqn_Thomas_5}
\end{equation}
\begin{equation}
  E^{*} _{min} := \frac{3}{2} E^{*} = \frac{A P}{2 V l r_{B}}\cdot \frac{1}{\rho c_{p}(T_{m}-T_{0})}
  \label{Eqn_Thomas_6}
\end{equation}
\end{subequations}
Here, the dimensionless thickness of the powder layer $l^{*}$ is defined via the physical powder bed depth $l$ and the laser beam radius $r_B$. Also the hatch spacing $h$ is normalized by the laser beam radius. Assuming a powder bed porosity of $0.67$, the term $0.67 \rho c_{p}(T_{m}\!-\!T_{0})$ represents the energy density required to increase the temperature of powder with effective density $0.67 \rho$ and heat capacity $c_{p}$ from the initial temperature $T_{0}$ to the melt temperature $T_m$. Again, $A P$ represents the effectively absorbed laser power. The normalization of the laser power with the product $2 V l r_{B}$ yields an energy per unit volume. By assuming a latent heat of melting in the range of $H_m \approx 0.5 \rho c_{p}(T_{m}\!-\!T_{0})$ for typically employed metals and preheating temperatures, Thomas postulated the energy density required to increase the powder temperature from $T_0$ to $T_m$ and to melt the powder subsequently as the value $0.67 \rho c_{p}(T_{m}\!-\!T_{0})+H_m=\rho c_{p}(T_{m}\!-\!T_{0})$. Consequently, the last dimensionless parameter $E^{*} _{min}$ represents the ratio of incident energy density to the energy density required for heating and melting the powder. The parameter combinations found in different experimental contributions from the literature have been plotted in a double-logarithmic $(E^{*} _{min})\!-\!(1/h^*)$ diagram as illustrated in Figure~\ref{Fig_Thomas_1_2016}.\\

\begin{table}[h!!!]\scriptsize
\begin{center}
\begin{tabular}{@{}p{1.6cm} @{}p{1.75cm} @{}p{1.3cm} @{}p{.1cm} @{}p{1.6cm} @{}p{1.2cm} @{}p{1.3cm} @{}p{1.7cm} @{}p{1.9cm} @{}p{1.6cm} p{1.6cm}} \hline
\setlength{\tabcolsep}{4pt} 
   & \multicolumn{2}{@{}l}{Heat source (type)} & & \multicolumn{6}{@{}l}{Processing parameters} & Thermophysical \\
   \cline{2-3} \cline{5-10}
   & AM platform & Alloy system & & Bed Temp., $T_{0}$ \newline [$K$] & Power, $P$ \newline[$W$] & Velocity, $V$ \newline[$m/s$] & Layer height, $l$ \newline[$m$] & Hatch spacing, $h$ \newline[$m$] & Beam radius, $r_{B}$ \newline[$m$] & Properties \\
  \hline\hline
  Thomas et. al. & Electron Beam\newline \em Arcam A2\em & Ti--6Al--4V & & $923$ & $600$ & -- & -- & -- & $150 x 10^{-6}$ & Al-Bermani \newline et al. \\ \hline
  Juechter et al. & Electron Beam\newline \em Arcam S12\em & Ti--6Al--4V & & $923$ & $60-$\newline$1400$ & $0.2-6.4$ & $50 x 10^{-6}$ & $100 x 10^{-6}$ & $150 x 10^{-6}$ & Al-Bermani \newline et al. \\ \hline
  Vranken et al. & Laser\newline (SMYb:YAG) \newline \em In-house LM-Q\em & Ti--6Al--4V & & $298$ & $250$ & $1.6$ & $30 x 10^{-6}$ & $60 x 10^{-6}$ & $52 x 10^{-6}$ & ASM \newline International \\ \hline
  Qui et al. & Laser\newline \em Concept Laser\em & Ti--6Al--4V & & $298$ & $150-$\newline$200$ & $0.8-1.5$ & $20 x 10^{-6}$ & $75 x 10^{-6}$ & $75 x 10^{-6}$ & ASM \newline International \\ \hline
  Xu et al. & Laser\newline \em SLM 250 HL\em & Ti--6Al--4V & & $498$ & $175-$\newline$375$ & $0.7-$\newline$1.029$ & $30-$\newline$90 x 10^{-6}$ & $120-$\newline$180 x 10^{-6}$ & $70-$\newline$120 x 10^{-6}$ & ASM \newline International \\ \hline
  Kamath et al. & Laser\newline \em Concept Laser\em \newline M2 & Ti--6Al--4V & & $298$ & $150-$\newline$400$ & $0.5-2.5$ & $30 x 10^{-6}$ & $112 x 10^{-6}$ & $27 x 10^{-6}$ & ASM \newline International \\ \hline
  Ziolkowski \newline et al. & Laser\newline \em SLM Realizer II\em & 316L SS & & $298$ & $97$ & $0.2$ & $50 x 10^{-6}$ & $125 x 10^{-6}$ & $100 x 10^{-6}$ & ASM \newline International \\ \hline
  Unpublished Data & Laser & 316L SS & & $298$ & -- & -- & -- & -- & -- & Al-Bermani \newline et al. \\ \hline
  Cooper et al. & Laser\newline \em In-house\em & Ti--6Al--4V & & $298$ & $800-$\newline$1000$ & $0.1-0.12$ & $100 x 10^{-6}$ & $500 x 10^{-6}$ & $850 x 10^{-6}$ & ASM \newline International \\ \hline
  Boswell et al. & Laser\newline \em SLM 280HL\em & Ti--6Al--4V & & $298$ & -- & -- & -- & -- & -- & ASM \newline International \\ \hline
  Carter et al. & Laser\newline \em Concept Laser\newline M2\em & Ti--6Al--4V & & $298$ & $100-$\newline$200$ & $0.4-2.0$ & $20 x 10^{-6}$ & $75 x 10^{-6}$ & $75 x 10^{-6}$ & Mukai et al. \newline Avala et al. \\ \hline
  Boswell et al. & Laser\newline \em EOS M270\em & Ti--6Al--4V & & $298$ & -- & -- & -- & -- & -- & Mukai et al. \newline Avala et al. \\ \hline
  Brif et al. & Laser\newline \em AM 125\em & Ti--6Al--4V & & $298$ & $200$ & $0.33-0.60$ & $20-50 x 10^{-6}$ & $90 x 10^{-6}$ & $25 x 10^{-6}$ & Brif et al. \\ \hline
\end{tabular}
\end{center}
\caption{Literature survey of SLM / EBM process parameters required for normalization and creation of process maps,~\cite{Thomas2016}.}
\label{Tab_Thomas_1_2016}
\end{table}

\begin{figure}[h!!!]
\begin{centering}
\includegraphics[width=1.0\textwidth]{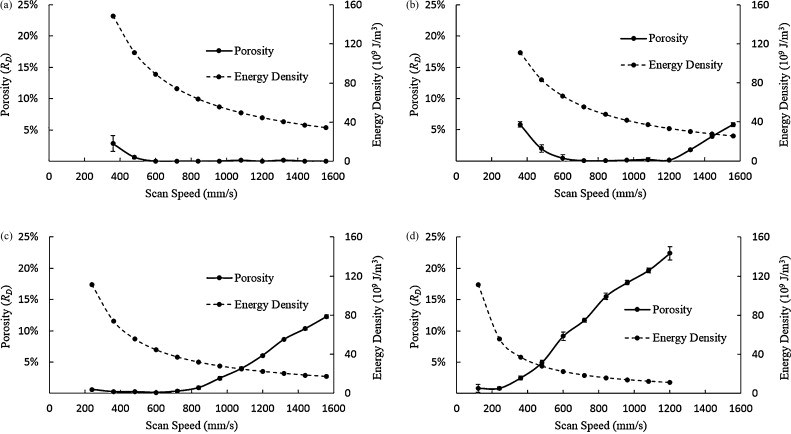}
\caption{Porosity in SLM of Ti-6Al-4V vs. energy density for different laser powers: (a) $P\!=\!160W$, (b) $P\!=\!120W$, (c) $P\!=\!80W$, (d) $P\!=\!40W$,~\cite{Gong2014}.}
\label{Fig_Gong_2_2014}
\end{centering}
\end{figure}

In order to gain further understanding concerning this specific choice of diagram axes, a slightly different version of normalized hatch spacing $\tilde{h}^*\!:=\!0.5h^{*} \!=\! h / (2r_{B})$ shall be considered in the following. In this context, $\tilde{h}^*=0$ represents full overlap and $\tilde{h}^*=1$ represents vanishing overlap of two successive scan tracks. In a $(E^{*} _{min})\!-\!(1/\tilde{h}^*)$ diagram following this modified definition, constant ratios $\tilde{E}^{*} _0\!:=\!E^{*} _{min}/\tilde{h}^*\!=\!\text{const.}$ are represented by straight lines with slope $-1$. For a powder bed surface of dimensions $L_v$ in scan direction and $L_h$ in hatch direction, $L_h/h$ scan tracks are required in order to scan the entire surface and $L_v/V$ represents the time of scanning one track. The total amount of energy absorbed in one powder layer is given by $L_h/h \cdot L_v/V \cdot AP$. On the other hand, based on the assumptions in~\cite{Thomas2016}, the amount of energy required to heat and melt the entire powder bed is given by $(L_hL_vl)\cdot\rho c_{p}(T_{m}\!-\!T_{0})$. It can easily be verified that the ratio $\tilde{E}^{*} _0\!=\!E^{*} _{min}/\tilde{h}^*$ exactly represents the ratio of these two energies, i.e. of the energy absorbed in the entire powder bed and the energy required to heat and melt the entire powder bed. The statements made in~\cite{Thomas2016} on the basis of the alternative dimensionless hatch spacing ${h}^*=1$ are equivalent, when considering the relation $\tilde{h}^*\!=\!0.5h^{*}$ or $\tilde{E}^{*} _0\!=\!2{E}^{*} _0$. Consequently, the constants ${E}^{*} _0$, denoted as isopleths and plotted in Figure~\ref{Fig_Thomas_1_2016}, have to be multiplied by a factor of 2 in order to yield the practically relevant ratio of absorbed energy to required energy for heating and melting the entire powder bed. For completeness, the experimental raw data from the literature underlying the data points in Figure~\ref{Fig_Thomas_1_2016} is summarized in Table~\ref{Tab_Thomas_1_2016} focusing mostly on Ti-6Al-4V and 316L SS.\\

Based on the failure mechanisms reported in the literature and included into Figure~\ref{Fig_Thomas_1_2016}, excessively high energy deposition may result in overheating, cracking and part swelling whereas insufficient energy deposition may yield an incomplete melting and incorporation of the powder bed and consequently undesired porosity and void formation. From these graphical results, Thomas et al. deduce that across different processes and alloys, SLM/EBM is most effective in the realm of $2 < E_0^{*} < 8$. Thus, the amount of effectively absorbed energy per powder layer was typically chosen to be at least by a factor of \textit{four} higher than the minimally required energy for heating and melting the powder of one layer (since $E_0^{*}>2$ is equivalent to $\tilde{E}^{*} _0>4$; see also the discussion in the paragraph above). The factor of four might become reasonable when considering that substantial parts of neighboring tracks and underlying layers have to be remelted in order to provide good adhesion as well as additional effects such as thermal losses due to emission, convection and evaporation on the powder bed surface and heat conduction across the build platform. Thomas et al. went on to conduct experiments using EBM of Ti-6Al-4V to develop a material specific processing map. They found that high quality (absence of defect voids) is obtained at highest $E_0^{*}$ of the study trade space. This region sacrifices production speed with slow scanning speed and more scans per a given part planar area (low $h^{*}$ and $v^{*}$). At higher speeds and wider hatch spacings the Vickers hardness number (VHN) has been observed to increase (as consequence of higher cooling rates and the possible creation of non-equilibrium phases) and structural integrity to be compromised by the presence of undesirable microstructural features such as lack of fusion defects. While the isopleth lines of constant ${E}^{*}$ might provide a rough process window as well as an evaluation of the process energy efficiency, this approach does not allow precise assessment of the stability and surface quality of individual melt tracks. For example, it is obvious that in practice an excessively too high energy deposition per melt track cannot be compensated by a larger hatch spacing, which, however, would be suggested by the definition of $E_0^{*}$.\\

Similar to the discussion of $\tilde{E}_0^{*}$ above, the dimensionless energy $E^{*} _{min}$, represented by vertical lines in Figure~\ref{Fig_Thomas_1_2016}, can be shown to represent the ratio of absorbed energy per melt track to the energy required for heating and melting a track of cross-section $2r_B \cdot l$. However, even a comparison of results related to constant values of $E^{*} _{min}$ exhibit difficulties. Considering for example the mechanisms of powder bed radiation transfer discussed in Section~\ref{sec:powder_modeling}, it becomes obvious that an excessively high beam power can typically not be compensated by a very high powder layer thickness. Due to the typical radiation transfer depth determined by the powder layer structure and the very low thermal conductivity of the powder, it is rather likely that the intended compensation results in excessive evaporation at the powder bed surface in combination with unmolten powder at the bottom of the too thick powder layer.\\

\begin{figure}[h!!!]
\centerline{
\begin{minipage}{0.72\textwidth}
\captionsetup{justification=centering}
\includegraphics[width=1.0\textwidth]{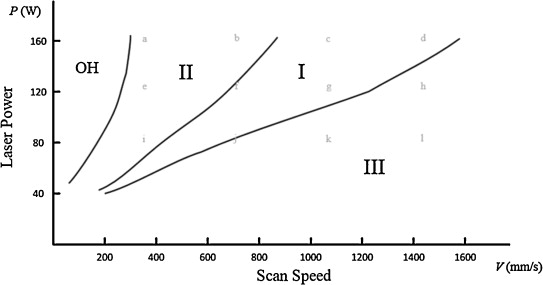}
\caption{Processing regimes for SLM of Ti-6Al-4V depending on laser power and scanning velocity: (I) "processing window", (II) "incomplete melting", (III) "overheating", (OH) "severe overheating",~\cite{Gong2014}.}
\label{Fig_Gong_1_2014}
\end{minipage}}
\end{figure}

For specific materials and process parameter sets, direct optimization of the scan parameters and their relation e.g. to porosity can readily be performed.  Gong et al.~\cite{Gong2014} varied the energy density for SLM-processing Ti-6Al-4V via laser beam power and velocity, however, \textit{at constant powder bed depth and density}. In Figure~\ref{Fig_Gong_2_2014}, each subplot representing a constant laser power shows the relation between incident energy density and part quality using porosity as the metric of quality. Accordingly, Gong et al. identified four broad regimes of process parameters through comparing porosity resulting from differing combinations of beam speed and beam power. These regimes are shown in Figure~\ref{Fig_Gong_1_2014} and denoted as $I$, $II$, $III$ and $OH$. Region $I$ is described as the "processing window" producing fully dense SLM parts. Process parameters in regions II and III produce porosity defects, due to underheating and incomplete melting of powder particles, or due to overheating, evaporation and gas cavitation, respectively. The region labeled as OH, referring to the "overheating" regime, induces such severe accumulated thermal strains and part distortions so as to inhibit operation of the SLM recoater. A similar approach has been taken in~\cite{Kamath2014} for 316L stainless steel.\\

\begin{figure}[h!!]
 \centering
  \includegraphics[height=0.27\textwidth]{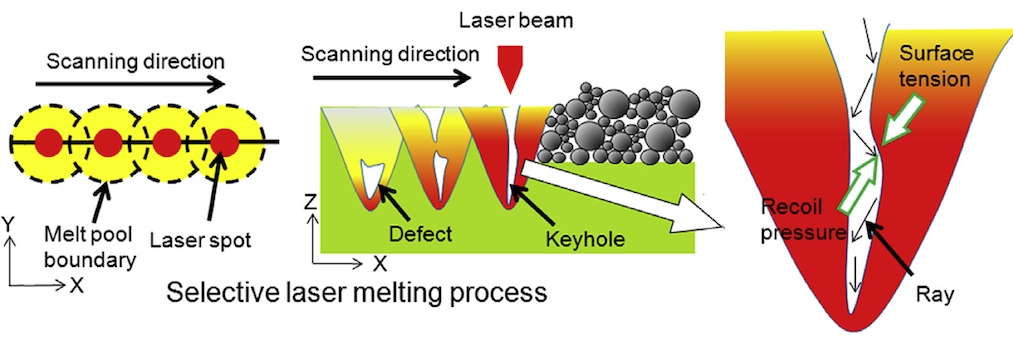}
 \caption{Visualization of physical mechanisms associated with melt pool dynamics and keyhole formation, ~\cite{Liu2016}.}
 \label{fig:Fig_zy6}
\end{figure}

\begin{figure}[b!!]
 \centerline{
 \begin{minipage}{0.93\textwidth}
  \includegraphics[width=1.0\textwidth]{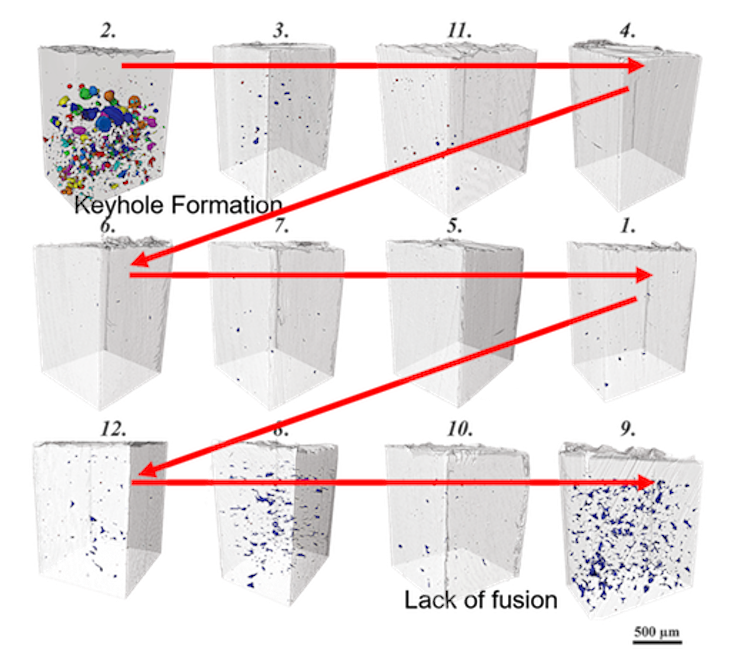}
 \caption{Results of x-ray micro computed tomography of Ti-6Al-4V samples processed by SLM, ordered by increasing effective track penetration depth (red arrows). The scans show the transition from large and numerous pores generated by keyhole formation to highly dense parts to highly porous samples generated by lack of fusion. It has to be noted that the inhomogeneity observed in these scans stems from the use of a different set of machine processing parameters ("top skin") for high part surface quality,~\cite{Cunningham2017}.}
 \label{fig:Fig_Rollett}
\end{minipage}}
\end{figure}

Commonly, porosities $< 0.5\%$ are desired for SLM processes. A high residual porosity may be regarded as a significant defect of SLM-manufactured parts for many applications because pores facilitate crack formation and propagation, which is particularly deleterious to strain-to-failure and cyclic fatigue life. In this context, the formation of gas-induced pores typically occurs at too high energy densities and can be explained by the so-called keyhole mechanism as illustrated in Figure~\ref{fig:Fig_zy6}. It is generally believed that the recoil pressure resulting from excessive heating and attended evaporation is able to form a deep depression of the melt pool directly below the laser beam (see also the discussion in Section~\ref{sec:mesoscopicmodels}). With increasing depth of this depression, the incident laser radiation is reflected on the sides of the depression, leading to a more concentrated energy input and higher temperatures at the bottom of the depression, further evaporation at this location and consequently to the creation of even deeper, narrow corridors. Typically, this keyhole mechanism can lead to considerable penetration depths of the laser beam into the underlying material. Due to the more focused energy source, SLM commonly results in deeper keyholes than EBM, whereas EBM yields larger melt pool sizes~\cite{Liu2016}. When the laser beam moves on, also the recoil pressure as driving force diminishes and, at some point, the keyhole will collapse under the action of surface tension and gravity (which, however, is typically negligible for the relevant range of length scales). The incomplete collapse of the vapor cavity at the bottom of the keyhole typically leaves voids with a characteristic spherical shape.\\

Recent micro x-ray computed tomography ($\mu$SXCT) studies have clearly shown the relationship between SLM processing energy density and the resulting porosity in the as-built SLM part. Cunningham et al.~\cite{Cunningham2017} performed $\mu$SXCT using the Advanced Photon Source at Argonne National Laboratory to image Ti-6Al-4V samples fabricated by SLM as shown in Figure~\ref{fig:Fig_Rollett}. The 12 samples studied were processed on an EOS M290 machine with varying laser powers (100 - 370 $W$), scan speeds (400 - 1200 $mm/s$), and hatch spacings (40 - 240 $\mu m$) to achieve widely varying melt-pool penetration overlap depths $d$ (0 - 128 $\mu m$ as calculated in the idealized geometry of two hemispherical melt tracks separated by the hatch spacing). As expected, lack-of-fusion defects were effectively eliminated for layer thicknesses~$t$ less than $d$. These results correlate well with the density and porosity results shown above by Gong et al.~\cite{Gong2014}.

\subsection{Microstructure evolution during the SLM process}
\label{sec:microstructure}

In this section, we summarize representative studies of the process-microstructure relationships of the materials Ti-6Al-4V, Inconel 718 as well as stainless steels. On the one hand, these materials are commonly studied in the context of SLM. On the other hand, they exhibit importantly different thermomechanical behavior.\\


\begin{figure}[h!!!]
\begin{centering}
\includegraphics[width=1.0\textwidth]{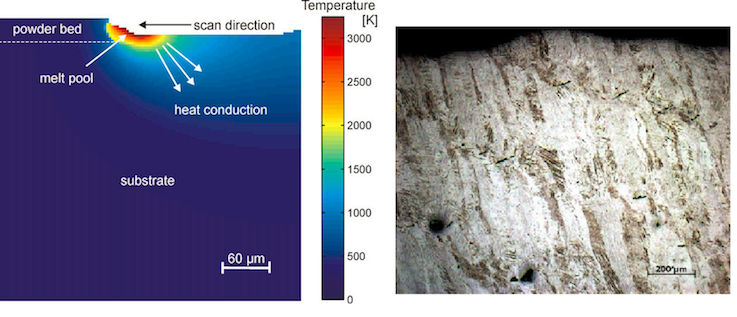}
\caption{(a) Temperature distribution profile in the xz-plane for a single Ti-6Al-4V powder layer on top of a Ti-6Al-4V substrate derived by means of a macroscopic SLM simulation model. (b) SEM micrograph (side view) of a Ti-6Al-4V reference sample processed with the same scan pattern, i.e. from right to left, indicating grain growth along the direction of highest temperature gradients,~\cite{Thijs2010}.}
\label{Fig_zy8}
\end{centering}
\end{figure}

Thijs et al.~\cite{Thijs2010} studied the microstructural evolution of Ti-6Al-4V during selective laser melting. Due to the high cooling rates and high-temperature gradients during the SLM process, an acicular martensitic phase, i.e. a hexagonal close-packed structure, was revealed in as-melted Ti-6Al-4V layers. They also found that the grains grew epitaxially, i.e. the material from a crystalline overlayer has a well-defined orientation with respect to the crystal structure of the underlying substrate, and observed elongated grains in the side and front views of the samples. It is interesting to note that the direction of the elongated grains depends on the local heat transfer condition, i.e. the grains are oriented towards the melt pool as shown in Figure~\ref{Fig_zy8}. Based on a simple thermal model, they indicated that the grain direction is parallel to the local conductive heat transfer given by the highest thermal gradient. Since the thermal field can be controlled by the scan strategy, also the resulting grain orientation can be influenced. Another interesting phenomenon occurring due to fast solidification during the SLM process is the segregation of aluminum. It is distinguishable as dark bands, which are believed to represent an intermetallic Ti-3Al phase. Further, when a higher energy density (e.g. due to higher laser power) is applied to the material, the volume of precipitates, i.e. second phases, will increase as consequence of lower cooling rates and longer diffusion times for the formation of second phases. Although the microstructure of Ti-6Al-4V is dominated by martensite due to rapid cooling rates, post heat treatment can transform the alpha prime martensite into alpha and beta phases. Owing to the prevalent thermal profiles, in particular through cycled heating, the SLM process can allow the formation of alpha and beta phases, and may form a graded microstructure in build direction. Recently, Xu et al.~\cite{Xu2015} observed the decomposition of alpha prime martensite into ultra-fine lamellar (alpha and beta) phases from the bottom to the top of a SLM-fabricated Ti-6Al-4V sample, which, in turn, might result in a decreasing material strength from the bottom to the top.\\


\begin{figure}[h!!!]
\begin{centering}
\includegraphics[width=0.9\textwidth]{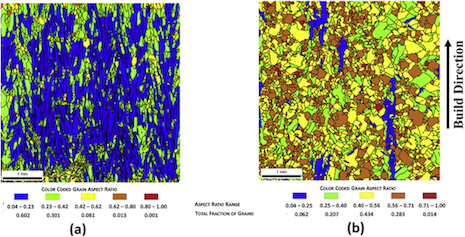}
\caption{Grain morphology of Inconel 718 samples built by EBM with (a) 10 mA and (b) 20 mA beam settings,~\cite{Raghavan2016}.}
\label{Fig_zy11}
\end{centering}
\end{figure}

Raghavan et al.~\cite{Raghavan2016} analyzed the microstructures resulting from EBM-processing of Inconel 718. As shown in Figure ~\ref{Fig_zy11}, the sample processed by means of a beam current of $10mA$ exhibits directional grains (columnar grains) along the build direction, while the one with an employed beam current of $20mA$ shows equiaxed grains. The reason for this observation is that a higher beam current induces a higher sample temperature (acting as local preheat temperature in the unmolten material regions), and therefore, the thermal gradient is lower, consequently resulting in equiaxed grains.\\

Dehoff et al.~\cite{Dehoff2015} demonstrate how to use computer-aided design to change the local crystal orientation. Three scanning strategies, relying on different beam powers and velocities, were applied in order to vary thermal gradients, cooling rates and solidification front velocities, which eventually allows to achieve tailored grain orientations. The working principle of this strategy has been proven by "writing" prescribed letters in a block of Inconel 718 such that these letters are represented by misoriented equiaxed grains while the surrounding bulk materials exhibits oriented [001] columnar grains. This methodology provides a fundamentally new design tool to tailor microstructures.\\

Also several classes of stainless steel alloys are of practical interest for SLM processing. Most steel parts manufactured by means of SLM satisfy typical requirements of general-purpose applications, especially in terms of high-strength and high-hardness, for example for mould and tool applications. Steel grades available for AM are typically austenitic stainless steels such as 304L and 316L. They commonly exhibit a completely austenitic microstructure after the AM process characterized by elongated and textured grains. For example, in 304L stainless steel, a typical microstructure with elongated grains in build direction has been observed. The precipitation of chromium carbides at grain boundaries, which would be observable for lower cooling rates prevalent in traditional processes such as casting, can typically be avoided by means of SLM. Similar observations have also been made for SLM-processed 316L stainless steel. Moreover, in~\cite{Garibaldi2016}, a gradient microstructure has been observed for SLM-manufactured 316L parts. Accordingly, an increasing distance from the build platform resulted in coarser microstructures due to lower cooling rates.

%
%
\section{Future directions and practical implementation}
\label{sec:practicalimplementationandquestions}
%
%

Indisputably, modeling and simulation approaches will play a key role in enabling SLM and EBM powder bed processes to build highly accurate complex parts, and to achieve stringent quality specifications.  Current implementations of macroscopic, mesoscopic, and microscopic models have shed further light on the underlying physical mechanisms of SLM and relations between the process parameters and the part characteristics, e.g. in terms of residual porosity, residual stresses or metallurgical microstructure, resulting on these different length scales. The long-term goal must be to combine these three model classes, based on a proper exchange of information, to yield an integrated modeling scheme that is capable of predicting final part characteristics on all of these three length scales (see also Figure~\ref{fig:sketch5}). Clearly, computational efficiency and robustness, not only of the sub-models but also of the coupling schemes, will play a key role in this regard. With such an integrated process model at hand, not only the prediction of process output for given input data, but also the inverse problem of determining locally optimal process parameters in order to optimize the resulting part characteristics, measured by a properly defined objective function and realized via methods of numerical optimization, is of highest practical interest (see e.g.~\cite{Calvello2004} for an overview on inverse analysis schemes).\\

The combination of integrated simulation and optimization methodologies with the capability of SLM to locally control temperature gradients and cooling rates opens the door for improved dimensional quality as well as engineered microstructures. This includes, but is not limited to, interest in metallurgical microstructure designs that are optimal with respect to prescribed requirements such as: inhomogeneous and anisotropic distributions of material strength and ductility, part density and porosity, material dissipation resulting in mechanical or acoustic damping, thermal or electrical conductivity, or internal flow/cooling capacity.\\

Of course, the predictive accuracy of the respective modeling approaches is strongly related to their degree of abstraction. In this context, correlation of in situ metrology data to models is essential to build both accurate and computationally efficient representations of the process physics with a proper degree of abstraction. Furthermore, by employing real-time in situ measurements as a means of fault indentification, close-loop process control can be achieved~\cite{Everton2016}. Eventually, a comparison of real-time in situ data for relevant physical fields (e.g. the temperature field) with the desired/optimal evolutions of these fields as predicted by numerical simulation opens the door for a proper manipulation of the process input parameters according to the principles of control theory.\\

Both, simulations and experiments must be pursued to understand the fundamental limits to SLM processes.  For instance, the process and machine parameters, as well as the heating and cooling rates resulting from the laser-material interaction and boundary conditions of the melt pool, impose a range of accessible heating and cooling rates which in turn limit the potential for microstructure and stress control. Also, the ability to perform SLM with powders having a broader candidate set of shape and size distributions is of practical interest, to increase the range of processible materials and geometrical resolutions and to access powder production techniques that have lower intrinsic cost. Currently, two main approaches of powder manufacturing are typically applied: Water-atomized powder is favorable in terms of lower production costs. However, water-atomized powder has irregular particle shape, and therefore typically leads to powder with poor flow and packing characteristics. Gas-atomized powder is more costly, yet is advantageous in producing spherical particles with higher purity~\cite{irrinki2016}. Consideration of powder purity and transient composition changes is critical, because oxidation on powder and substrate surfaces might considerably decrease the wetting behavior of the liquid melt, possibly leading to melt pool balling and reduced surface quality, defects such as pores and inclusions, and reduced layer-to-layer adhesion, which might induce delamination~\cite{Das2003}. Furthermore, in practice, SLM powder must be kept dry since evaporation of water contents during the melt process induces undesirable recoil pressure and fosters surface oxidation. The water content is commonly reduced as far as possible by combined approaches of preheating and a controlled inert gas flow flushing out possible sources of contamination from the build chamber.\\

A further paramount need is to assist quality control and certification of the resulting parts and materials. First, the complexity of the SLM process and the underlying physical phenomena leads to a higher sensitivity of part quality with respect to process parameter variations than is the case for rather traditional manufacturing technologies such as casting or forging. Second, in combination with the limited practical experience in this comparatively new technology, the frequency of defective parts is considerably higher. Third, the types of defects as well as the resulting mechanisms of failure are specific for the physical phenomena governing the SLM process, and consequently, require sophisticated surface and bulk metrology such as X-ray tomography. In this context, in-situ measurement and recording of certain physical fields, e.g. the temperature field, might be useful in locating different types of defects within SLM parts.\\

The success of SLM as manufacturing technology is of course also strongly determined by economic aspects. Currently, the limited understanding of underlying physical phenomena, leads to strong restrictions on the SLM process in form of tight process windows. For example, a maximum scan velocity may not be exceeded in order to avoid melt pool instabilities resulting in high surface roughness and possible porosity of the final part. Further, excessive energy input as consequence of high power densities can lead to overheating and decreased surface quality, high residual stresses, and crack propagation. However, for high process throughput, increased scan velocities, beam spot sizes and power densities would be desirable. In this regard, the study of novel scanning strategies, laser configurations, and beam shaping approaches seem to be very promising in order to find regimes of stable processing yet higher production throughput. A rough estimate for the lower energy bound of SLM per layer is given by the amount of energy required to heat and melt the powder material of one layer due to its heat capacity. The amount of energy required for the actual manufacturing process is considerably higher because substantial parts of neighboring tracks and underlying layers have to be remelted in order to provide good adhesion, and because of various thermal losses, e.g. due to emission, convection and evaporation on the powder bed surface as well as heat conduction across the build platform. Energy consumption is insignificant to the overall cost of SLM at present, yet will become critical as the other cost drivers, namely material and equipment cost, improve.  Besides the pure energy balance, the avoidance of evaporation is also essential in terms of material savings, melt track quality and avoidance of machine and instrument contamination due to condensation of metallic vapor, which is especially undesirable when remelted during the processing of different alloys in the same machine. Conductive and emissive losses have to be minimized via proper thermal insulation and shielding of build platform and machine casing, while research in the field of powder material production may yield improved grain surface properties in terms of heat conduction and radiation absorption. All in all, it is certain that SLM and related AM technologies present a rich set of multidisciplinary scientific and practical challenges, along with opportunity to transform the design of advanced products and manufacturing systems.

\section*{Acknowledgements}
\label{sec:Acknowledgements}

Financial support for preparation of this article was provided by the German Academic Exchange Service (DAAD) to C. Meier; Honeywell Federal Manufacturing \& Technologies to R. Penny and A.J. Hart; the Swiss National Science Foundation Early Postdoctoral Mobility Fellowship to Y. Zou (Grant: P2EZP2 165278), and United States Navy and the Engineering Duty Officer program to J.S. Gibbs. This work has been partially funded by Honeywell Federal Manufacturing \& Technologies, LLC which manages and operates the Department of Energy's Kansas City National Security Campus under Contract No. DE-NA-0002839.  The United States Government retains and the publisher, by accepting the article for publication, acknowledges that the United States Government retains a nonexclusive, paid up, irrevocable, world-wide license to publish or reproduce the published form of this manuscript, or allow others to do so, for the United States Government purposes.


\section*{Nomenclature}

\subsubsection*{Abbreviations}
\begin{center}
\begin{tabular}{p{0.12\textwidth}p{0.8\textwidth}}
$AM$    & Additive Manufacturing\\
$CA$    & Cellular Automation\\
$CAD$   & Comuter Aided Design\\
$DEM$   & Discrete Element Method\\
$DIC$   & Digital Image Correlation\\
$EBM$   & Electron Beam Melting\\
$EBW$   & Electron Beam Welding\\
$FDM$   & Finite Difference Method\\
$FEM$   & Finite Element Method\\
$FSI$   & Fluid Structure Interaction\\
$FVM$   & Finite Volume Method\\
$HAZ$   & Heat Affected Zone\\
$LBW$   & Laser Beam Welding\\
$LWIR$  & Long Wave Infrared\\
$NIR$   & Near Infrared\\
$PFM$   & Phase Field Method\\
$RTE$   & Radiation Transport Equation\\
$SLM$   & Selective Laser Melting\\
$SLS$   & Selective Laser Sintering\\
$VHN$   & Vickers Hardness Number\\
$VOF$   & Volume of Fluid Method\\
$XFEM$  & Extended Finite Element Method
\end{tabular}
\end{center}

\subsubsection*{SLM process}
\begin{center}
\begin{tabular}{p{0.12\textwidth}p{0.8\textwidth}}
$L / l$  		& Powder bed thickness of process material\\
$R$  		& Radius of spherical particle\\
$P$  		& Laser power\\
$V$  		& Laser velocity\\
$h$  		& Hatch spacing\\
$r_B$  		& Laser beam radius\\
$(.)*$		& Associated dimensionless quantity\\
$L_v$  		& Powder bed dimension in scan direction\\
$L_h$  		& Powder bed dimension in hatch direction\\
$A$  		& Factor describing fraction of effectively absorbed laser energy
\end{tabular}
\end{center}

\subsubsection*{Mechanical problem}
\begin{center}
\begin{tabular}{p{0.12\textwidth}p{0.8\textwidth}}
$\boldsymbol{\sigma}$  		& Cauchy stress tensor\\
$\mb{b}$  		& Vector of volume forces\\
$\mb{u}$  		& Displacement vector\\
$\mb{\epsilon}$  		& Solid strain tensor\\
$\tilde{\mb{\epsilon}}$  		& Rate-of-deformation tensor\\
$\gamma$			  & Surface tension\\
$\theta_0$								   & Wetting angle\\
$\kappa_I$								   & Interface curvature\\
$p$								   & Pressure in melt pool\\
$p_{ref}$								   & Pressure of ambient gas\\
$\lambda_{ev}$								   & Evaporation energy per particle\\
$A_s$								   & Sticking coefficient\\
$R_g$								   & Gas constant\\
$M$								   & Molar mass
\end{tabular}
\end{center}

\subsubsection*{Thermal problem}
\begin{center}
\begin{tabular}{p{0.12\textwidth}p{0.8\textwidth}}
$T$								           & Temperature of processes material\\
$T_{ref}$								    & Temperature of ambient atmosphere\\
$T_{0}$								    & Initial temperature\\
$T_{m}$								    & Melting temperature\\
$T_{b}$								    & Boiling temperature\\
$T_{l}$								    & Liquidus temperature\\
$T_{s}$								    & Solidus temperature\\
$\boldsymbol{\Omega}$			& Unit vector describing direction of radiation transfer\\
$\mb{x}$									& Spatial position vector\\
$I$									       & Radiation intensity\\
$\sigma$								   & Scattering coefficient\\
$\kappa$								   & Absorption coefficient\\
$\beta$								   & Extinction coefficient\\
$\omega$								   & Albedo\\
$S_c$								          & Scattering phase function\\
$k_{SB}$								   & Stefan-Boltzmann constant\\
$\mb{q}$									& General heat flux density\\
$\mb{q}_r$									& Radiation heat flux density\\
$\mb{q}_e$									& Emissive heat flux density\\
$\mb{q}_v$									& Evaporation heat flux density\\
$\mb{q}_{\lambda}$			  & Conductive heat flux density\\
$\mb{q}_{c}$			  & Convective heat flux density\\
$u_s$									& Incident energy density\\
$\alpha_S$								   & Perpendicular component of polarized absorptivity\\
$\alpha_P$								   & Parallel component of polarized absorptivity\\
$n$								              & Complex refraction index\\
$\theta$								        & Angle of radiation incidence\\
$\mb{k}$								        & Direction of electro-magnetic wave propagation\\
$w$								             & wave length of electro-magnetic wave\\
$\mb{e}$								        & Direction of electric field\\
$\mb{n}$								        & Surface normal vector / general unit vector\\
$\lambda / k$								  & Thermal conductivity\\
$\lambda_e$								  & Effective thermal conductivity\\
$a$								            & Thermal diffusivity\\
$\alpha$								            & General absorptivity\\
$\alpha_0$								            & Flat surface absorptivity\\
$\mb{H}_m$								  & Latent heat of melting\\
${\Phi}$											& Powder density\\
$n$											& Coordination number\\
$r_c$											& Radius of spherical contact area\\
$x$											& Dimensionless contact size\\
$\rho$								   & Density\\
$c_p$								   & Specific heat at constant pressure\\
$t$								      & Time\\
$\mb{v}$								   & Velocity vector\\
$\mb{k}$								   & Thermal conductivity tensor\\
$\Omega$			  & Problem domain\\
$\Gamma$			  & Problem boundary\\
$c$			  & Convection coefficient
\end{tabular}
\end{center}

\begin{center}
\begin{tabular}{p{0.12\textwidth}p{0.8\textwidth}}
$\epsilon$			  & Radiation emissivity\\
${\phi}$											& Phase field\\
$c$											& Concentration\\
$\boldsymbol{\Psi}$			& Rotation vector describing grain orientation\\
$\Pi$			& Free energy\\
$\tilde{\Pi}$			& Volume-specific free energy\\
$c_{\phi}$			& Boundary layer thickness parameter in phase field method
\end{tabular}
\end{center}

%
\bibliographystyle{abbrv}
\bibliography{review_heat_transfer_final_09012017.bib}
%
%
\end{document}